\documentclass[fleqn,usenatbib]{mnras}

\usepackage{newtxtext,newtxmath}

\usepackage[T1]{fontenc}

\DeclareRobustCommand{\VAN}[3]{#2}
\let\VANthebibliography\thebibliography
\def\thebibliography{\DeclareRobustCommand{\VAN}[3]{##3}\VANthebibliography}

\usepackage{graphicx}	
\usepackage{amsmath}	
\usepackage[dvipsnames]{xcolor}
\usepackage{tabularx}
\usepackage{subcaption}
\usepackage[nameinlink,capitalize]{cleveref}
\usepackage{enumitem}
\usepackage{siunitx}
\newcommand{\ifm}[1]{\relax\ifmmode#1\else$\mathsurround=0pt #1$\fi}

\newcommand{\software}{\textsc}
\newcommand{\run}{\texttt}

\DeclareSIUnit{\Msun}{{\rm M}_{\sun}}
\DeclareSIUnit{\Lsun}{{\rm L}_{\sun}}
\DeclareSIUnit{\kms}{{\rm km}\,{\rm s}^{-1}}
\DeclareSIUnit{\Mpc}{{\rm Mpc}}
\DeclareSIUnit{\kpc}{{\rm kpc}}
\DeclareSIUnit{\pc}{{\rm pc}}
\DeclareSIUnit{\Gyr}{{\rm Gyr}}
\DeclareSIUnit{\Myr}{{\rm Myr}}
\DeclareSIUnit{\yr}{{\rm yr}}
\DeclareSIUnit{\K}{{\rm K}}
\DeclareSIUnit{\Mpcc}{{\rm Mpc}^{-3}}
\DeclareSIUnit{\cmc}{{\rm cm}^{-3}}
\DeclareSIUnit{\cms}{{\rm cm}^{-2}}
\DeclareSIUnit{\ergs}{{\rm erg}\,{\rm s}^{-1}}
\DeclareSIUnit{\sy}{M_{\sun}\,{\rm yr}^{-1}}
\DeclareSIUnit{\skpc}{M_{\sun}\,{\rm kpc}^{-1}}

\DeclareSIUnit{\hmpc}{\,\ifm{h^{-1}}{\rm Mpc}}
\DeclareSIUnit{\hkpc}{\,\ifm{h^{-1}}{\rm kpc}}

\newcommand{\Omegax}[1]{\Omega_{\rm #1}}

\newcommand{\dotM}[1]{\dot{M}_{\rm #1}}
\newcommand{\Mx}[1]{M_{\rm #1}}

\newcommand{\tx}[1]{t_{\rm #1}}
\newcommand{\Rx}[1]{R_{\rm #1}}

\newcommand{\vx}[1]{v_{\rm #1}}

\newcommand{\nx}[1]{n_{\rm #1}}
\newcommand{\fx}[1]{f_{\rm #1}}

\newcommand{\Halpha}{H$\alpha$}

\newcommand{\zyn}[1]{{\color{black} #1}}

\title[Size Evolution at Cosmic Dawn]{The \software{MAGE} Project: Size Evolution and Disk Formation in Massive Galaxies at Cosmic Dawn}

\author[Yao et al.]{
Zhiyuan Yao$^{1}$\thanks{E-mail: zhiyuan.yao@mail.huji.ac.il},
Avishai Dekel$^{1,2}$,
Romain Teyssier$^{3}$,
Rachel S. Somerville$^{4}$,
Nir Mandelker$^{1}$,
\newauthor
Zachary Andalman$^{3}$,
Ulrich P. Steinwandel$^{4,5}$,
Zhaozhou Li$^{6,7}$
\\
$^{1}$Racah Institute of Physics, The Hebrew University, Jerusalem 91904, Israel\\
$^{2}$SCIPP, University of California, Santa Cruz, CA 95064, USA\\
$^{3}$Department of Astrophysical Sciences, Princeton University, 4 Ivy Lane, 08540, Princeton, NJ, USA\\
$^{4}$Center for Computational Astrophysics, Flatiron Institute, New York, NY 10010, USA\\
$^{5}$Max-Planck-Institut für Astrophysik, Karl-Schwarzschild-Straße 1, D-85741 Garching, Germany\\
$^{6}$School of Astronomy and Space Science, Nanjing University, Nanjing 210093, People’s Republic of China\\
$^{7}$Key Laboratory of Modern Astronomy and Astrophysics, Nanjing University, Ministry of Education, Nanjing 210093, People’s Republic of China
}
\date{Accepted XXX. Received YYY; in original form ZZZ}

\pubyear{\the\year}

\begin{document}
\label{firstpage}
\pagerange{\pageref{firstpage}--\pageref{lastpage}}
\maketitle

\begin{abstract}
We investigate size evolution and disk formation with a new suite of eleven cosmological zoom-in simulations, Massive Assembly of Galaxies in the Early universe (\software{MAGE}), run with \software{RAMSES} at \ifm{\sim\SI{10}{pc}} resolution until \ifm{z\simeq9}. The target galaxies span stellar masses of \ifm{\Mx{\star}\simeq2\times10^8\text{--}3\times10^{10}\,\si{\Msun}} at \ifm{z=9}. The stellar half-mass radius grows steeply with stellar mass at low masses, turns over near \ifm{\Mx{\star}\simeq\SI{5e9}{\Msun}}, and evolves weakly with redshift. The intrinsic UV sizes in MAGE overlap the compact end of observed distributions, including the compact strong-line systems, but not the more extended population at $z>9$. Dust, PSF-limited size measurements, and our overdense environments may contribute to this offset. In the simulations, stars dominate the mass within the half-mass radius above \ifm{\Mx{\star}\simeq\SI{5e8}{\Msun}}, where the central circular speed exceeds a characteristic stellar-feedback scale. Outflows are then increasingly retained within the galaxy, consistent with gas recycling on galactic rather than halo scales, which sustains central gas supply and drives a compaction-like central buildup. The turnover in size at \ifm{\Mx{\star}\simeq\SI{5e9}{\Msun}} coincides with rapid growth of rotational support to \ifm{v/\sigma\simeq1.5\text{--}3} in star-forming gas and stars younger than \SI{10}{Myr}, as the central circular speed exceeds the halo virial speed, which the star-forming gas velocity dispersion tracks. The subsequent low central gas fraction, declining central star formation, and younger outskirts suggest the onset of inside-out quenching.
\end{abstract}



\section{Introduction}
\label{sec:intro}

The James Webb Space Telescope (JWST) has made it possible to measure the sizes and internal structure of galaxies during the first few hundred million years of cosmic history. Galaxy sizes encode how stellar mass is distributed within the gravitational potential \citep[for reviews, see][]{Conselice.14,Naab.Ostriker.17}, and thus constrain the angular momentum of the accreted gas \citep{Fall.Efstathiou.80,Mo.etal.98,Dutton.vandenBosch.12,Somerville.etal.18}, the spatial distribution of star formation \citep{Nelson.etal.16}, and the strength of feedback \citep{Sales.etal.10,El-Badry.etal.16}.

At fixed luminosity or stellar mass, galaxies become progressively smaller towards higher redshift, with effective radii scaling roughly as \ifm{(1+z)^{-1}} over \ifm{z\sim0\text{--}10} \citep{Shibuya.etal.15,Yang.etal.25}. JWST has extended these measurements to \ifm{z\gtrsim9}, where galaxies have UV half-light radii of a few hundred parsecs \citep[][hereafter M24]{Ono.etal.23,Ono.etal.25,Morishita.etal.24}. Over \ifm{5<z<14}, however, M24 find a much slower average size evolution, \ifm{\Rx{e}\propto(1+z)^{-0.4\pm0.2}}, suggesting that the decline in size flattens at cosmic dawn. This average conceals a compact population: \ifm{\sim13} per cent of their sample is only marginally resolved in NIRCam imaging, with \ifm{\Rx{e}\lesssim\SI{100}{pc}} and star-formation rate (SFR) surface densities of \ifm{\Sigma_{\rm SFR}>10\,\si{\Msun\,yr^{-1}\,kpc^{-2}}}, a regime occupied by fewer than 0.01 per cent of local star-forming galaxies.

Subsequent spectroscopic studies have examined this compact population more systematically and suggest that the size distribution of luminous galaxies at \ifm{z>10} may be bimodal \citep{Harikane.etal.25,Roberts-Borsani.etal.26}. The most compact of these sources preferentially show strong high-ionization UV lines, more negative UV continuum slope \ifm{\beta}, and high SFR surface densities \citep{Harikane.etal.25,Roberts-Borsani.etal.26}. Rather than treating them as a distinct class, \citet{Roberts-Borsani.etal.26} interpret these compact strong-line sources as a short-lived phase near the peak of a central starburst along a common evolutionary pathway, during which galaxies temporarily occupy the compact end of the size distribution. 

The size--mass relation offers a complementary constraint on structural evolution. M24 find a positive UV size--mass relation at \ifm{5<z<14}, with the compact systems at \ifm{\Mx{\star}\lesssim\SI{1e9}{\Msun}} forming a low-mass tail that lies \ifm{\gtrsim1.5\sigma} below the main relation. These compact sources are more prevalent in their lensing-cluster field, where magnification aids the detection of small, low-mass galaxies. Lensed faint galaxies can be compact down to tens of parsecs, comparable to individual star-cluster complexes \citep{Bouwens.etal.17,Adamo.etal.24,Mowla.etal.24}, and their size--luminosity relation is much steeper than the normal one \citep{Bouwens.etal.17}. In the rest-frame optical, JWST samples show a positive size--mass relation at \ifm{2<z<8} with a non-evolving slope of \ifm{\simeq0.2\text{--}0.25} \citep{Miller.etal.25,Yang.etal.25}. In addition, a population of blue, barely resolved sources accounts for roughly 10 per cent of the sample at \ifm{z\gtrsim5\text{--}6} \citep{Miller.etal.25,Yang.etal.25}. At the high-mass end, \citet{ChenLongyue.etal.26} find that the optical size--mass relation flattens above a nearly redshift-independent stellar mass of \ifm{\sim\SI{1e10}{\Msun}} over \ifm{0.5<z<6}, and the bending becomes progressively stronger at higher redshifts. 

Cosmological simulations diverge in their predicted size--mass relation at high redshift, as well as in their treatment of compact objects at the low-mass end. A positive stellar half-mass radius--mass relation has been found in \software{FIRE-2} at \ifm{z=5\text{--}10} \citep{Ma.etal.18}, in the \software{SERRA} galaxies at \ifm{z\simeq7.7} \citep{Pallottini.etal.22}, and in \software{THESAN-ZOOM} at \ifm{z=3\text{--}13} \citep{McClymont.etal.25}. Of these, the corresponding UV half-light size--mass \zyn{relations} in \software{FIRE-2} and \software{THESAN-ZOOM} are closer to the observed one for typical galaxies, while \software{SERRA} agrees better with lensed galaxies at \ifm{6\leq z\leq8}. Moreover, many other simulations have difficulty reproducing a positive size--mass relation at high redshift, let alone the compact galaxies at the low-mass end. A negative stellar half-mass radius--mass relation above \ifm{\Mx{\star}\simeq\SI{5e8}{\Msun}} has been reported in \software{IllustrisTNG} \citep{Genel.etal.18}, \software{BLUETIDES} \citep{Marshall.etal.22}, \software{FLARES} \citep{Roper.etal.22}, \software{THESAN-1} \citep{Shen.etal.24}, and \software{FirstLight} \citep{Cataldi.etal.25}. These studies commonly attribute the negative relation to wet compaction at high redshift, although a positive UV half-light radius--mass relation can be recovered once dust effects are included, as in \software{THESAN-1} \citep{Shen.etal.24} and \software{FirstLight} \citep{Ceverino.etal.26}.

Wet compaction connects gas inflow, central star formation, and structural evolution, and is widely used to explain the negative trend in the size--mass relation at high redshift. \citet{Dekel.Burkert.14} first proposed that gas-rich disks can contract into compact star-forming ``blue nuggets'' when the inflow time is shorter than the star-formation time, allowing gas to reach the centre before being converted into stars. Cosmological simulations by \citet{Zolotov.etal.15} trace this dissipative buildup and the subsequent central gas depletion and inside-out quenching, with lower-mass systems experiencing repeated compaction episodes before sustained quenching. \citet{Tacchella.etal.16} further show how cycles of gas compaction, depletion, and replenishment produce excursions around the star-forming main sequence. \citet{Dekel.etal.19} link the preferred mass scale of major compaction to the combined effects of supernova feedback in low-mass haloes and virial shock heating in more massive haloes. \citet{Lapiner.etal.23} identify major compaction events near \ifm{\Mx{\star}\sim\SI{1e10}{\Msun}}, accompanied by a transition from dark-matter to baryon dominance in the centre. These compaction events can be triggered by wet mergers, counter-rotating streams, violent disk instability, and gas recycling \citep{Zolotov.etal.15,Dekel.etal.19,Lapiner.etal.23}.

Compaction can promote disk formation and inside-out quenching. In cosmological simulations, gas depletion after the central starburst suppresses star formation in the compact core, while an extended star-forming disk can grow around it \citep{Zolotov.etal.15,Tacchella.etal.16,Lapiner.etal.23}. In \software{TNG50}, the compact centres that compaction builds in massive galaxies are themselves flattened and rotation supported, resembling nuclear disks and bars rather than classical bulges, and extended disks assemble around them before the centres become dispersion dominated at \ifm{z\lesssim1} \citep{Luo.etal.26}. Controlled simulations and analytical models suggest that a concentrated central potential can help gas settle into coherent rotation \citep{Hopkins.etal.23,Semenov.26}. These results suggest that size evolution, disk formation, and inside-out quenching may be coupled through central stellar buildup and the subsequent redistribution of star formation.

Observations also provide evidence for a connection between compact central structures and rotation at early epochs. \citet{Baker.etal.25} find a dense core surrounded by an extended star-forming component at \ifm{z=7.43}, with a radial increase in specific SFR indicating inside-out growth. Resolved gas kinematics establish a dynamically cold disk in REBELS-25 at \ifm{z=7.31} \citep{Rowland.etal.24,Rowland.etal.26}, while the Cosmic Grapes galaxy at \ifm{z=6.072} shows that numerous star-forming clumps can coexist with a rotating gas disk \citep{Fujimoto.etal.25}. At \ifm{z=10.6}, recent spectroscopy of GN-z11 supports a substantial contribution from young massive stars to its compact emission \citep{ChenZuyi.etal.26,Nakane.Ouchi.26}, and its kinematics show tentative rotational support with \ifm{v/\sigma\simeq3} \citep{Xu.etal.24}. \zyn{At \ifm{z=14.18}, JADES-GS-z14-0 also shows a tentative velocity gradient in [\ion{O}{iii}] \SI{88}{\micro\metre} emission \citep{Scholtz.etal.25}.} More indirectly, in a photometric sample of 169 candidates at \ifm{z\sim10\text{--}16}, \citet{Ono.etal.25} find a log-normal size distribution, a nearly uniform distribution of projected axis ratios over \ifm{q\simeq0.2\text{--}0.8}, approximately exponential profiles with S\'{e}rsic indices close to unity, and a characteristic size-to-halo-radius ratio of \ifm{\Rx{e}/\Rx{v}\simeq0.015}, similar to the nearly constant ratio of \ifm{\simeq0.01\text{--}0.035} found over \ifm{z\sim0\text{--}8} \citep{Shibuya.etal.15,Yang.etal.25}. These properties are consistent with the classical disk-formation picture, in which disk sizes are set by the angular momentum that galaxies inherit from their haloes and therefore scale with the halo virial radius \citep{Fall.Efstathiou.80,Mo.etal.98}, although morphology alone does not establish rotational support. Together, these results motivate a study of central buildup and disk formation beyond \ifm{z=9}, where it remains an open question whether this connection is already in place.

In this paper, we introduce \software{MAGE} (Massive Assembly of Galaxies in the Early universe), a new suite of 11 cosmological zoom-in simulations run with the \software{RAMSES} adaptive mesh refinement (AMR) cosmological hydrodynamics code, to examine these connections over \ifm{9\lesssim z\lesssim15}. With \ifm{\sim\SI{10}{pc}} resolution and the star-formation and feedback model of \citet{Andalman.etal.25}, the simulations allow us to resolve compact galaxies more reliably and to study their connection to disk formation at \ifm{z>9}.

This paper is organized as follows. \Cref{sec:methods} describes the simulations and analysis methods. \Cref{sec:sizeevo} presents the size evolution of representative galaxies and the statistical size--mass and size--magnitude relations. \Cref{sec:in-situsf} examines in-situ mass fractions and the spatial distributions of stellar ages and star formation. \Cref{sec:diskformation} connects the emergence of rotational support to central mass buildup and gas flows. \Cref{sec:discuss} discusses wet compaction, the conditional disk-formation diagnostic, feedback dependence, and comparisons with simulations and observations. We summarize the results in \cref{sec:summary}.

\begin{table*}
    \centering
    \caption{Summary of the \software{MAGE} sample. Name: ``MAGE'' + ID, ordered by decreasing halo mass at \ifm{z=9} with increasing ID; $L_\mathrm{box}$: box size of the cosmological simulation; $m_\mathrm{dm,min}$: minimum dark matter particle mass in the zoom-in region; $m_\mathrm{b,min}$: minimum baryon mass in the zoom-in region; $\Delta x_\mathrm{min}$: minimum cell size in the zoom-in region. The following properties are measured at \ifm{z=9}. $\Mx{v}$: virial mass of the halo; \ifm{\Mx{\star}}: total stellar mass within the aperture \ifm{\Rx{gal}\equiv0.15\,\Rx{v}}; \ifm{\Mx{g}}: total gas mass within \ifm{\Rx{gal}}; \ifm{\Rx{v}}: virial radius of the halo; \ifm{\Rx{e}}: mean projected 2D UV half-light radius of the galaxy measured along three orthogonal directions; \ifm{\Rx{1/2}}: 3D half-mass radius of the galaxy within \ifm{\Rx{gal}}; $\mathrm{SFR}_{50}$: SFR within \ifm{\Rx{gal}} over the past 50 Myr; $(v/\sigma)_\mathrm{SFgas}$: rotation-to-dispersion ratio for the star-forming gas weighted by SFR; $(v/\sigma)_{\star,10}$: rotation-to-dispersion ratio for the stars younger than 10 Myr; $(v/\sigma)_{\star}$: rotation-to-dispersion ratio for the full stellar population. The slight differences in \ifm{m_\mathrm{dm,min}}, \ifm{m_\mathrm{b,min}}, and $\Delta x_\mathrm{min}$ between \run{MAGE04} and \run{MAGE07} (\zyn{re-simulated from the initial conditions of the fiducial model in \citet{Andalman.etal.25}}) arise from the use of different cosmological parameters.}
    \label{tab:sims}
    \small
    \setlength{\tabcolsep}{2.5pt}
    \begin{tabular*}{\textwidth}{@{\extracolsep{\fill}}l*{14}{r}@{}}
        \hline
        Name & $L_\mathrm{box}$ & $m_\mathrm{dm,min}$  & $m_\mathrm{b,min}$ & $\Delta x_\mathrm{min}$
             & $M_\mathrm{v}$   & $M_\star$   & $M_\mathrm{g}$  & $\Rx{v}$ & $\Rx{e}$ & $\Rx{1/2}$
             & $\mathrm{SFR}_{50}$  & $(v/\sigma)_\mathrm{SFgas}$ & $(v/\sigma)_{\star,10}$ & $(v/\sigma)_{\star}$  \\
        & {\scriptsize [$h_0^{-1}\,\mathrm{cMpc}$]}
        & {\scriptsize [\ifm{\SI{1e4}{\Msun}}]}
        & {\scriptsize [\ifm{\SI{1e3}{\Msun}}]}
        & {\scriptsize [pc]}
        & {\scriptsize [\ifm{\SI{1e11}{\Msun}}]}
        & {\scriptsize [\ifm{\SI{1e9}{\Msun}}]}
        & {\scriptsize [\ifm{\SI{1e9}{\Msun}}]}
        & {\scriptsize [kpc]}
        & {\scriptsize [kpc]}
        & {\scriptsize [kpc]}
        & {\scriptsize [\ifm{\si{\Msun\,yr^{-1}}}]}
        &
        &
        & \\
        \hline

        \run{MAGE01}  & 250 & 4.72 & 8.80 & 8.80  & 6.36 & 31.56 & 7.80 & 26.70 & 0.16 & 0.22 & 209.47 & 1.35 & 3.52 & 0.80\\
        \run{MAGE02}  & 250 & 4.72 & 8.80 & 8.80  & 3.45 & 14.03 & 6.73 & 21.76 & 0.26 & 0.53 & 119.01 & 1.17 & 1.50 & 0.78\\
        \run{MAGE03}  & 250 & 4.72 & 8.80 & 8.80  & 3.14 & 13.12 & 5.74 & 21.14 & 0.23 & 0.33 & 104.69 & 3.50 & 3.20 & 1.00\\
        \run{MAGE04}  & 100 & 2.42 & 4.50 & 7.03  & 2.23 & 10.93 & 3.46 & 18.84 & 0.21 & 0.27 & 62.10  & 1.70 & 2.50 & 1.10\\
        \run{MAGE05}  & 250 & 4.72 & 8.80 & 8.80  & 2.10 & 3.46  & 2.59 & 18.45 & 0.19 & 0.31 & 31.38  & 1.08 & 1.37 & 0.56\\
        \run{MAGE06}  & 250 & 4.72 & 8.80 & 8.80  & 1.33 & 2.92  & 2.48 & 15.89 & 0.22 & 0.34 & 18.16  & 2.00 & 2.40 & 1.59\\
        \run{MAGE07}  & 100 & 2.04 & 4.36 & 6.81  & 0.95 & 3.67  & 3.13 & 14.17 & 0.18 & 0.24 & 45.90  & 1.26 & 1.69 & 1.16\\
        \run{MAGE08}  & 250 & 4.72 & 8.80 & 8.80  & 0.44 & 0.99  & 0.61 & 11.08 & 0.04 & 0.10 & 2.19   & 2.06 & 1.13 & 0.46\\
        \run{MAGE09}  & 250 & 4.72 & 8.80 & 8.80  & 0.36 & 0.38  & 0.39 & 10.28 & 0.11 & 0.21 & 2.83   & 0.87 & 1.60 & 0.64\\
        \run{MAGE10}  & 250 & 4.72 & 8.80 & 8.80  & 0.27 & 0.50  & 0.35 & 9.47  & 0.04 & 0.06 & 2.31   & 0.47 & 0.69 & 0.83\\
        \run{MAGE11}  & 250 & 4.72 & 8.80 & 8.80  & 0.27 & 0.23  & 0.07 & 9.42  & 0.02 & 0.03 & 0.89   & 0.57 & 1.25 & 0.42\\
        \hline
    \end{tabular*}
\end{table*}

\section{Methods}
\label{sec:methods}

\subsection{Numerical simulations}
The \software{MAGE} suite comprises 11 cosmological zoom-in simulations of massive haloes evolved to $z\simeq9$ with \software{RAMSES}. The suite extends the fiducial simulation of \citet{Andalman.etal.25} using the same star-formation and feedback prescriptions. We summarize the numerical methods here and refer to that work for implementation details.

We assume a \textit{Planck} cosmology \citep[\ifm{\Omegax{m}=0.308, \Omegax{\Lambda}=0.692, \Omegax{b}=0.0484, H_0=67.81\,\si{km\,s^{-1}\,Mpc^{-1}};}][]{Planck.16}, except for \run{MAGE07}, \zyn{re-simulated from the initial conditions of the fiducial run of \citet{Andalman.etal.25}}, which adopts slightly different cosmological parameters \ifm{\Omegax{m}=0.276, \Omegax{\Lambda}=0.724, \Omegax{b}=0.049, H_0=70\,\si{km\,s^{-1}\,Mpc^{-1}}}. The target haloes are selected from large parent boxes of size \SI{100}{\ifm{h_0^{-1}}cMpc} and \SI{250}{\ifm{h_0^{-1}}cMpc} (\ifm{h_0\equiv H_0/(100\,\si{km\,s^{-1}\,Mpc^{-1}})}), chosen to host haloes with \ifm{\Mx{v} > \SI{1e10}{\Msun}} by \ifm{z = 9}. To construct zoom-in initial conditions, we trace the dark matter particles within three times the virial radius of each target halo back to their positions at \ifm{z \simeq 124}, the starting redshift of the parent boxes, and define the convex hull enclosing these particles as the zoom region using \software{MUSIC} \citep{Hahn.Abel.11}. The zoom-in initial conditions are generated at \ifm{z=50}, and the zoom-in simulations are started from that epoch. The zoom-in simulations are run with the AMR code \software{RAMSES} \citep{Teyssier.02}, which solves the Euler equations with a second-order Godunov scheme, the HLLC Riemann solver, and MUSCL--Hancock reconstruction. Dark matter and star particles are evolved in a Lagrangian framework, while gas is modelled on an Eulerian grid that refines dynamically according to mass thresholds and local Jeans criteria. This approach yields an effective physical resolution of \ifm{\sim \SI{10}{pc}} at the highest refinement level throughout the simulations. The main properties of each zoom-in halo are summarized in \cref{tab:sims}.

To capture unresolved turbulent motions, we employ a Large Eddy Simulation (LES) framework that evolves the turbulent kinetic energy (TKE) using a sub-grid model based on \citet{Kretschmer.Teyssier.20}. The sub-grid model decomposes density into a volume-averaged component and fluctuations, while velocity and temperature are decomposed using mass-weighted (Favre) averages \citep{Schmidt.Federrath.11}. The sub-grid turbulence model tracks the evolution of the TKE with creation and dissipation source terms determined from the local velocity gradients and a dissipation time-scale set by the grid scale and turbulent velocity. Turbulent pressure contributes to supporting gas against collapse and is particularly relevant in regulating star-formation efficiency (SFE) in massive, rapidly accreting systems at high redshift.

Star formation proceeds according to a multi-freefall model \citep{Federrath.Klessen.12} as implemented in \software{RAMSES} by \citet{Kretschmer.Teyssier.20}. The local star-formation rate density follows the Schmidt law, $\dot{\rho}_* = \epsilon_{\rm ff}\rho / t_{\rm ff}$, with the star-formation efficiency per free-fall time $\epsilon_{\rm ff}$ depending on the local gas properties and sub-grid turbulence. In contrast to conventional fixed-efficiency models, this framework allows $\epsilon_{\rm ff}$ to vary dynamically in response to the turbulent state of the gas. The fiducial model assumes fully compressive turbulence forcing, which leads to elevated $\epsilon_{\rm ff}$ in dense regions.

Supernova (SN) feedback is implemented via a hybrid thermal and kinetic injection model. Each SN injects thermal energy locally and radial momentum through the neighbouring cell faces, with the momentum calibrated using cooling-radius scaling relations from \citet{Martizzi.etal.15}. The feedback is purely local and does not resolve the Sedov--Taylor expansion in detail, but approximates the net momentum transfer over unresolved stages of remnant evolution. The number of SNe per star particle is set by two parameters: the fraction of each star particle's mass in SN progenitors, 0.2, and the mean progenitor mass, \SI{10}{\Msun}. Each solar mass of stars formed therefore produces \ifm{0.2/10=0.02} SNe. Metal enrichment is included with a yield of \ifm{y=0.1}, so each SN returns 10 per cent of the progenitor mass, \SI{1}{\Msun}, as metals.

In addition to SN feedback, an early feedback model accounts for photoionization heating from massive stars. Gas in cells containing young star particles (age $< 20$~Myr) is heated to $T_{\rm phot} = 2 \times 10^4$~K to mimic the thermal pressure of ionized \ion{H}{ii} regions. This heating is implemented by setting the thermal energy corresponding to a sound speed of $c_\mathrm{s} \simeq 22$\,km\,s$^{-1}$, assuming an ideal monatomic gas. While this model neglects radiation pressure and stellar winds (see \cref{subsec:caveats}), it captures the rapid thermal response of star-forming gas to early stellar feedback.

Gas cooling is computed using equilibrium chemistry for hydrogen and helium \citep{Katz.etal.96}, with a metallicity-dependent cooling curve from \citet{Sutherland.Dopita.93}. A uniform, time-evolving UV background \citep{Haardt.Madau.96} is switched on at $z=10$, with self-shielding that suppresses it in gas denser than $n_{\rm H}\sim0.01\,\mathrm{cm^{-3}}$. It is therefore active only during the final $\simeq70$\,Myr of our analysis ($9\lesssim z<10$). We find no discontinuity at $z=10$ in the gas content of galaxies at fixed halo mass, and the background is expected to have little effect on the star-forming gas over this short interval. A temperature floor is imposed at the redshift-dependent cosmic microwave background (CMB) temperature as an approximate treatment of heating by CMB photons. The initial gas metallicity is set to $Z=10^{-3}Z_\odot$, so the simulations model enriched, Population II star formation without explicitly following Population III enrichment. Limitations of the sub-grid prescriptions are discussed by \citet{Andalman.etal.25}.

\subsection{Halo finders, merger trees, and galaxy properties}
\label{subsec:halofinder}
Halo catalogues and merger histories are constructed with the six-dimensional phase-space halo finder \software{ROCKSTAR} \citep{Behroozi.etal.13a}, coupled with the gravitationally consistent merger tree algorithm \software{CONSISTENT TREES} \citep{Behroozi.etal.13b}. The target halo is traced along the main branch of its merger tree, which extends to $z\sim20$--24.

Each simulation stores 27--49 outputs (426 in total) between the initial conditions at $z=50$ and the final output at $z\simeq9$. Only a few outputs are written above $z\simeq16$, whereas at $z\lesssim16$ they are spaced by a median of $\simeq10$\,Myr (16th--84th percentiles 6--13\,Myr). We analyse the target galaxies only from $z\sim15$--16 onwards: at earlier times the progenitors have masses of just $\sim10^{6}$--$10^{7}\,\mathrm{M}_\odot$ and the outputs are sparse, so halo and galaxy properties are not robust. This yields 20--39 analysed snapshots per target galaxy and 318 in total.

The halo catalogues are constructed with \software{ROCKSTAR} applied to the dark matter particles alone, so the virial mass and radius it returns describe only the dark matter component of the hydrodynamical simulations. We therefore recompute both quantities from the total matter distribution. Around the \software{ROCKSTAR} centre of each distinct (host) halo, we sum the enclosed mass of dark matter particles, star particles and gas cells as a function of radius. We define the virial radius \ifm{\Rx{v}} as the radius at which the mean enclosed total density first falls below $\Delta_{\rm vir}(z)\,\rho_{\rm crit}(z)$, where $\Delta_{\rm vir}$ follows \citet{Bryan.Normal.98}, the same virial definition adopted by \software{ROCKSTAR}. Because $\Omega_{\rm m}(z)\simeq1$ at the redshifts considered here, this threshold corresponds to $\simeq178$ times the mean matter density. The virial mass is then \ifm{\Mx{v}=(4\pi/3)\,\Delta_{\rm vir}\rho_{\rm crit}\Rx{v}^3}, i.e. all dark matter, stars and gas within \ifm{\Rx{v}}, so that the virial radius and mass are defined self-consistently.

The host and subhalo relationships are taken from \software{CONSISTENT TREES}. Merger mass ratios are defined as the ratio of the infalling halo mass to the host halo mass, evaluated at the final snapshot before the infalling halo becomes a subhalo. This minimizes the impact of subsequent tidal stripping.

For the statistical sample, the galaxy centre is determined with an iterative shrinking-sphere method \citep{Power.etal.03}. Starting from a sphere of radius $0.3\,R_\mathrm{v}$ centred on the halo centre returned by \software{ROCKSTAR}, we compute the mass-weighted centre of the enclosed star particles, recentre the sphere on this point, and shrink its radius by a factor of $0.7$ at each iteration. We repeat this procedure down to a radius of $0.02\,R_\mathrm{v}$, terminating early if fewer than $20$ star particles remain enclosed, and adopt the final mass-weighted stellar centre as the galaxy centre. This procedure reduces the influence of satellites and offsets between the stellar and halo centres. Visual inspection nevertheless reveals occasional miscentring, particularly during major mergers when multiple dense stellar components lie within \ifm{0.3\,\Rx{v}}. For the 11 target central galaxies, we visually inspect the centres and manually fix selected snapshots to follow the main stellar component consistently. These corrections affect only \ifm{\sim5} per cent of the snapshots, so we expect the statistical results to be largely unaffected.

We define the galaxy boundary as $R_\mathrm{gal} \equiv0.15\,R_\mathrm{v}$. Unless a different aperture is specified, stellar mass, gas mass, SFR, and the 3D stellar half-mass radius \ifm{\Rx{1/2}} are computed within this boundary at each snapshot. The enclosed mass, \ifm{M(<r)}, is the cumulative mass within a sphere of radius \ifm{r}. We define the circular speed as \ifm{\vx{c}(r)=\sqrt{GM(<r)/r}} and the halo virial speed as \ifm{\vx{v}=\sqrt{G\Mx{v}/\Rx{v}}}. Component circular speeds use the corresponding enclosed mass. We also tested a larger galaxy boundary of \ifm{R_\mathrm{gal}=0.3\,R_\mathrm{v}}. The derived stellar properties are largely insensitive to this choice, whereas the gas properties exhibit a markedly stronger dependence on the adopted boundary \citep[see also][]{Stevens.etal.14}.

Gas inflow and outflow rates across \ifm{\Rx{1/2}}, \ifm{\Rx{gal}}, and \ifm{\Rx{v}} are estimated by multiplying the volume average of \ifm{\rho v_r} in a spherical shell of thickness \ifm{\Delta R=0.1\,R} centred on each radius \ifm{R} by the corresponding spherical area, \ifm{4\pi R^{2}}. Inward- and outward-moving gas are treated separately, and both rates are plotted as positive magnitudes. Radial velocities are measured relative to the mass-weighted stellar bulk velocity within \ifm{\Rx{1/2}}. SFRs are estimated by summing the formed masses of stars younger than the adopted 5 or 50 Myr window within the present aperture and dividing by that window. If \ifm{m_i} is its current mass and \ifm{t_i} its age, we adopt \ifm{m_{{\rm form},i}=m_i/f_{\rm SN}(t_i)}, where
\begin{equation}
 f_{\rm SN}(t)=
 \begin{cases}
 1, & t\leq\SI{3}{Myr},\\
 1-0.2\,(t-\SI{3}{Myr})/(\SI{17}{Myr}), & \SI{3}{Myr}<t\leq\SI{20}{Myr},\\
 0.8, & t>\SI{20}{Myr}.
 \end{cases}
 \label{eq:photometric_mass_correction}
\end{equation}
is the  mean fraction of its initial mass that a particle retains under our SN feedback model, which returns 20 per cent of it to the gas at a constant rate between 3 and 20 Myr. This is the same estimate \software{RAMSES} uses to set each particle's SN rate. Because individual SNe are drawn stochastically, a particle's true initial mass differs from \ifm{m_{{\rm form},i}} by \ifm{\lesssim2} per cent, which is negligible once summed over a galaxy. The \SI{5}{Myr} average emphasizes recent bursts and overlaps the onset of SN feedback, whereas the \SI{50}{Myr} average smooths the burst--lull cycle. The two intervals also correspond to the time-scales probed by nebular emission lines and by the rest-frame UV continuum, respectively \citep{Kennicutt.Evans.12,FloresVelazquez.etal.21}, so their ratio is the simulation analogue of the H\ifm{\alpha}-to-UV burstiness indicator used at high redshift \citep{Faisst.Morishita.24,Dome.etal.24}. 

We classify stars by their birth positions relative to the main progenitor: a star is in situ if it formed within the galaxy aperture (\ifm{0.15\Rx{v}}) at its birth time, and ex situ otherwise. This radial criterion misclassifies stars formed in companion galaxies that have already entered the aperture during mergers. To correct for this, we identify spatially distinct, persistent star-forming companions within the aperture by applying the density-based spatial clustering of applications with noise (DBSCAN) algorithm \citep{Ester.etal.96} to the birth positions of stars formed in short ($\simeq2$\,Myr) time windows, short enough that a companion does not move appreciably within a window. The main galaxy is the cluster dominating the central few hundred parsecs, and stars born in other sufficiently massive clusters that persist over consecutive windows are reclassified as ex situ. Small star-forming clumps within the disk are thus not counted as companions. This step can only move stars from in situ to ex situ. Its effect is largest during mergers, but at $z=9$ it reduces the in-situ fraction by a median of 0.04 and changes the half-mass radius of the in-situ population by at most 10 per cent.

For each in-situ star we record its birth radius, \ifm{\Rx{birth}}, defined as the distance between its birth position and the centre of the main progenitor at its formation time, \ifm{\tx{birth}}, together with the properties of the main progenitor at that epoch: its stellar mass within \ifm{\Rx{gal}}, \ifm{\Mx{\star}(\tx{birth})}, its stellar half-mass radius, \ifm{\Rx{1/2}(\tx{birth})}, and its virial radius, \ifm{\Rx{v}(\tx{birth})}. Birth positions are stored by the simulation at the time of formation. Because the outputs are spaced by \ifm{\simeq\SI{10}{Myr}}, the progenitor centre and properties at \ifm{\tx{birth}} are linearly interpolated in time between the two bracketing outputs.

We employ all galaxies located within the zoom-in region for our statistical analysis, provided that their host haloes are distinct in the \software{CONSISTENT TREES} merger trees. Specifically, in addition to the principal target central galaxy of each zoom-in simulation, we include all galaxies that satisfy the following criteria:
\begin{enumerate}[label=(\roman*),itemindent=1em,align=left]
    \item \ifm{\log_{10}(\Mx{\star}/\si{\Msun})\gtrsim7.5};
    \item \ifm{9\lesssim z \lesssim15};
    \item central galaxies of distinct haloes;
    \item a fraction of high-resolution dark-matter particles within \ifm{\Rx{v}} such that \ifm{N_\mathrm{dm,highres}/N_\mathrm{dm,tot}\geq0.9};
    \item a fraction of the total stellar mass beyond half the galactic radius satisfying \ifm{f(R>\Rx{gal}/2)<0.25}.
\end{enumerate}
A central galaxy in criterion (iii) is the galaxy located within \ifm{\Rx{gal}=0.15\,\Rx{v}} of the centre of a distinct host halo in the merger trees. Galaxies hosted by subhaloes are treated as satellites and excluded. The high-resolution particles in criterion (iv) have \ifm{m_{\rm dm}<\SI{5e4}{\Msun}}. The stellar-mass threshold in criterion (i) selects systems with more than 3000 star particles within \ifm{\Rx{gal}}. Below this threshold, measured sizes approach the numerical grid scale and the gravitational softening length of star particles, which is linked to the local AMR grid in \software{RAMSES}. The threshold of 0.25 in criterion (v) is motivated by the conventional 1:4 stellar-mass ratio used to define major mergers \citep{Puskas.etal.25}. We adopt this value as an empirical cut on the outer stellar-mass fraction after testing different thresholds and inspecting their effectiveness in removing systems with multiple spatially separated stellar components. Our relatively large aperture, \ifm{\Rx{gal}=0.15\Rx{v}}, places \ifm{\Rx{gal}/2} well beyond the typical stellar half-mass radius, so the cut removes multi-component systems without penalizing extended single galaxies. Excluding criterion (v) only increases the upper scatter of \ifm{\Rx{1/2}} and does not significantly affect the median in any stellar-mass or redshift bin.

\begin{figure}
    \centering
    \includegraphics[width=0.97\columnwidth]{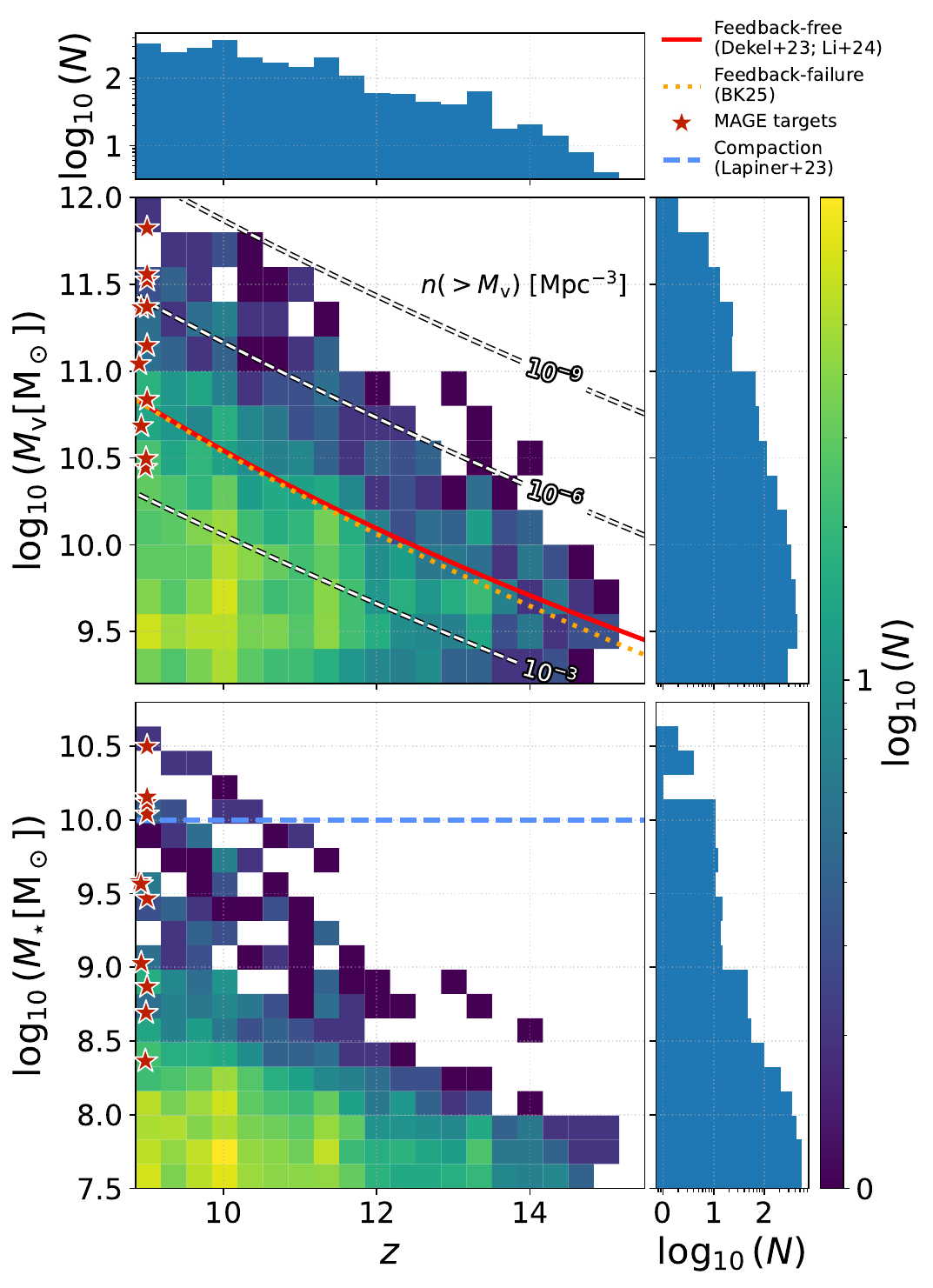}
    \caption{Distribution of the \software{MAGE} galaxy sample in redshift, halo mass, and stellar mass. The sample contains 2415 galaxy--snapshot measurements of central galaxies with \ifm{\Mx{\star}>10^{7.5}\,\si{\Msun}}, drawn from all snapshots of the 11 \software{MAGE} zoom-in regions at \ifm{9\lesssim z\lesssim15}. \textit{Top:} redshift distribution of the sample. \textit{Middle:} two-dimensional histogram of host-halo virial mass \ifm{\Mx{v}} against redshift; the panel on the right shows the marginal \ifm{\Mx{v}} distribution. White dashed contours mark constant comoving number densities of haloes above \ifm{\Mx{v}}, \ifm{n(>\Mx{v})=10^{-9}}, \ifm{10^{-6}}, and \ifm{10^{-3}\,\si{\Mpcc}}, computed from the virial halo mass function of \citet{Despali.etal.16} for the \textit{Planck} cosmology adopted in the simulations \citep{Planck.16}. The red solid line is the threshold for feedback-free starbursts, \ifm{\Mx{v}=10^{10.8}\,[(1+z)/10]^{-6.2}\,\si{\Msun}} \citep{Dekel.etal.23,Li.etal.24}. The orange dotted line is the feedback-failure threshold of \citet{Boylan-Kolchin.25}, where the central acceleration of a Navarro--Frenk--White (NFW) halo reaches \ifm{g_{\rm crit}/G=3100\,\si{\Msun\,pc^{-2}}}. \textit{Bottom:} the same sample in the stellar-mass--redshift plane, with the marginal \ifm{\Mx{\star}} distribution on the right. The blue dashed line marks the characteristic stellar mass of major wet-compaction events in cosmological simulations, \ifm{\Mx{\star}\simeq\SI{1e10}{\Msun}} \citep{Lapiner.etal.23}. Both two-dimensional histograms share the colour scale, where colour gives the number of galaxy--snapshot measurements per bin. Red stars mark the target halo of each simulation (the most massive halo at the final snapshot, \ifm{z\simeq9}) and its central galaxy. The targets span \ifm{\Mx{v}\simeq10^{10.4}\text{--}10^{11.8}\,\si{\Msun}} and \ifm{\Mx{\star}\simeq10^{8.4}\text{--}10^{10.5}\,\si{\Msun}}, at number densities of \ifm{n(>\Mx{v})\simeq10^{-7.6}\text{--}10^{-3.4}\,\si{\Mpcc}}. The seven most massive targets (\run{MAGE01}--\run{MAGE07}) lie above both thresholds at \ifm{z\simeq9}, whereas the four least massive (\run{MAGE08}--\run{MAGE11}) lie below them. The four most massive targets (\run{MAGE01}--\run{MAGE04}) reach the compaction mass by \ifm{z\simeq9}.}
    \label{fig:host_gal_dists}
\end{figure}

The selection yields 2415 galaxy--snapshot measurements over the analysed redshift range. These are repeated measurements of evolving systems, rather than 2415 independent galaxies. Their distributions in the halo-mass--redshift and stellar-mass--redshift planes are shown in \cref{fig:host_gal_dists}. The non-target galaxies primarily populate the low-mass regime, whereas the high-mass end remains dominated by the target central galaxies, marked by red stars at \ifm{z\simeq9}. For statistics presented as a function of stellar mass, we divide the sample into three redshift intervals over \ifm{9\lesssim z\lesssim15} and bin the stellar mass with a width of \ifm{\Delta\log_{10}(M_{\star}/\si{\Msun})=0.5}. For statistics presented as a function of redshift, we divide the sample into three stellar-mass intervals over \ifm{7.5\lesssim\log_{10}(M_{\star}/\si{\Msun})\lesssim10.5} and bin redshift with a width of \ifm{\Delta z=0.5}. We show bins containing at least seven galaxy--snapshot measurements. Unless otherwise specified, curves show medians and error bars span the 16th--84th percentiles of these measurements.

For reference, the middle panel of \cref{fig:host_gal_dists} places the sample relative to theoretical thresholds for efficient star formation at cosmic dawn. Feedback-free starbursts (FFBs) are expected in haloes above \ifm{\Mx{v}\simeq10^{10.8}\,[(1+z)/10]^{-6.2}\,\si{\Msun}}, where gas densities exceed \ifm{\sim3\times10^{3}\,\mathrm{cm^{-3}}} and free-fall times become shorter than the \ifm{\sim}\SI{1}{Myr} delay before stellar winds and SNe act \citep{Dekel.etal.23,Li.etal.24}. This threshold nearly coincides with the halo mass above which central dark-matter accelerations exceed the critical value at which stellar feedback can no longer overcome gravity, \ifm{\Mx{v}\simeq6.5\times10^{10}\,[(1+z)/10]^{-6}\,\si{\Msun}} \citep{Boylan-Kolchin.25}. In a closely related formulation, \citet{Somerville.etal.25} let the cloud-scale star-formation efficiency rise smoothly with gas surface density, approaching unity above \ifm{\Sigma_{\rm crit}\simeq2\times10^{3}\,\si{\Msun\,pc^{-2}}}, where gravity overcomes the momentum injected by stellar winds and radiation pressure \citep[see also][]{Grudic.etal.18,Menon.etal.24}. Implemented in a semi-analytic model, this prescription raises the predicted abundance of UV-luminous galaxies at \ifm{z\simeq12\text{--}14} by one to two orders of magnitude and reproduces the observed UV luminosity functions. At \ifm{z=9}, the haloes of \run{MAGE01}--\run{MAGE07} (\ifm{\Mx{v}\simeq1\text{--}6\times10^{11}\,\si{\Msun}}, see \cref{tab:sims}) lie above both halo-mass thresholds, whereas those of \run{MAGE08}--\run{MAGE11} (\ifm{\Mx{v}\simeq3\text{--}4\times10^{10}\,\si{\Msun}}) lie below them but above the halo mass at which the virial velocity reaches the characteristic SN-feedback scale \ifm{\vx{v}\simeq\SI{100}{km/s}} (\citealt{Dekel.Silk.86}, see \cref{sec:diskformation}). The middle panel also shows contours of constant comoving number density, \ifm{n(>\Mx{v})}, from the virial halo mass function of \citet{Despali.etal.16}. The bulk of the sample lies at halo masses with \ifm{n(>\Mx{v})\gtrsim10^{-3}\,\si{\Mpcc}}, whereas the target haloes at \ifm{z\simeq9} are rare objects with \ifm{n(>\Mx{v})\simeq10^{-7.6}\text{--}10^{-3.4}\,\si{\Mpcc}}.

The bottom panel of \cref{fig:host_gal_dists} marks the characteristic stellar mass of major wet-compaction events in cosmological simulations at cosmic noon, \ifm{\Mx{\star}\simeq\SI{1e10}{\Msun}} \citep{Lapiner.etal.23}, a scale we return to in \cref{subsec:compaction}. Only the four most massive targets (\run{MAGE01}--\run{MAGE04}) reach this mass by \ifm{z\simeq9}, and galaxy--snapshot measurements above it are confined to \ifm{z\lesssim10.5}. The bulk of the sample lies one to two decades below it, so the sample crosses the compaction scale only in its high-mass tail.

\subsection{Kinematics of star-forming gas and stars}
\label{subsec:kinematics}

We characterize rotational support with the rotation-to-dispersion ratio, \ifm{v/\sigma} (often denoted \ifm{\vx{rot}/\sigma} or \ifm{V/\sigma} in the literature), measured within \ifm{\Rx{gal}=0.15\Rx{v}} for three tracers: the star-forming gas, stars younger than \SI{10}{Myr}, and the full stellar population. All three are measured with the same procedure, which differs only in the weight \ifm{w} assigned to each resolution element (the SFR for gas cells and the mass for star particles) and, for the gas, an additional pressure term. Each tracer defines its own frame from its \ifm{w}-weighted bulk velocity and total angular momentum. After subtracting the bulk velocity, we align the angular-momentum vector with the \ifm{z}-axis and calculate the cylindrical velocity components \ifm{v_R}, \ifm{v_\phi}, and \ifm{v_z}. The rotation speed and effective one-dimensional velocity dispersion are
\begin{equation}
 v=\langle v_\phi\rangle_w,\qquad
 \sigma=\left[\frac{\sigma_R^2+\sigma_\phi^2+\sigma_z^2}{3}+\sigma_{\rm P}^2\right]^{1/2},
 \label{eq:kinematic_definition}
\end{equation}
where \ifm{\langle\cdot\rangle_w} denotes the \ifm{w}-weighted mean, \ifm{\sigma_k^2=\langle(v_k-\langle v_k\rangle_w)^2\rangle_w} is the variance of the resolved velocity component \ifm{v_k}, and \ifm{\sigma_{\rm P}} accounts for pressure support that is not carried by resolved motions. For stars, \ifm{\sigma_{\rm P}=0}.

For the star-forming gas, SFR weighting selects the gas that is forming stars at that instant, the component traced observationally by nebular lines. Because this gas is supported by thermal and unresolved turbulent pressure in addition to its resolved bulk motions, we set \ifm{\sigma_{\rm P}^2=\langle c_{\rm s}^2+\sigma_{\rm turb}^2\rangle_w}, where \ifm{c_{\rm s}} is the thermal sound speed and \ifm{\sigma_{\rm turb}} the one-dimensional sub-grid turbulent velocity of the LES model (\cref{sec:methods}). We denote the resulting quantities \ifm{\vx{rot,SFgas}} and \ifm{\sigma_{\rm SFgas}}, and their ratio \ifm{(v/\sigma)_{\rm SFgas}}.

The two stellar tracers bracket the kinematic memory of the stellar population. Young stars trace the recently formed stellar component and initially inherit the resolved velocities of their parent gas, therefore they are the collisionless counterpart of the star-forming gas, without the pressure term. The full population spans many orbits and several burst--lull cycles, and tests whether rotational support survives phase mixing and feedback. For our simulated tracers, \ifm{v/\sigma} is defined by \cref{eq:kinematic_definition}, whereas quoted observational values retain their authors' tracers and dispersion definitions.

\subsection{Rest-frame UV surface brightness}
\label{subsec:uvsb}

For each selected central galaxy, we construct intrinsic (dust-free, noiseless) JWST/NIRCam F200W images using the Flexible Stellar Population Synthesis code \citep[\software{FSPS};][]{Conroy.etal.09,Conroy.Gunn.10}. We adopt simple stellar populations (SSPs) with a Chabrier initial mass function \citep{Chabrier.03}, MIST isochrones, and the MILES stellar spectral library. Nebular emission and internal dust attenuation are omitted. F200W samples rest-frame UV wavelengths of approximately \SI{2000}{\angstrom} at \ifm{z=9} and \SI{1250}{\angstrom} at \ifm{z=15}. Photometry is tabulated over stellar age and metallicity, with each particle's metallicity mapped to the nearest available FSPS index. The SSP flux density, normalized per unit formed stellar mass, is scaled by the particle's birth mass \ifm{m_{\mathrm{form},i}}. Thus the particle flux weight is \ifm{f_{\nu,i}=f_{\nu,\rm SSP}(t_i,Z_i,z)\,m_{{\rm form},i}/\si{\Msun}}, with the SSP flux obtained from its AB magnitude \citep{Oke.Gunn.83}.

We select stars within the three-dimensional \ifm{\Rx{gal}} sphere and sum their flux weights onto a \ifm{256\times256} grid spanning \ifm{-\Rx{gal}} to \ifm{+\Rx{gal}} in each projected coordinate. The measurement field therefore has width \ifm{2\Rx{gal}} and pixel size \ifm{2\Rx{gal}/256}, which ranges from a few parsecs for the least massive haloes to \ifm{\simeq30\,\si{pc}} for the most massive target halo at \ifm{z=9}, comparable to the \ifm{\sim\SI{10}{pc}} resolution of the simulations. It is several to tens of times finer than the NIRCam short-wavelength pixel scale of \ifm{0.031\arcsec} (\ifm{\simeq130\,\si{pc}} at \ifm{z\simeq10}) and far below the F200W point spread function (PSF) full width at half-maximum (FWHM) of \ifm{\simeq0.066\arcsec} (\ifm{\simeq280\,\si{pc}}). These images are therefore intrinsic surface-brightness maps, intended for illustration and for comparing intrinsic sizes among simulations, rather than mock observations. Reproducing NIRCam measurements would require dust attenuation, PSF convolution, noise, observational pixelization, and the source-detection and profile-fitting procedures of the observations. We will present mock observations in more bands, including the effects of dust, in a follow-up study (Nakazato et al., in preparation).

\begin{figure*}
    \centering	\includegraphics[width=\textwidth]{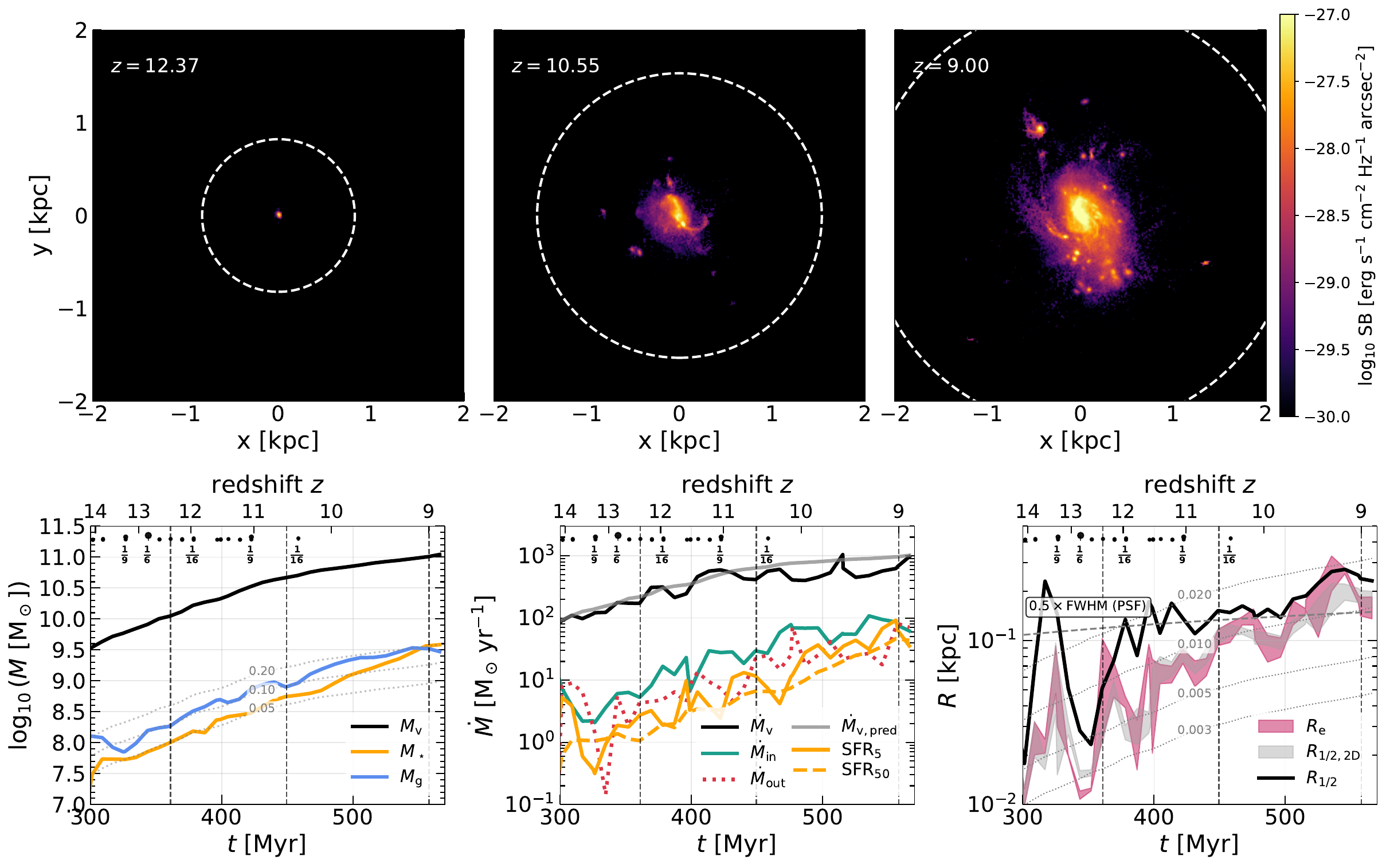}
   \caption{The morphological transition from compact to extended structures in the smooth-accretion system \run{MAGE07}. \textit{Top:} Intrinsic rest-frame UV surface-brightness images of the simulated galaxy in the NIRCam F200W band at $z = 12.37$, $z = 10.55$, and $z = 9$, shown within a $4 \times 4\,\mathrm{pkpc}^2$ region centred on the galaxy along the simulation \ifm{z}-axis. The colour scale shows the logarithmic surface brightness, with pixels below its lower limit shown in black. The white dashed circle marks the galaxy boundary, defined as $R_\mathrm{gal}=0.15\,R_\mathrm{v}$ at each redshift. The projection depth is $2\,R_\mathrm{gal}$. The galaxy evolves from a compact progenitor at high redshift into a more extended morphology through sustained smooth gas accretion and in-situ star formation. \textit{Bottom:} Time evolution of the mass components (left), mass flow rates (middle), and galaxy size (right) for the same system. In the left panel, the black, orange, and blue curves show the virial, stellar, and gas mass, respectively. The grey lines indicate different fractions of $f_\mathrm{b}M_\mathrm{v}$, where \ifm{\fx{b}\equiv\Omegax{b}/\Omegax{m}} is the baryon fraction, corresponding to different global star-formation efficiencies (SFEs). In the middle panel, the black curve shows the virial mass accretion rate, the grey line the theoretical rate expected in a \ifm{\Lambda}CDM cosmology in the Einstein--de Sitter (EdS) regime, the orange curves the star-formation rate (SFR) averaged over $5\,$Myr (solid) and $50\,$Myr (dashed), and the solid teal and dotted red curves the gas inflow and outflow rates at \ifm{\Rx{gal}}, respectively. In the right panel, the magenta shaded region denotes the projected half-light radius, $R_\mathrm{e}$, measured along three orthogonal projections, the grey shaded region shows the projected stellar half-mass radius, \ifm{\Rx{1/2,2D}}, measured along the same projections with the same source definition and estimator as \ifm{\Rx{e}}, and the black curve shows the 3D stellar half-mass radius, \ifm{\Rx{1/2}}. The thick dotted grey line marks half the JWST/NIRCam F200W PSF FWHM as an observational reference scale, and the thin dotted grey lines indicate fixed fractions of the virial radius. Vertical dashed lines in all three lower panels mark the redshifts corresponding to the images above. All halo mergers onto the main branch during $9<z<14$ with mass ratios greater than 0.01 are indicated by lollipops, with symbol sizes proportional to the merger mass ratios; values are labelled for mergers with ratios larger than 0.05. This system experiences no major halo mergers below $z=14$, nor any major galaxy mergers below $z=13$.}
   \label{fig:MockImage_MAGE07}
\end{figure*}

\subsection{UV half-light and half-mass radii}
\label{subsec:uvradius}

We measure intrinsic UV sizes from projected F200W images, which include neither noise nor PSF convolution. Sources are identified with \software{photutils} \citep{bradley.etal.24} as regions of at least five connected pixels above a threshold of one standard deviation above the median of the image. Because the images are noiseless, this threshold acts as a surface-brightness contour that delineates individual galaxies and clumps rather than as a signal-to-noise cut. For the target-galaxy histories we select the source whose centroid is closest to the image centre, and for the statistical catalogue the source with the largest area. Absolute UV magnitudes are calculated from the FSPS rest-frame spectra near \SI{1500}{\angstrom}, summing the contributions of all particles assigned to the same selected source. 

We define the UV half-light radius, \ifm{\Rx{e}}, and the projected stellar half-mass radius, \ifm{\Rx{1/2,2D}}, using the same non-parametric estimator, the circular half-light radius (\texttt{rhalf\_circ}) from \software{statmorph} \citep{Rodriguez-Gomez.etal.19}, applied to the F200W image and to a projected stellar-mass image constructed on the same pixel grid, respectively. In each case it is the radius of the circle, centred on the asymmetry-minimizing centre of the image, that encloses half of the light (or mass) within the outer radius \texttt{rmax\_circ}. Both measurements use the source and segmentation map defined from the F200W image: emission from other detected sources is excluded, while faint emission below the detection threshold is included. The two radii therefore differ only in their weighting by UV light or stellar mass.

Our primary intrinsic size is the three-dimensional stellar half-mass radius, \ifm{\Rx{1/2}}: the radius of the sphere, centred on the galaxy centre of \cref{subsec:halofinder}, that encloses half of the stellar mass within \ifm{\Rx{gal}}. Unlike \ifm{\Rx{e}} and \ifm{\Rx{1/2,2D}}, it depends on neither the viewing direction nor the source segmentation. The three radii thus form a sequence: \ifm{\Rx{1/2}} measures the intrinsic mass distribution, \ifm{\Rx{1/2,2D}} the projected mass distribution of the segmented source, and \ifm{\Rx{e}} the projected UV light of the same source.

\section{Size Evolution Across Cosmic Time}
\label{sec:sizeevo}
We first compare the size histories of the smooth-accretion system \run{MAGE07} and the major-merger system \run{MAGE04} (\cref{subsec:individual}). We then examine the size--mass and size--redshift relations of the full statistical sample (\cref{subsec:size-mass}), followed by the UV size--magnitude relation and its observational comparison (\cref{subsec:size-magnitude}).

\subsection{Size evolution of two example galaxies}
\label{subsec:individual}

\run{MAGE07} and \run{MAGE04} illustrate contrasting assembly histories at cosmic dawn. The former grows mainly through smooth accretion, whereas the latter has undergone recent major mergers. Despite these differences, they reach similar rest-frame UV morphologies and sizes by \ifm{z\simeq9}.

\cref{fig:MockImage_MAGE07} traces the evolution of \run{MAGE07} through its rest-frame UV morphology, mass buildup, mass flow rates, and characteristic size. The top panels present intrinsic F200W-band surface-brightness maps at \ifm{z=12.37}, \ifm{z=10.55}, and \ifm{z=9}. At the earliest time, the emitting region spans only a few image pixels, with most of the UV light concentrated near the centre. As the system evolves, the emitting region becomes progressively more extended and develops multiple star-forming clumps within the galaxy. By \ifm{z=9}, \run{MAGE07} has transformed from a compact progenitor into an extended galaxy.

The lower panels of \cref{fig:MockImage_MAGE07} show that this morphological growth is accompanied by a relatively smooth buildup of mass. Halo mergers onto the main branch with mass ratios greater than \ifm{0.01} are indicated by lollipops at the top of each panel. No major halo merger occurs for this system at \ifm{z<14}, although a major galaxy merger at \ifm{z\simeq13.5} leaves a visible imprint, as discussed below. The halo mass therefore grows steadily, while the gas and stellar masses within \ifm{\Rx{gal}} increase overall with time. By \ifm{z=9}, the stellar mass reaches \ifm{\Mx{\star}\simeq0.22\,\fx{b}\Mx{v}}, where \ifm{\fx{b}} is the cosmological baryon fraction, slightly above the uppermost grey line, which marks a global star-formation efficiency of 0.2. The mild, transient dip in stellar mass at \ifm{t\simeq\SI{385}{Myr}} is caused by orbiting satellites temporarily moving outside \ifm{\Rx{gal}} at this snapshot. The rapid consumption of gas at \ifm{t\simeq\SI{325}{Myr}} is driven by a major galaxy merger, an event accompanied by a transient peak in \ifm{\Rx{1/2}}. Despite this depletion, the system remains gas-rich throughout most of its evolution, maintaining a gas mass comparable to its stellar mass until \ifm{z=9}. The virial mass accretion rate is generally consistent with the theoretical prediction for a \ifm{\Lambda}CDM cosmology in the Einstein--de Sitter (EdS) regime \citep{Dekel.etal.13}
\begin{equation}
\label{eq:mdotv}
    \frac{\dotM{v}}{\Mx{v}}\simeq s(1+z)^{5/2},
\end{equation}
where \ifm{s\simeq\SI{0.030}{Gyr^{-1}}}. The baryonic mass-flow history at \ifm{\Rx{gal}} is similarly smooth: the gas inflow rate gradually rises and remains larger than the outflow rate for most of the period shown, sustaining the gas reservoir within the galaxy. The SFR, averaged over both \SI{5}{Myr} and \SI{50}{Myr}, increases steadily with time. By \ifm{z=9}, the SFR reaches \ifm{\sim\SI{50}{\Msun\,yr^{-1}}}, comparable to both \ifm{\dotM{in}} and \ifm{\dotM{out}}.

The projected UV half-light radius, \ifm{\Rx{e}} (magenta), grows from compact values at \ifm{z\gtrsim12} to \ifm{\sim\SI{0.2}{kpc}} by \ifm{z=9}, corresponding to \ifm{0.01\text{--}0.02\,\Rx{v}}. The projected stellar half-mass radius, \ifm{\Rx{1/2,2D}} (grey), measured with the same source definition and estimator (\cref{subsec:uvradius}), tracks \ifm{\Rx{e}} closely and is only marginally smaller at most epochs. The UV light and the stellar mass of the source are therefore distributed over similar scales, and the small residual offset is consistent with recent star formation extending slightly beyond the region containing half of the projected stellar mass. For comparison, \citet{Somerville.etal.18} adopt projected mass-to-light size ratios of 0.83 and 0.87 for low-redshift optical disks and spheroids, respectively, similar to the ratio seen here. These optical values provide context rather than a conversion applied to our UV measurements, whose mass-to-light size ratio can change during individual star-formation episodes.

The 3D stellar half-mass radius, \ifm{\Rx{1/2}} (black), is generally larger than its projected counterpart because projection reduces apparent stellar distances \citep{Somerville.etal.18,Behroozi.etal.22}. The shaded ranges in \cref{fig:MockImage_MAGE07} show the viewing-angle dependence. The offsets among the three radii reflect stellar-population and geometric effects, as well as the aperture and source definitions described in \cref{subsec:uvradius}. The 3D radius includes all stars within the spherical \ifm{\Rx{gal}} aperture, whereas image segmentation identifies the central source and can separate disconnected companions in the projected measurements. During the major galaxy merger at \ifm{z\simeq13.5}, the separation between the merging components produces a pronounced peak in \ifm{\Rx{1/2}}, while the projected radii remain more closely associated with the central component. This peak in \ifm{\Rx{1/2}} traces the spatial extent of the combined stellar system during the interaction, rather than the size of an individual galaxy. Following the three radii together therefore helps distinguish the growth of an individual galaxy from the temporary contribution of companions within an aperture. The thick dotted grey line in the lower-right panel marks half the JWST/NIRCam F200W PSF FWHM. The projected UV radius becomes comparable to or larger than this reference scale near \ifm{z\sim9}, although its observational measurability also depends on signal-to-noise, PSF knowledge, and the assumed light profile.

\begin{figure*}
    \centering	\includegraphics[width=\textwidth]{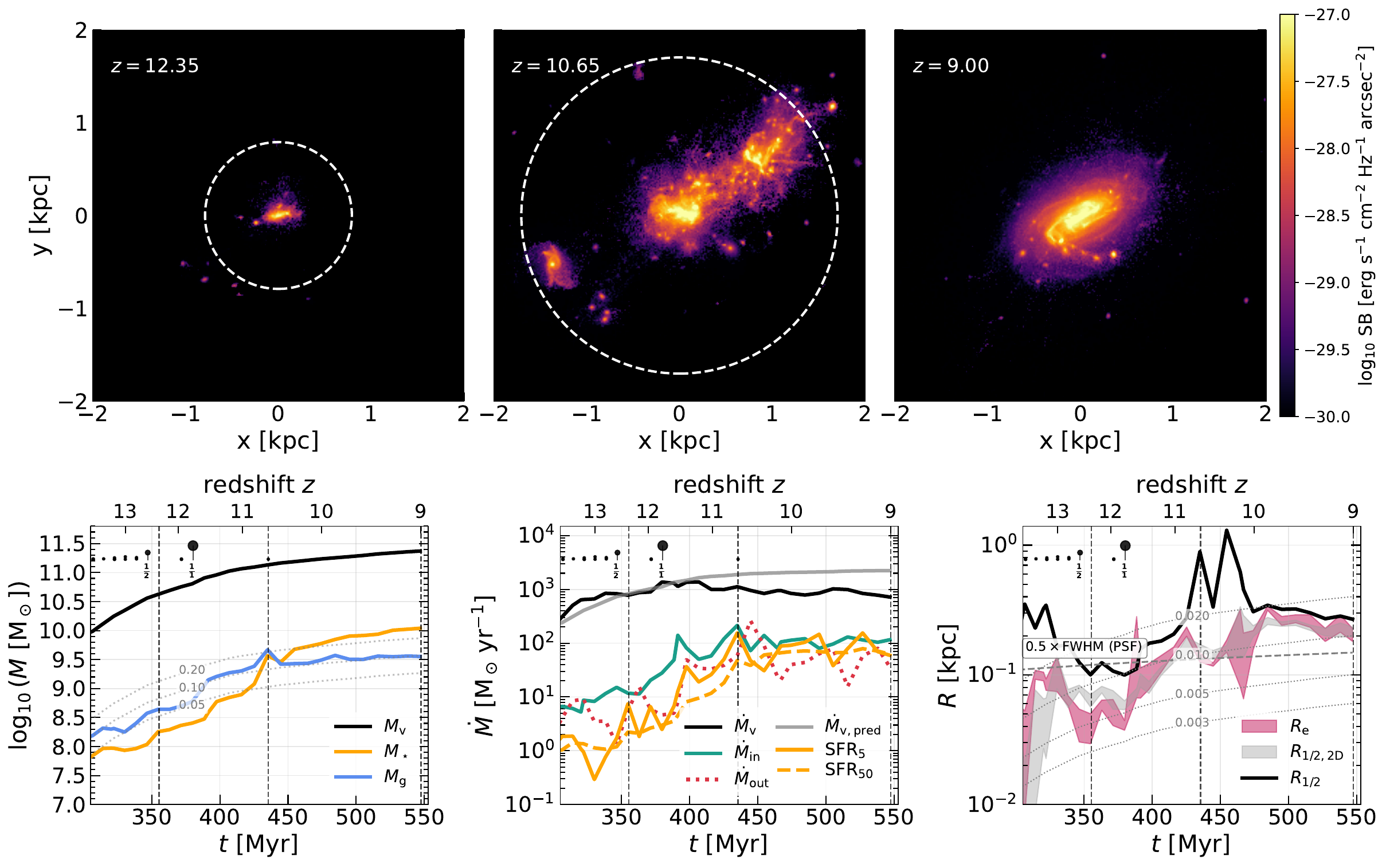}
   \caption{Similar to \cref{fig:MockImage_MAGE07}, but for the recent major-merger system \run{MAGE04}. The upper panels show the galaxy at \ifm{z=12.35}, \ifm{z=10.65}, and \ifm{z=9}. Despite exhibiting a morphology broadly similar to \run{MAGE07} at \ifm{z=9}, this system follows a markedly different evolutionary pathway dominated by recent mergers. Two major halo mergers occur around \ifm{z\sim12} (indicated by the lollipop markers in the lower panels), followed by \zyn{two major galaxy mergers} at \ifm{z\sim10.6} visible in the top-middle panel. Subsequently, the galaxy undergoes structural contraction. At \ifm{z=9}, the white dashed galaxy-boundary circle lies outside the entire displayed field.}
   \label{fig:MockImage_MAGE04}
\end{figure*}

\begin{figure*}
    \centering	\includegraphics[width=\textwidth]{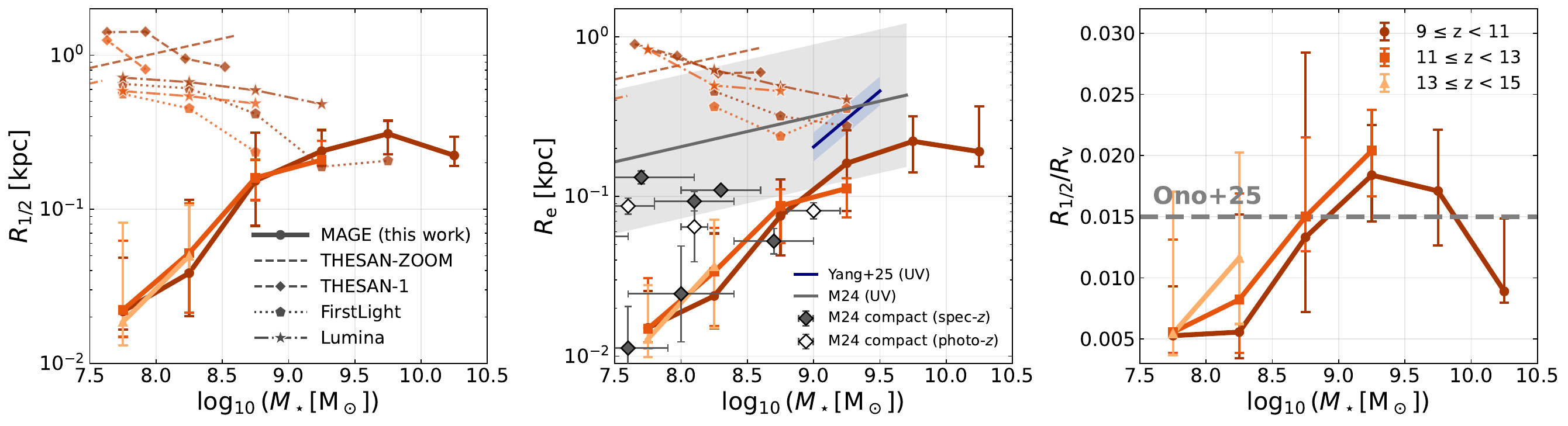}
   \caption{The 3D half-mass radius (\textit{left}), 2D projected UV half-light radius (\textit{middle}), and ratio of half-mass to virial radius (\textit{right}) as a function of stellar mass for all selected central galaxies. Darker shades of orange denote lower redshift bins. Error bars indicate the 16th to 84th percentiles. Dashed lines without markers show the corresponding scaling relations for each redshift bin from \software{THESAN-ZOOM} \citep{McClymont.etal.25}, and dashed lines with diamonds show the binned medians from \software{THESAN-1} \citep{Shen.etal.24}. Dotted lines with pentagons and dot-dashed lines with stars show the binned medians from \software{FirstLight} \citep{Ceverino.etal.26} and \software{Lumina} \citep[][Liu et al., to be submitted]{Zier.etal.26}, respectively, using the same redshift bins and shading as our sample. In the middle panel all comparison sizes from simulations are intrinsic UV half-light radii computed without dust attenuation, matching our measurements. The navy shaded band shows JWST observational constraints on the effective radius at \ifm{9<z<10} in the rest-frame UV bands \citep{Yang.etal.25}, while the grey line and shaded band show the corresponding size--mass relation at \ifm{10<z<13} in the F200W UV band \citep[M24;][]{Morishita.etal.24}. Grey diamonds show the compact sources of M24 at \ifm{z>9}, defined as galaxies that are marginally resolved, with \ifm{\Rx{e}} below half the PSF FWHM, or that lie more than \ifm{1.5\sigma} below their size--mass relation; filled and open symbols denote spectroscopic and photometric redshifts, respectively, and error bars show the quoted uncertainties. Stellar masses and sizes are from M24, which lists two of the ten sources as spectroscopically confirmed; five more have since been confirmed with NIRSpec: UNCOVER-37126 (\ifm{z=10.02}), GHZ9 (\ifm{z=10.14}), and GHZ7 (\ifm{z=10.48}) by \citet{Roberts-Borsani.etal.26}, and UNCOVER-13151 (\ifm{z=9.80}) and RUBIES-EGS-952693 (\ifm{z=9.03}) in the DAWN JWST Archive (DJA v4.4; \citealt{Heintz.etal.25,deGraaff.etal.25}). One compact source at \ifm{\Mx{\star}=10^{7.4}\,\si{\Msun}} lies below the plotted mass range. M24 sizes are major-axis effective radii measured in F200W. In the right panel, the horizontal grey dashed line marks the characteristic UV half-light-to-virial-radius ratio \ifm{\Rx{e}/\Rx{v}\simeq0.015} inferred for galaxies at \ifm{z\sim10\text{--}16} by \citet{Ono.etal.25}; note that this is a half-light ratio, whereas our points are half-mass ratios.}
   \label{fig:sizemass}
\end{figure*}

\cref{fig:MockImage_MAGE04} presents the contrasting case of \run{MAGE04}. Unlike \run{MAGE07}, this system experiences a strongly merger-driven assembly history, as indicated by the lollipops in the bottom panels. Two major halo mergers occur around \ifm{z\sim12}, followed by two major galaxy mergers at \ifm{z\sim10.6}, which are visible in the top-middle panel as multiple bright components within the galactic region. Subsequently, the stellar mass increases rapidly, and the gas mass drops sharply, indicating efficient star formation triggered by wet mergers. By \ifm{z=9}, the stellar mass reaches \ifm{\simeq0.31\,\fx{b}\Mx{v}}, compared with \ifm{\simeq0.22\,\fx{b}\Mx{v}} for \run{MAGE07}.

The virial mass accretion rate is broadly consistent with the theoretical expectation for a median halo mass accretion history (shown by the grey line in the middle panel) prior to the final major halo merger at \ifm{z\simeq11.7}. The corresponding gas inflow, outflow, and star-formation rates rise steadily, reaching \ifm{\sim\SI{100}{\Msun\,yr^{-1}}} during the major galaxy mergers at \ifm{z\simeq10.6}. Afterwards, the rates fluctuate around approximately constant levels, so that the final mass flow rates at \ifm{z=9} are comparable to those of \run{MAGE07}. The size evolution follows a similar pattern: \ifm{\Rx{e}} and \ifm{\Rx{1/2,2D}}, which again remain close to each other, increase steadily before the major merger phase, with \ifm{\Rx{e}} exceeding \ifm{\SI{0.1}{kpc}} already by \ifm{z\simeq11}. Subsequently, the galaxy size plateaus or undergoes net contraction, reflecting a wet compaction episode triggered by mergers that results in a final size at \ifm{z\sim9} comparable to that of \run{MAGE07}. The transient peaks in \ifm{\Rx{1/2}} again reflect multiple stellar components within \ifm{\Rx{gal}}. Criterion (v) in \cref{subsec:halofinder} reduces the contribution of such configurations to the statistical sample. 

These two systems show that comparable sizes at \ifm{z\simeq9} can arise from different assembly histories. \run{MAGE07} grows relatively smoothly, whereas \run{MAGE04} undergoes rapid merger-driven growth followed by a phase of gradual size contraction. Both end with \ifm{\Rx{e}\sim\SI{0.2}{kpc}}, \ifm{\Rx{1/2}\sim0.2\text{--}0.3\,\mathrm{kpc}}, and comparable inflow, outflow, and star-formation rates. In both systems, the inflow and outflow rates at \ifm{\Rx{gal}} rise more steeply than the halo accretion rate, so the galaxy processes a growing fraction of the cosmological baryon supply, \ifm{\fx{b}\dotM{v}}, without exceeding it. This suggests either more efficient penetration of fresh gas or the return through \ifm{\Rx{gal}} of gas ejected earlier. Meanwhile, the outflow rate falls from exceeding the SFR at early times to comparable values by \ifm{z=9}, so the declining outflow-to-SFR ratio indicates that star formation becomes increasingly important relative to outward gas transport. This convergence motivates examining how the internal potential, feedback, and spatial distribution of star formation regulate sizes across the full sample.

\subsection{The size--mass relation across redshift}
\label{subsec:size-mass}
We now examine the size--mass relation of galaxies at cosmic dawn using our statistical sample. \cref{fig:sizemass} shows the median 3D half-mass radius, $R_{1/2}$ (left panel), the 2D projected UV half-light radius, $\Rx{e}$ (middle panel), and the corresponding ratio $R_{1/2}/R_{\rm v}$ (right panel) as a function of stellar mass in three redshift bins, with error bars spanning the 16th--84th percentiles. At a fixed stellar mass, we find little variation in \ifm{\Rx{1/2}} and \ifm{\Rx{e}} across the three redshift bins, although the higher-redshift bins contain fewer data points at high stellar masses. The ratio \ifm{\Rx{1/2}/\Rx{v}} is slightly elevated at higher redshifts, primarily driven by a correspondingly smaller \ifm{\Rx{v}}. These redshift dependencies are weaker than the mass dependence.

We find a clear positive correlation between \ifm{\Rx{1/2}} (as well as \ifm{\Rx{e}} and \ifm{\Rx{1/2}/\Rx{v}}) and \ifm{\Mx{\star}} for \ifm{\Mx{\star}<\SI{5e9}{\Msun}}. At \ifm{\Mx{\star}\simeq\SI{5e9}{\Msun}}, the sizes reach \ifm{\Rx{1/2}\simeq \SI{300}{pc}}, \ifm{\Rx{e}\simeq \SI{200}{pc}}, and \ifm{\Rx{1/2}/\Rx{v}\simeq0.02}. Above this mass, the half-mass and UV half-light size--mass relations become approximately flat or even turn over. This turnover is predominantly driven by the central galaxies in the four most massive targets (\run{MAGE01}--\run{MAGE04}), as other galaxies in the sample do not reach \ifm{\Mx{\star}>\SI{5e9}{\Msun}} by \ifm{z\simeq9}. The system \run{MAGE04}, shown in \cref{fig:MockImage_MAGE04}, is one such example. While \ifm{\Rx{1/2}} and \ifm{\Rx{e}} subsequently decrease by at most a factor of 2, the drop in \ifm{\Rx{1/2}/\Rx{v}} can exceed a factor of 2 due to the concurrent increase in \ifm{\Rx{v}} with stellar mass. At high mass, typical half-mass and UV half-light radii are \ifm{200\text{--}400\,\mathrm{pc}} and \ifm{100\text{--}300\,\mathrm{pc}}, respectively, for luminous objects at cosmic dawn.

\begin{table*}
\centering
\caption{Power-law fits, \cref{eq:sizemassfit}, to the unbinned size--mass relation of central galaxies for both the 3D half-mass radius (\ifm{\Rx{1/2}}) and 2D UV half-light radius (\ifm{\Rx{e}}), restricted to the adopted stellar-mass range \ifm{\log_{10}(M_\star/\si{\Msun}) \geq 7.5}. The quoted parameter uncertainties are formal standard errors from unweighted linear regression on galaxy--snapshot measurements in log--log space; they do not include temporal correlations or uncertainty from the finite number of independent haloes. The residual scatter is given by \ifm{\sigma_{\log R}}, and \ifm{N_{\rm meas}} counts galaxy--snapshot measurements.}
\label{tab:sizemassfit}
\begin{tabular}{lccccc}
\hline
Sample & \ifm{\log_{10}(M_\star/\si{\Msun})} & \ifm{N_{\rm meas}} & \ifm{\alpha} & \ifm{R_0} [pc] & \ifm{\sigma_{\log R}} [dex] \\
\hline
\multicolumn{6}{c}{\textbf{3D Half-Mass Radius (\ifm{\Rx{1/2}})}} \\
\hline
\ifm{9 \le z<11}, single & 7.5--10.5 & 1585 & \ifm{0.54 \pm 0.01} & \ifm{127 \pm 4} & 0.28 \\
\ifm{9 \le z<11}, lower-mass interval & 7.5--9.25 & 1520 & \ifm{0.62 \pm 0.02} & \ifm{154 \pm 8} & 0.28 \\
\ifm{9 \le z<11}, higher-mass interval & 9.25--10.5 & 65 & \ifm{-0.12 \pm 0.04} & \ifm{340 \pm 26} & 0.10 \\
\ifm{11 \le z<13} & 7.5--9.25 & 615 & \ifm{0.66 \pm 0.04} & \ifm{183 \pm 18} & 0.28 \\
\hline
\multicolumn{6}{c}{\textbf{2D UV Half-Light Radius (\ifm{\Rx{e}})}} \\
\hline
\ifm{9 \le z<11}, single & 7.5--10.5 & 1585 & \ifm{0.54 \pm 0.01} & \ifm{79 \pm 2} & 0.22 \\
\ifm{9 \le z<11}, lower-mass interval & 7.5--9.25 & 1520 & \ifm{0.55 \pm 0.02} & \ifm{81 \pm 3} & 0.22 \\
\ifm{9 \le z<11}, higher-mass interval & 9.25--10.5 & 65 & \ifm{0.04 \pm 0.07} & \ifm{198 \pm 25} & 0.17 \\
\ifm{11 \le z<13} & 7.5--9.25 & 615 & \ifm{0.59 \pm 0.03} & \ifm{97 \pm 8} & 0.22 \\
\hline
\end{tabular}
\end{table*}

\begin{figure*}
    \centering	\includegraphics[width=\textwidth]{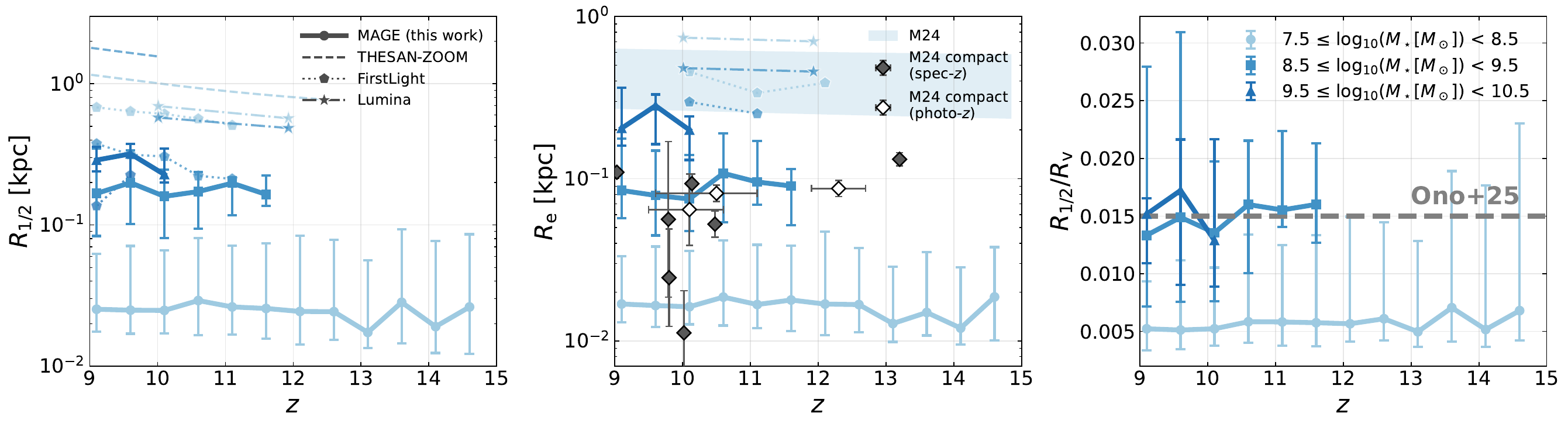}
   \caption{The 3D half-mass radius (\textit{left}), 2D projected UV half-light radius (\textit{middle}), and ratio of half-mass to virial radius (\textit{right}) as a function of redshift for all selected central galaxies. Darker shades of blue denote higher stellar mass bins. Error bars indicate the 16th to 84th percentiles. In the left panel, dashed lines show the fitting relations at each stellar mass bin from the \software{THESAN-ZOOM} simulations \citep{McClymont.etal.25}. In the left and middle panels, dotted lines with pentagons and dot-dashed lines with stars show the binned medians in the same stellar-mass bins from \software{FirstLight} \citep{Ceverino.etal.26} and \software{Lumina} \citep[][Liu et al., to be submitted]{Zier.etal.26}, respectively, with the same shading as our sample; the middle-panel values are intrinsic UV half-light radii without dust attenuation, as for our sample. The faint blue shaded region denotes the JWST size--redshift relation at \ifm{\Mx{\star}=\SI{1e8}{\Msun}} \citep[M24;][]{Morishita.etal.24}, and grey diamonds show the ten compact sources of M24 at \ifm{z>9}, with \ifm{\Mx{\star}\simeq10^{7.4}\text{--}10^{9.0}\,\si{\Msun}}; filled and open symbols denote spectroscopic and photometric redshifts, and sources are plotted at their spectroscopic redshift where available (see the caption of \cref{fig:sizemass}). In the right panel, the horizontal grey dashed line marks the characteristic UV half-light-to-virial-radius ratio \ifm{\Rx{e}/\Rx{v}\simeq0.015} inferred at \ifm{z\sim10\text{--}16} by \citet{Ono.etal.25}, as in \cref{fig:sizemass}.}
   \label{fig:sizeevolution3d}
\end{figure*}

To quantify the relation, we fit power laws of the form
\begin{equation}
R = R_0 \left(\frac{M_\star}{10^{9}\,\si{\Msun}}\right)^{\alpha}
\label{eq:sizemassfit}
\end{equation}
directly to the galaxy--snapshot measurements (an unbinned fit) over the adopted stellar-mass range \ifm{\log_{10}(M_\star/\si{\Msun}) \geq 7.5} for both \ifm{\Rx{1/2}} and \ifm{\Rx{e}}. The best-fit parameters are listed in \cref{tab:sizemassfit}. A single power law does not capture the high-mass behaviour. We therefore report separate fits over the intervals \ifm{7.5\leq\log_{10}(\Mx{\star}/\si{\Msun})<9.25} and \ifm{9.25\leq\log_{10}(\Mx{\star}/\si{\Msun})\leq10.5}. These are independent fits over chosen mass intervals, rather than a fitted continuous broken power law. 

In the left panel, we also compare our 3D half-mass radius--mass relation to those from \software{THESAN-ZOOM} \citep{McClymont.etal.25}, \software{THESAN-1} \citep{Shen.etal.24}, \software{FirstLight} \citep{Ceverino.etal.26}, and \software{Lumina} \citep[][Liu et al., to be submitted]{Zier.etal.26}. Our galaxies are systematically smaller, and the offset grows towards low stellar mass. At \ifm{\Mx{\star}\simeq10^{7.75}\,\si{\Msun}}, our median \ifm{\Rx{1/2}\simeq\SI{20}{pc}} lies more than an order of magnitude below the \ifm{0.5\text{--}1.5\,\si{kpc}} found in the four comparison suites, whereas by \ifm{\Mx{\star}\simeq10^{9}\text{--}10^{9.5}\,\si{\Msun}} our medians agree with \software{FirstLight} and lie within a factor of \ifm{\sim2} of \software{Lumina}. At \ifm{\Mx{\star}\lesssim10^{7.5}\,\si{\Msun}}, our half-mass radii are on the order of the spatial resolution of \ifm{\simeq\SI{10}{pc}}, which motivates the stellar-mass cut (criterion i in \cref{subsec:halofinder}) that selects only \ifm{\Mx{\star}\gtrsim10^{7.5}\,\si{\Msun}}. The nature of these compact systems and the possible origins of the offset are discussed in \cref{subsubsec:lowmasscompact}. Furthermore, our positive correlation at the low-mass end broadly aligns with \software{THESAN-ZOOM}, contrasting with the negative trend seen in \software{THESAN-1}, \software{FirstLight}, and \software{Lumina} over \ifm{10^{7.75}\lesssim\Mx{\star}/\si{\Msun}\lesssim10^{9.25}}. In our sample, the half-mass relation turns over near \ifm{\Mx{\star} \simeq \SI{5e9}{\Msun}}, suggesting that the characteristic transition mass varies strongly between simulations (\cref{subsubsec:turnovercomparison}). We explore the implications of both this transition scale and our sub-grid feedback choices further in \cref{sec:discuss}.

In the middle panel, we compare our intrinsic 2D projected UV half-light radii with the corresponding dust-free values from \software{THESAN-ZOOM}, \software{THESAN-1}, \software{FirstLight}, and \software{Lumina}. Dust-attenuated sizes are not shown because they would not be comparable to our measurements. Even without dust, the comparison simulations lie within or above the range of JWST measurements (indicated by the navy and grey bands from \citealt{Yang.etal.25} and M24, respectively), whereas our sizes fall below it for \ifm{\Mx{\star}\lesssim10^{9.5}\,\si{\Msun}}, with the same growing offset towards low mass as in the half-mass radii. The observed population is not confined to these bands, however. M24 identify a compact subpopulation of 44 galaxies (\ifm{\simeq13} per cent of their sample) that are marginally resolved or lie more than \ifm{1.5\sigma} below their size--mass relation, with star-formation-rate surface densities of \ifm{\Sigma_{\rm SFR}>10\,\si{\Msun\,yr^{-1}\,kpc^{-2}}}. The ten such sources at \ifm{z>9} (diamonds, seven with spectroscopic redshifts) have \ifm{\Rx{e}\simeq10\text{--}130\,\si{pc}} at \ifm{\Mx{\star}\simeq10^{7.4}\text{--}10^{9}\,\si{\Msun}} and fall between our medians and the mean relation.\footnote{\label{fn:u37126}We adopt the M24 sizes and stellar masses for all compact sources, so that the two quantities derive from the same photometry. For UNCOVER-37126 (A2744-26963 in M24), M24 report \ifm{\Rx{e}=\SI{11}{pc}}, whereas \citet{Roberts-Borsani.etal.26} report \ifm{\Rx{e}\simeq\SI{61}{pc}}, the value \zyn{plotted in \cref{fig:sizemag}}. The difference illustrates the uncertainty of size measurements for marginally resolved sources (\cref{subsubsec:observedsize}).} About half of them, including both at \ifm{\Mx{\star}\gtrsim10^{8.5}\,\si{\Msun}}, lie within or at the edge of our 16th--84th percentile range, and the most compact, with \ifm{\Rx{e}\simeq10\text{--}25\,\si{pc}}, coincide with our medians at \ifm{\Mx{\star}\simeq10^{7.6}\text{--}10^{8}\,\si{\Msun}}, where the mean relation lies an order of magnitude above them. Our intrinsic sizes therefore overlap the compact tail of the observed distribution rather than its mean relation. The effects of dust attenuation and PSF on these comparisons, which can be substantial for compact galaxies, are discussed in \cref{subsubsec:observedsize,subsec:caveats}.

The redshift evolution is further detailed in \cref{fig:sizeevolution3d}, where $R_{1/2}$ (left panel), $\Rx{e}$ (middle panel), and \ifm{\Rx{1/2}/\Rx{v}} (right panel) are shown as a function of $z$ across different stellar-mass bins. At a fixed stellar mass, $R_{1/2}$, $\Rx{e}$, and \ifm{\Rx{1/2}/\Rx{v}} exhibit minimal evolution with cosmic time, reflecting the weak redshift dependence observed in \cref{fig:sizemass}. This stability broadly aligns with other simulations (\software{THESAN-ZOOM}, \software{FirstLight}, \software{Lumina}) and JWST constraints (M24), though our intrinsic galaxy sizes remain systematically more compact. The offset is again largest in the lowest-mass bin: for \ifm{7.5\leq\log_{10}(\Mx{\star}/\si{\Msun})<8.5}, \software{FirstLight} and \software{Lumina} maintain \ifm{\Rx{1/2}\simeq0.5\text{--}0.7\,\si{kpc}} over \ifm{9\lesssim z\lesssim12}, compared with \ifm{\simeq\SI{25}{pc}} in \software{MAGE}, whereas for \ifm{8.5\leq\log_{10}(\Mx{\star}/\si{\Msun})<9.5} our half-mass radii approach those of \software{FirstLight} at \ifm{z\gtrsim10}. The compact sources of M24 at \ifm{z>9} (diamonds in the middle panel) have \ifm{\Rx{e}\simeq10\text{--}130\,\si{pc}} over \ifm{z\simeq9\text{--}13}, within the range spanned by our two lower-mass bins at the same redshifts and well below the band for their full sample at \ifm{\Mx{\star}=\SI{1e8}{\Msun}}. The medians in the two higher stellar-mass bins indicate a typical size ratio of \ifm{\Rx{1/2}/\Rx{v}\simeq0.015} for galaxies with \ifm{\Mx{\star}>10^{8.5}\,\si{\Msun}}, comparable to \ifm{\Rx{e}/\Rx{v}\simeq0.015} of \citet{Ono.etal.25} at \ifm{z\sim10\text{--}16}, shown as the grey dashed line in the right panels of \cref{fig:sizemass,fig:sizeevolution3d}, as well as the nearly constant half-light ratios of \ifm{\Rx{e}/\Rx{v}\simeq0.01\text{--}0.035} measured from \ifm{z\sim0} to \ifm{z\sim10} \citep{Shibuya.etal.15,Yang.etal.25} and \ifm{\simeq0.015\text{--}0.02} inferred at \ifm{z\lesssim3} \citep{Kravtsov.13,Somerville.etal.18}, despite using different size definitions.

The principal result is a positive low-mass size--mass relation with weak redshift evolution, followed by a flattening or turnover in the half-mass and UV half-light size--mass relations at high mass. The transition occurs at a higher stellar mass than in several comparison simulations, while our intrinsic sizes have a lower normalization, most strongly at \ifm{\Mx{\star}\lesssim10^{8}\,\si{\Msun}}, where our galaxies are compact and star-cluster-like rather than extended. Relative to observations, our intrinsic UV sizes lie below the mean size--mass relation but overlap the compact subpopulation of M24. 

\begin{figure}
    \centering	\includegraphics[width=\columnwidth]{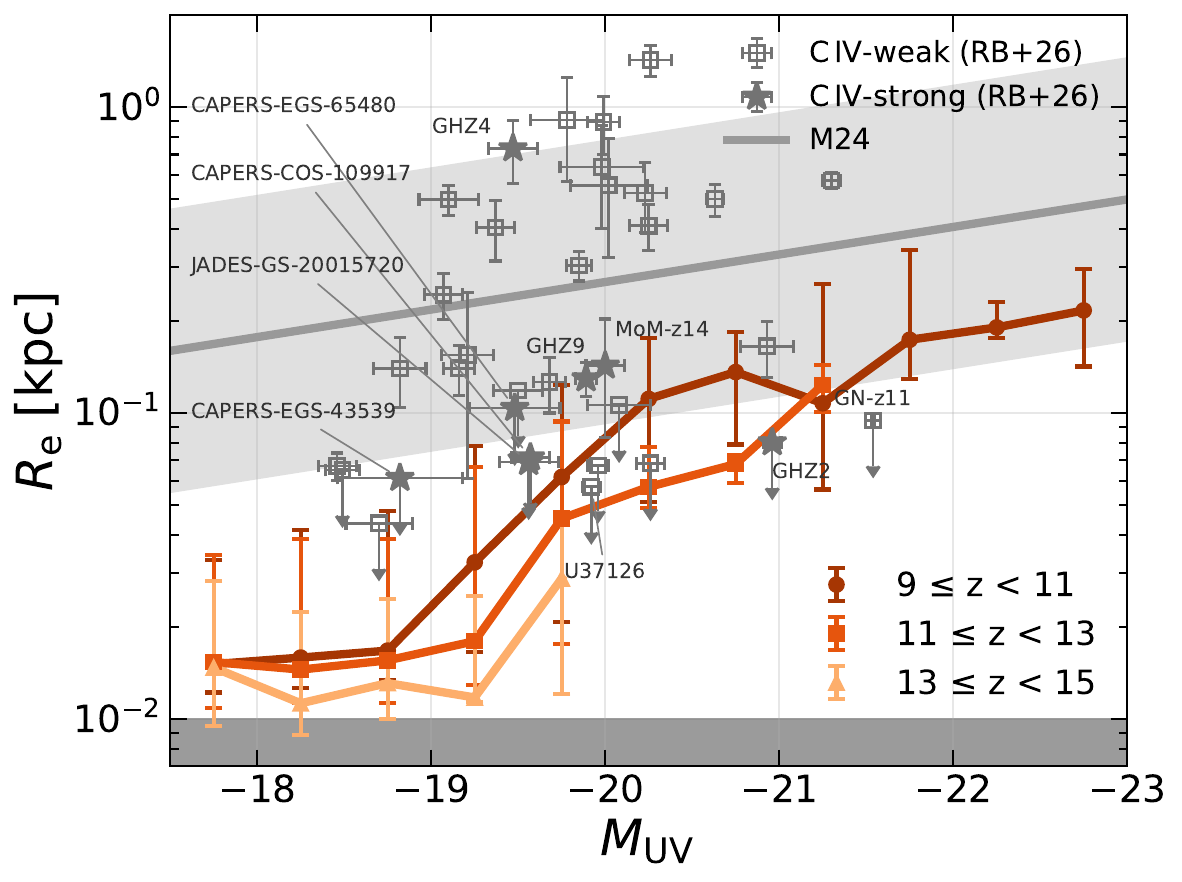}
   \caption{Rest-frame UV size--magnitude relations for central galaxies, compared with observational constraints. Orange symbols show simulation medians in three redshift bins: \ifm{9\leq z<11} (circles), \ifm{11\leq z<13} (squares), and \ifm{13\leq z<15} (triangles), with darker shades indicating lower redshifts. Error bars span the 16th--84th percentiles of the simulated distribution. Absolute UV magnitudes are evaluated for the selected source at \ifm{\lambda=\SI{1500}{\angstrom}}. Grey symbols show the spectroscopically confirmed \ifm{z\geq10} galaxies of \citet{Roberts-Borsani.etal.26} with half-light radii measured in NIRCam F200W (35 of their 41 sources). Filled stars mark sources with \ion{C}{iv} emission detected in their NIRSpec prism spectra (\ion{C}{iv}-strong), which are labelled by name, and open squares mark the remaining (\ion{C}{iv}-weak) sources, among which U37126 and GN-z11 are also labelled. Downward arrows denote unresolved sources, whose radii are upper limits. The grey line and shaded band show the \ifm{M_{\rm UV}}--size relation \citep[M24;][]{Morishita.etal.24}, fit to their full \ifm{5<z<14} sample with a redshift-dependent intercept and evaluated at \ifm{z=11.2}, representative of their F115W-dropout window (\ifm{9.7<z<13}), where sizes are major-axis radii measured in F200W. The band spans \ifm{\pm1.5\sigma} of the intrinsic scatter (\ifm{\sigma_{\log R_{\rm e}}=0.31\,\si{dex}}), consistent with the original figure in M24. The dark grey shaded region marks the spatial resolution limit of the simulations, below which galaxy size measurements are unresolved.}
   \label{fig:sizemag}
\end{figure}

\subsection{The size--magnitude relation}
\label{subsec:size-magnitude}

Because stellar-mass estimates at \ifm{z\geq10} are uncertain, we now examine the size--magnitude relation at \ifm{9\leq z<15}, which can be compared more directly with observations. \cref{fig:sizemag} shows the median 2D projected UV half-light radius, $\Rx{e}$, as a function of rest-frame UV magnitude for all central galaxies across three redshift bins, compared with the spectroscopically confirmed \ifm{z\geq10} sample of \citet{Roberts-Borsani.etal.26}, split by whether \ion{C}{iv} emission is detected in their NIRSpec prism spectra, and with the empirical relation of M24. Absolute UV magnitudes in \software{MAGE} are evaluated at \ifm{\lambda=\SI{1500}{\angstrom}} from the selected source, using the aperture conventions described in \cref{subsec:uvradius}. These monochromatic magnitudes differ slightly from the F200W band used for the sizes, which samples a redshift-dependent rest-frame wavelength, but we expect this to have a negligible effect. They are intrinsic, with no dust attenuation, PSF effects, or noise included.

At fixed absolute UV magnitude, galaxies at higher redshift are more compact, a stronger redshift evolution than seen in the size--mass relation. This is consistent with \zyn{the younger, brighter stellar populations at higher redshift}, so a given \ifm{\Mx{UV}} corresponds to a lower stellar mass. In each redshift interval, brighter galaxies are generally larger. The slope is steep between \ifm{\Mx{UV}=-19} and \ifm{-21}, then flattens at the bright end, where it is more consistent with the slope of the typical sample of M24, although our galaxies lie close to the lower envelope of the distribution. At the faint end (\ifm{\Mx{UV}>-20}), our galaxies fall below the scatter of the typical sample and populate the compact tail of the observed size distribution, coinciding with the population of \ion{C}{iv}-strong sources \citep{Roberts-Borsani.etal.26}.

Part of the offset is definitional: M24 report major-axis effective radii, whereas both \citet{Roberts-Borsani.etal.26} and we use circularized half-light radii. For an axis ratio \ifm{q}, a major-axis radius exceeds the corresponding circularized radius by a factor of approximately \ifm{q^{-1/2}}, or \ifm{\gtrsim0.1\,\si{dex}} for \ifm{q\lesssim0.6}. This convention alone therefore shifts their relation upward relative to ours, although a precise correction would require matched radius estimators. Dust attenuation, the finite angular resolution of the observations, our limited sample size, and our exclusion of interacting systems also affect this comparison. Their effects on the observed sizes and magnitudes are discussed in \cref{subsubsec:observedsize}. Finally, for galaxies fainter than \ifm{\Mx{UV}\simeq-19}, the measured sizes approach the numerical resolution limit of the simulations.
\begin{figure*}
    \centering
    \includegraphics[width=\textwidth]{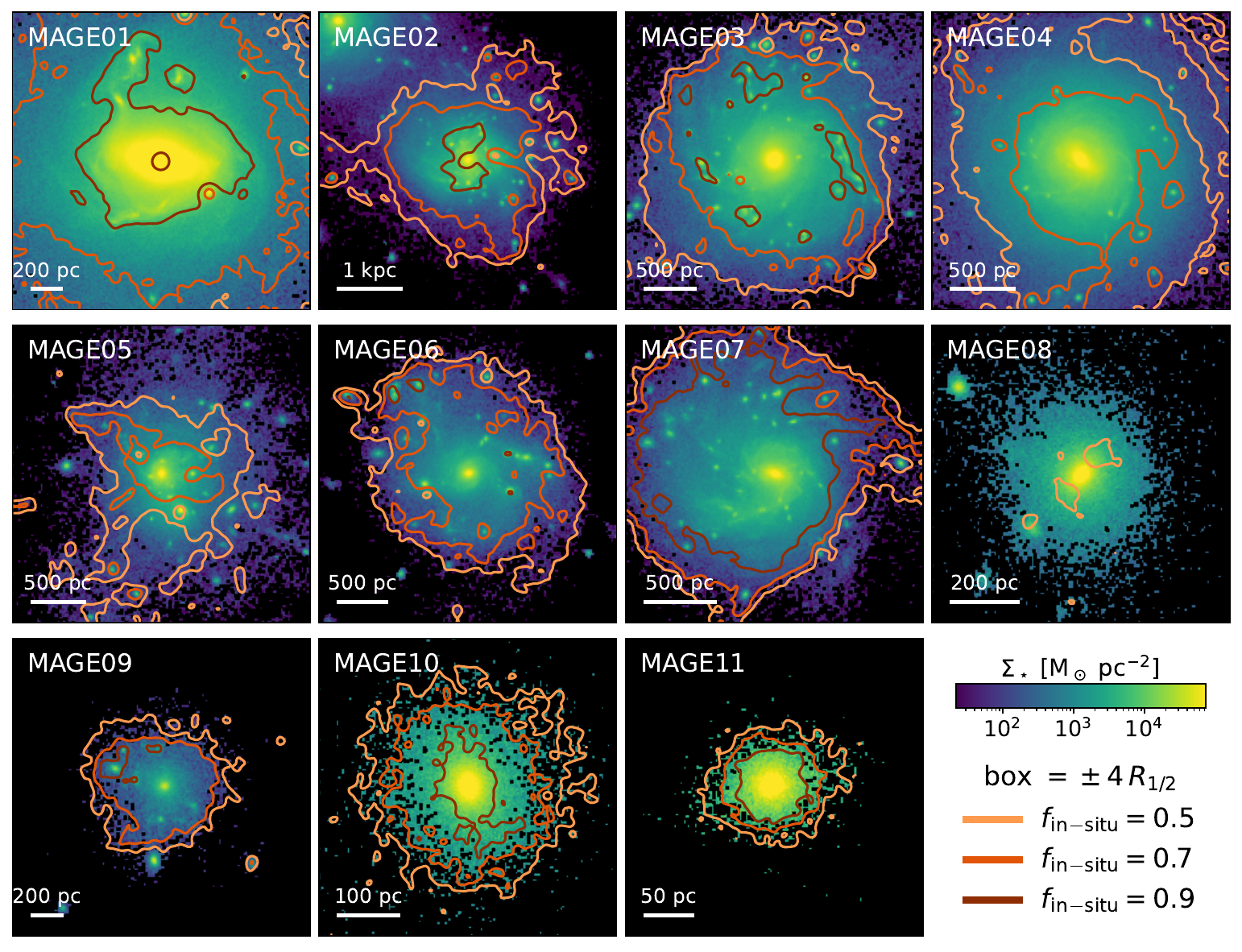}
    \caption{Face-on stellar surface density maps at \ifm{z=9} for the 11 central galaxies of our target zoom-in haloes, ordered by decreasing halo mass from \run{MAGE01} to \run{MAGE11}. Each panel spans \ifm{\pm4\,\Rx{1/2}} of the respective galaxy, with a projection depth of \ifm{2\Rx{gal}}, and the scale bar in each panel gives the physical scale. The colour scale, shared by all panels, shows the logarithmic stellar surface density. Contours in progressively darker shades of orange mark projected in-situ stellar mass fractions of 0.5, 0.7, and 0.9, respectively. The in-situ classification uses stellar birth positions and the companion-identification procedure described in \cref{subsec:halofinder}. In ten of the eleven galaxies, the 0.7 contour encloses the central stellar body, including the recent major-merger system \run{MAGE04}. The exception is \run{MAGE08}, observed within \SI{50}{Myr} of a major merger, where the in-situ fraction exceeds 0.5 only in small central patches. The 0.5 contour traces the outer boundary of the in-situ-dominated region and isolates satellites and tidal debris, and the small closed contour at the centre of \run{MAGE01} marks a localized central depression in the in-situ fraction.}
    \label{fig:proj_insitu}
\end{figure*}

\begin{figure*}
    \centering	\includegraphics[width=\textwidth]{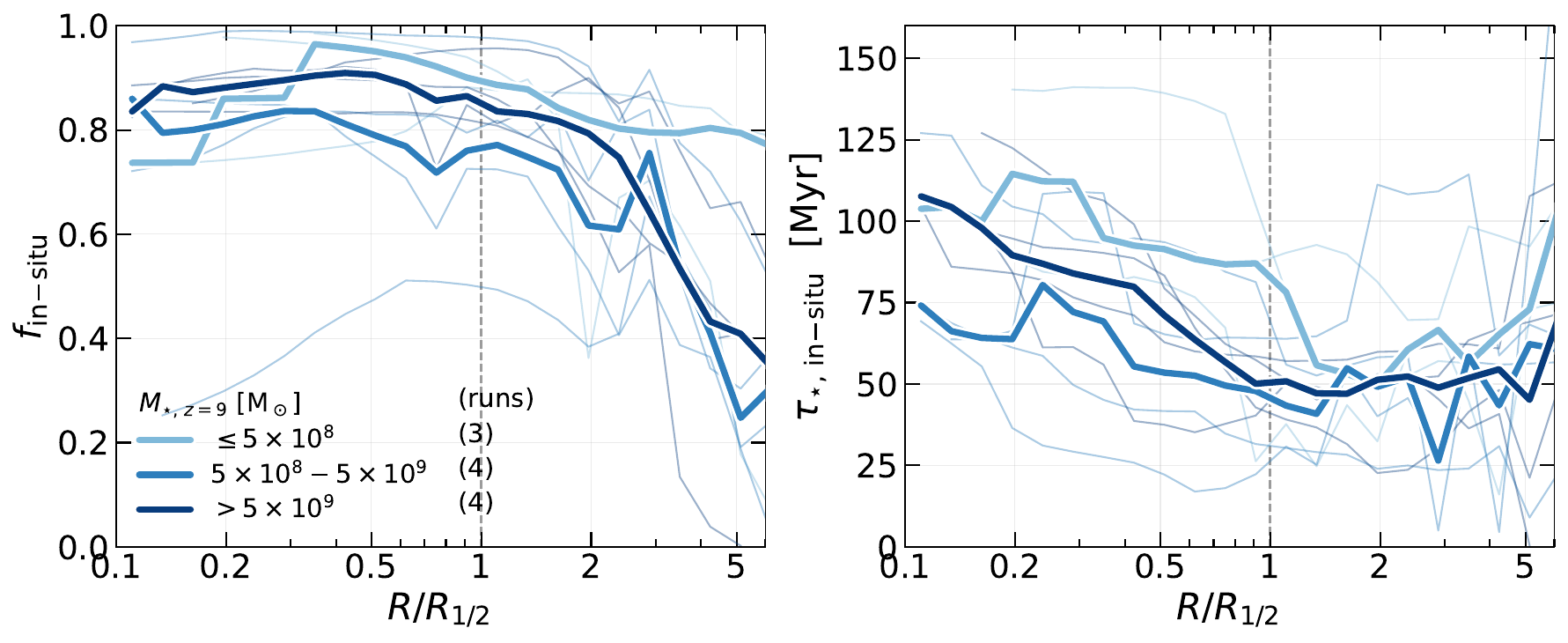}
   \caption{Radial profiles of the in-situ stellar mass fraction (\textit{left}) and the age of the in-situ stars (\textit{right}) at \ifm{z=9} for the 11 central galaxies of our target zoom-in haloes. The radial coordinate is normalized by the individual 3D half-mass radius, \ifm{\Rx{1/2}}, of each galaxy, and the vertical dashed line marks \ifm{R=\Rx{1/2}}. The galaxies are divided into three groups according to their stellar mass at \ifm{z=9}: \ifm{\Mx{\star}\leq\SI{5e8}{\Msun}} (light blue; 3 runs), \ifm{\SI{5e8}{\Msun}<\Mx{\star}\leq\SI{5e9}{\Msun}} (medium blue; 4 runs), and \ifm{\Mx{\star}>\SI{5e9}{\Msun}} (dark blue; 4 runs), corresponding to the three evolutionary phases identified in \cref{sec:diskformation}. Thick curves show the median of each group, and thin curves of the same colour show the individual galaxies.}
   \label{fig:Insitu}
\end{figure*}

\begin{figure*}
    \centering	\includegraphics[width=\textwidth]{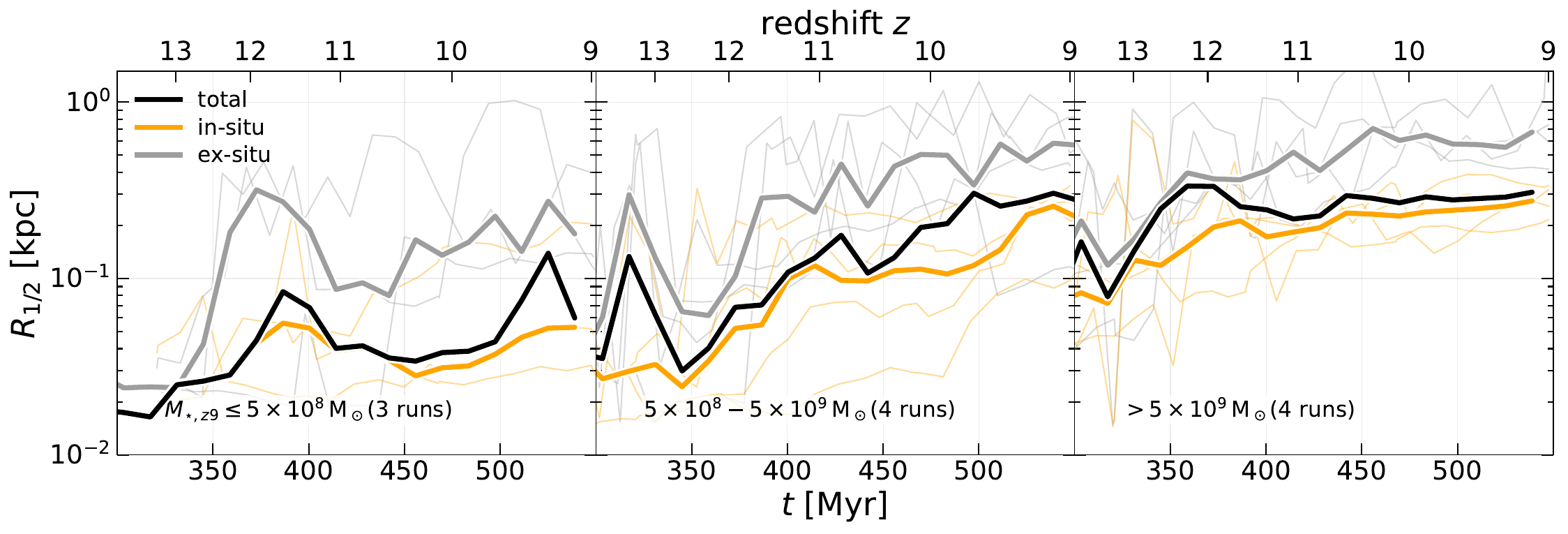}
   \caption{Time evolution of the 3D half-mass radius for the 11 central galaxies in our zoom-in suite, split into the same three mass groups as in \cref{fig:Insitu}: \ifm{\Mx{\star}\leq\SI{5e8}{\Msun}} (\textit{left}; 3 runs), \ifm{\SI{5e8}{\Msun}<\Mx{\star}\leq\SI{5e9}{\Msun}} (\textit{middle}; 4 runs), and \ifm{\Mx{\star}>\SI{5e9}{\Msun}} (\textit{right}; 4 runs), with \ifm{\Mx{\star}} measured at \ifm{z=9}. Colours indicate the stellar component: total (black), in-situ (orange), and ex-situ (grey). Thick lines trace group medians, while thin lines show individual galaxies (plotted for the in-situ and ex-situ components only). The upper axis gives the corresponding redshift.}
   \label{fig:sizegrowth}
\end{figure*}

\section{In-situ vs. ex-situ growth: how did early galaxies assemble their stars?}
\label{sec:in-situsf}
In \cref{sec:in-situsf,sec:diskformation}, we explore the origin of size evolution at cosmic dawn. We follow the central galaxies of the 11 target haloes, whose main-progenitor histories allow a more detailed analysis than the full statistical sample. Motivated by the turnover in galaxy size at \ifm{\Mx{\star}\simeq\SI{5e9}{\Msun}} found in \cref{subsec:size-mass}, and by a second characteristic mass at \ifm{\Mx{\star}\simeq\SI{5e8}{\Msun}} that emerges from the kinematic analysis in \cref{sec:diskformation}, we divide our sample of 11 central galaxies into three groups based on their final stellar mass at \ifm{z=9}. In order of increasing mass, these are a low-mass group (\ifm{\Mx{\star}\leq\SI{5e8}{\Msun}}; \run{MAGE09--11}), an intermediate-mass group (\ifm{\SI{5e8}{\Msun}<\Mx{\star}\leq\SI{5e9}{\Msun}}; \run{MAGE05--08}), and a high-mass group (\ifm{\Mx{\star}>\SI{5e9}{\Msun}}; \run{MAGE01--04}). This grouping follows each system according to its final mass. An individual progenitor can cross several instantaneous mass intervals during its evolution. We relate these intervals to the structural and kinematic transitions in \cref{sec:diskformation}.

We compare the spatial distributions of in-situ and ex-situ stars defined in \cref{subsec:halofinder} to assess their contributions to size evolution. \Cref{fig:proj_insitu} provides a visual overview at \ifm{z=9}, showing face-on stellar surface-density maps of the 11 target central galaxies within \ifm{\pm4\,\Rx{1/2}} and contours of the projected in-situ fraction at 0.5, 0.7, and 0.9. In ten of the eleven galaxies, the 0.7 contour encloses the central stellar body, and in several systems (e.g. \run{MAGE01}, \run{MAGE07}, \run{MAGE10}, and \run{MAGE11}) the in-situ fraction exceeds 0.9 over much of the inner region. In-situ stars therefore dominate the central stellar mass in ten of the eleven systems, spanning a broad range of masses and merger histories. The predominantly in-situ central mass of \run{MAGE04}, despite its recent major mergers, is consistent with the high gas fractions during the interaction (bottom-left panel of \cref{fig:MockImage_MAGE04}): stars formed locally can contribute more central mass than the directly accreted stellar population.

The one exception is \run{MAGE08}, which experienced a major merger within the last \SI{50}{Myr}. Its in-situ fraction exceeds 0.5 only in small central patches, so directly accreted stars dominate its stellar body at \ifm{z=9}. In the other systems dominated by in-situ stars, ex-situ stars concentrate in the outskirts, where the 0.5 contour isolates satellites and tidal debris, and in localized central dips in the in-situ fraction such as the small closed contour at the centre of \run{MAGE01}. These dips likely trace the remnants of accreted companions that have sunk to the core via dynamical friction \citep{Chandrasekhar.43,Barnes.Hernquist.92,Boylan-Kolchin.etal.08}. The projected maps thus show predominantly in-situ growth in the central regions, with ex-situ stars dominating the stellar body only in the one galaxy observed immediately after a major merger.

The left panel of \cref{fig:Insitu} presents the radial profiles of the 3D in-situ stellar mass fraction at \ifm{z=9} for all central galaxies, split into the three mass groups. In all three groups, the median in-situ fraction is \ifm{\fx{in\text{-}situ}\gtrsim0.7} within \ifm{\Rx{1/2}} and declines mildly towards the centre. This central dip is likely driven either by ex-situ stars that have sunk inward through dynamical friction, as suggested by the closed central contour of \run{MAGE01} in \cref{fig:proj_insitu}, or by more recent in-situ star formation near \ifm{\Rx{1/2}}, as shown in the right panel of \cref{fig:Insitu}. Beyond \ifm{\simeq1.5\text{--}2\,\Rx{1/2}}, the median of the low-mass group remains at \ifm{\simeq0.8} out to \ifm{5\,\Rx{1/2}}, albeit with only three galaxies and a large spread among them, whereas the in-situ fraction of the intermediate- and high-mass groups drops rapidly, reaching \ifm{\fx{in\text{-}situ}\simeq0.2\text{--}0.4} at \ifm{R\simeq5\,\Rx{1/2}}.

The galaxy-to-galaxy scatter within \ifm{\Rx{1/2}} also differs between the groups: the three low-mass galaxies span \ifm{\fx{in\text{-}situ}\simeq0.7\text{--}1}. The intermediate-mass group includes one galaxy (\run{MAGE08}) with \ifm{\fx{in\text{-}situ}\simeq0.3\text{--}0.5} throughout its stellar body due to a very recent major merger within \ifm{\SI{50}{Myr}}, and the high-mass group exhibits the smallest scatter, with all four galaxies confined to \ifm{\fx{in\text{-}situ}\simeq0.8\text{--}1}. The low scatter within the high-mass group is consistent with a rapid central stellar mass buildup (discussed below), which anchors their inner \ifm{\fx{in\text{-}situ}} near unity despite their different assembly histories in this sample. Finally, the increasing scatter at large radii reflects stochastic contributions from mergers, accreted satellites, and tidal debris, which predominantly populate the galactic outskirts.

\begin{figure*}
    \centering	\includegraphics[width=\textwidth]{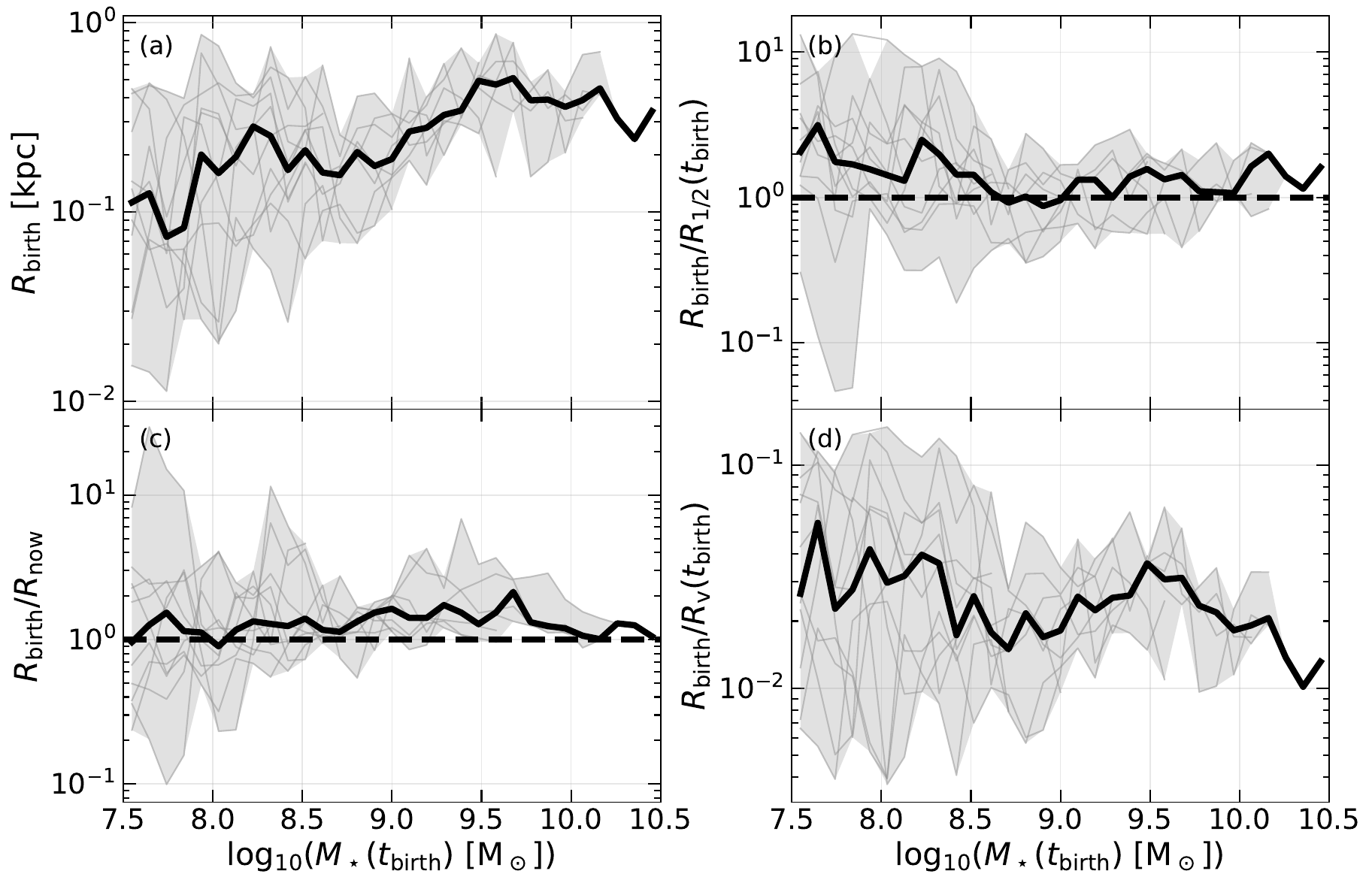}
   \caption{Birth properties of the in-situ stars at \ifm{z=9} in the 11 central target galaxies, shown as a function of the stellar mass of the host galaxy at the time of formation, \ifm{\Mx{\star}(\tx{birth})}. Panel (a) shows the absolute stellar birth radii, panel (b) their ratio to the half-mass radius at birth, panel (c) their ratio to each star's galactocentric radius at \ifm{z=9}, \ifm{\Rx{now}}, and panel (d) their ratio to the virial radius at birth. Thin grey curves show the medians of individual galaxies, the thick black curve shows the median of the full sample, and the grey shaded region spans the full range (minimum to maximum) of the individual galaxies' medians. Above \ifm{\Mx{\star}\gtrsim\SI{1.5e10}{\Msun}}, only \run{MAGE01} contributes. The horizontal dashed lines in panels (b) and (c) mark a ratio of unity.}
   \label{fig:sizebirth}
\end{figure*}

The fact that in-situ stars dominate the central stellar mass does not by itself establish that they control size evolution. To investigate this, we compare the half-mass radii of the total, in-situ, and ex-situ stellar populations in \cref{fig:sizegrowth}, showing the three mass groups in separate panels. In all three groups, the total \ifm{\Rx{1/2}} broadly follows the in-situ radius and is smaller than the ex-situ radius. The in-situ component exhibits a smoother growth history and closely tracks the total \ifm{\Rx{1/2}} for the majority of the evolution. In the high-mass group, the two become nearly indistinguishable after \ifm{t\simeq\SI{440}{Myr}}. In contrast, the ex-situ \ifm{\Rx{1/2}} is a factor of \ifm{\sim2\text{--}4} larger and highly stochastic, driving several transient peaks in the overall size (e.g., at \ifm{t\simeq\SI{525}{Myr}} in the low-mass group and at \ifm{t\simeq\SI{320}{Myr}} in the intermediate-mass group). This behaviour supports a scenario where in-situ star formation governs the baseline structural evolution. Ex-situ stars, accreted via mergers, settle into a more extended distribution and their smaller mass contribution generally limits their effect on the total \ifm{\Rx{1/2}}.

The growth histories differ markedly between the groups. The low-mass group grows only slowly, from \ifm{\simeq\SI{15}{pc}} to \ifm{\simeq\SI{50}{pc}}, sizes that are only a few times the \ifm{\sim\SI{10}{pc}} spatial resolution. The intermediate-mass group exhibits the steepest temporal evolution, growing steadily by an order of magnitude, from \ifm{\simeq\SI{30}{pc}} at \ifm{t\simeq\SI{300}{Myr}} to \ifm{\simeq\SI{250}{pc}} at \ifm{z=9}. In the high-mass group, the in-situ \ifm{\Rx{1/2}} grows from \ifm{\simeq\SI{80}{pc}} at \ifm{t\simeq\SI{300}{Myr}} to \ifm{\simeq\SI{250}{pc}} by \ifm{t\simeq\SI{440}{Myr}} and flattens thereafter, so that the intermediate-mass group catches up and reaches comparable sizes by \ifm{z=9}. These trends align with the size--mass relation presented in \cref{fig:sizemass}, which exhibits a steep slope at the low-mass end that flattens---and even turns over---above \ifm{\Mx{\star}\simeq\SI{5e9}{\Msun}}.

The radial ages of in-situ stars provide a complementary view of their formation history. The right panel of \cref{fig:Insitu} shows their radial age profiles at \ifm{z=9}. All three groups exhibit an inside-out formation signature, with the oldest stars at the centre and the youngest at \ifm{R\gtrsim\Rx{1/2}}, but the shape of the age profile changes systematically with mass. In the low-mass group, the median age is nearly flat at \ifm{\simeq90\text{--}110\,\si{\Myr}} throughout \ifm{R<\Rx{1/2}} and then drops abruptly to \ifm{\simeq50\text{--}65\,\si{\Myr}} outside \ifm{\Rx{1/2}}. The intermediate-mass group shows a smooth, shallow age gradient and has the lowest median stellar ages of the three groups at all radii within \ifm{\Rx{1/2}}. Its median age declines from \ifm{\simeq\SI{75}{Myr}} at the centre to \ifm{\simeq\SI{45}{Myr}} at \ifm{1\text{--}2\,\Rx{1/2}}, consistent with the steady, ongoing size growth of this group in \cref{fig:sizegrowth}. The high-mass group shows a smooth but steeper gradient. The median stellar age in its innermost region is \SI{110}{Myr}, corresponding to an epoch of \ifm{t\simeq\SI{440}{Myr}} (\ifm{z\simeq10.6}). This roughly coincides with the epoch when \run{MAGE04} experiences a merger-driven peak in star formation (\cref{fig:MockImage_MAGE04}). Moving outward, the median stellar age declines smoothly, reaches a minimum of \ifm{\simeq\SI{50}{Myr}} near \ifm{\Rx{1/2}}, and remains low out to \ifm{5\,\Rx{1/2}}, indicating a more recently formed population in this extended radial region. The older centre and younger outskirts of the high-mass group are consistent with more pronounced inside-out quenching following the rapid central stellar buildup: star formation declines in the compact core while continuing at larger radii. These galaxies, with \ifm{\Mx{\star}>\SI{5e9}{\Msun}}, have also undergone the size turnover and developed substantial rotational support (\cref{sec:diskformation}).

The birth radii of the in-situ stars present at \ifm{z=9} provide a retrospective measure of their formation sites (\cref{fig:sizebirth}). This analysis is restricted to stars retained within the selected galaxy at that epoch. We bin these stars by the stellar mass of the main progenitor at their formation time, \ifm{\Mx{\star}(\tx{birth})} (\cref{subsec:halofinder}). Since the 11 galaxies reach a given mass at different epochs, this places them on a common evolutionary sequence and allows a direct comparison with the size--mass relation (\cref{fig:sizemass}) and with the transition masses identified in \cref{sec:diskformation}. Panel (a) displays the absolute birth radii. The median \ifm{\Rx{birth}} is \ifm{\simeq100\text{--}200\,\si{\pc}} for \ifm{\Mx{\star}\lesssim\SI{1e9}{\Msun}}, rises to \ifm{\simeq400\text{--}500\,\si{\pc}} at \ifm{\Mx{\star}\simeq(3\text{--}5)\times10^{9}\,\si{\Msun}}, and then flattens, and even declines slightly, at higher masses, mirroring the turnover of the size--mass relation. Panel (b) compares the birth radii to the half-mass radii at the time of formation. At \ifm{\Mx{\star}\lesssim\SI{5e8}{\Msun}}, the median birth radius exceeds \ifm{\Rx{1/2}} by a factor of \ifm{\sim2}, with a galaxy-to-galaxy range spanning more than an order of magnitude: stars form predominantly beyond the existing stellar half-mass radius, and this geometric offset favours expansion of \ifm{\Rx{1/2}} and the steep slope of the size--mass relation at the low-mass end observed in \cref{fig:sizemass}. Above \ifm{\Mx{\star}\simeq\SI{5e8}{\Msun}}, the ratio converges to \ifm{\Rx{birth}/\Rx{1/2}\simeq1\text{--}1.5} and the galaxy-to-galaxy range narrows markedly, indicating that star formation tracks the half-mass radius more closely, although this convergence precedes the turnover near \ifm{\SI{5e9}{\Msun}}.

We explore stellar redistribution by comparing these birth radii to the galactocentric radii of the same selected stars at \ifm{z=9} (panel (c) of \cref{fig:sizebirth}). The median \ifm{\Rx{birth}/\Rx{now}} is generally \ifm{\simeq1\text{--}1.5}. It peaks at \ifm{\sim2} for stars formed at \ifm{\Mx{\star}\simeq\SI{5e9}{\Msun}} and returns to unity for those formed at \ifm{\Mx{\star}\gtrsim\SI{1e10}{\Msun}}. The galaxy-to-galaxy range is large for stars formed at \ifm{\Mx{\star}\lesssim\SI{1e8}{\Msun}}, consistent with the more chaotic assembly of the low-mass progenitors, but narrows considerably at higher masses. These ratios indicate modest net decreases in galactocentric radius for most cohorts, with larger decreases near the turnover. Finally, the normalized birth radius, \ifm{\Rx{birth}/\Rx{v}} (panel (d)), fluctuates around \ifm{\simeq0.02} without a systematic trend up to \ifm{\Mx{\star}\simeq\SI{5e9}{\Msun}}, and then declines to \ifm{\simeq0.01} at \ifm{\Mx{\star}\gtrsim\SI{1e10}{\Msun}}.

Overall, a coherent picture emerges along the stellar-mass sequence. Below \ifm{\Mx{\star}\simeq\SI{5e8}{\Msun}}, the star-forming radius inferred from stellar birth sites tracks the halo at \ifm{\simeq0.02,\Rx{v}} and lies well outside \ifm{\Rx{1/2}}, driving rapid expansion of the half-mass radius toward this outer boundary. Between \ifm{\Mx{\star}\simeq\SI{5e8}{\Msun}} and \ifm{\simeq\SI{5e9}{\Msun}}, the two radii have converged and grow together. Above \ifm{\Mx{\star}\simeq\SI{5e9}{\Msun}}, the star-forming radius saturates around \ifm{\sim\SI{400}{pc}} while the halo continues to grow, so that \ifm{\Rx{birth}/\Rx{v}}, and hence \ifm{\Rx{1/2}/\Rx{v}}, decline. This coincides with the turnover of the size--mass relation in \cref{fig:sizemass}. This sequence suggests a transition from size growth that tracks the halo to stronger internal regulation at high stellar mass. We return to the physical origin of these transition masses in \cref{sec:discuss}.

\begin{figure*}
    \centering
    \includegraphics[width=\textwidth]{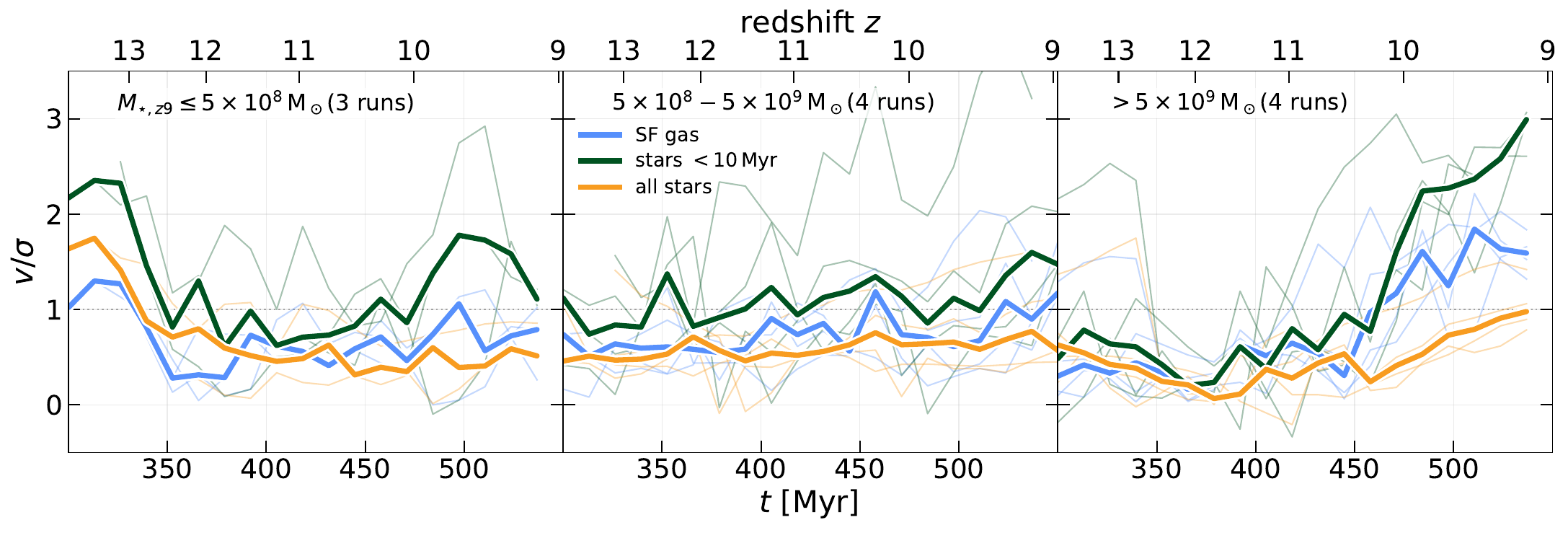}
    \caption{Time evolution of the rotation-to-dispersion ratio, \ifm{v/\sigma}, of the star-forming gas (blue), stars younger than \SI{10}{Myr} (green), and the full stellar population (orange) for the 11 central target galaxies in our zoom-in suite. The galaxies are split into the same three mass groups as in \cref{fig:Insitu}: \ifm{\Mx{\star}\leq\SI{5e8}{\Msun}} (\textit{left}; 3 runs), \ifm{\SI{5e8}{\Msun}<\Mx{\star}\leq\SI{5e9}{\Msun}} (\textit{middle}; 4 runs), and \ifm{\Mx{\star}>\SI{5e9}{\Msun}} (\textit{right}; 4 runs), with \ifm{\Mx{\star}} measured at \ifm{z=9}. Thick curves show the median of each group, and thin curves of the same colour show the individual galaxies. The stellar ratios are mass weighted, whereas the gas ratio is SFR weighted and its effective dispersion includes the thermal and unresolved turbulent contributions (\cref{subsec:kinematics}). The horizontal dotted line marks \ifm{v/\sigma=1}, and the upper axis gives the corresponding redshift. Only the high-mass group shows a sustained rise in all three tracers, beginning near \ifm{z\simeq10}.}
    \label{fig:v2sigma_centrals}
\end{figure*}

\begin{figure*}
    \centering
    \includegraphics[width=\textwidth]{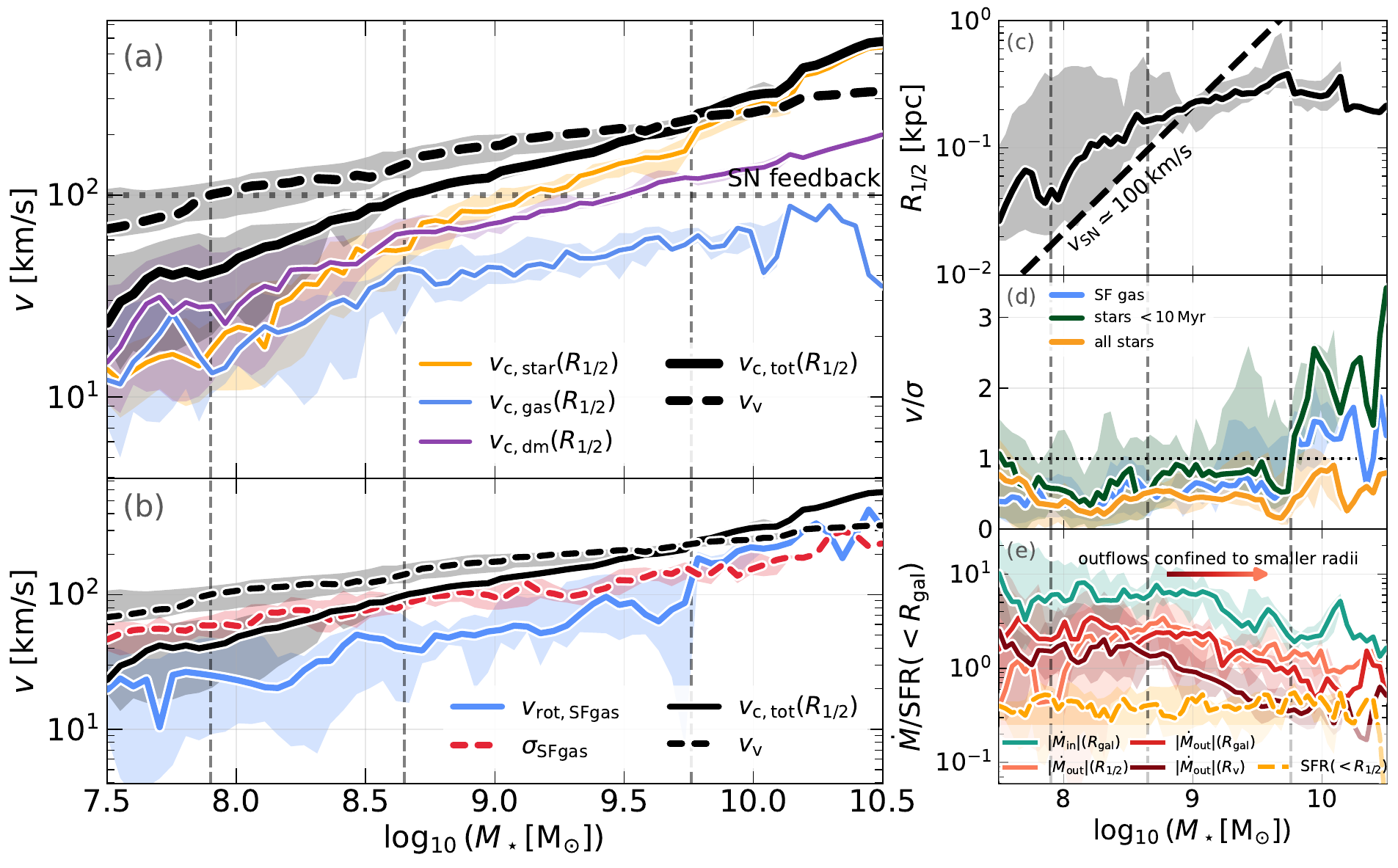}
    \caption{Kinematic and structural properties of the 11 central galaxies as a function of stellar mass. Panel (a) shows the circular velocity evaluated at \ifm{\Rx{1/2}} for the total mass (black), dark matter (purple), stars (orange), and gas (blue), together with the halo virial velocity, \ifm{\vx{v}} (dashed). The horizontal dotted line marks the critical velocity for effective supernova (SN) feedback, \ifm{\vx{SN}\simeq\SI{100}{km/s}} \citep{Dekel.Silk.86}. Panel (b) compares the kinematics of the star-forming gas with the same two gravitational scales: the SFR-weighted rotation speed, \ifm{\vx{rot,SFgas}} (solid blue), and the SFR-weighted effective one-dimensional velocity dispersion, \ifm{\sigma_{\rm SFgas}} (dashed red), which includes the thermal and unresolved turbulent contributions (\cref{subsec:kinematics}), are shown together with \ifm{\vx{c,tot}(\Rx{1/2})} (solid black) and \ifm{\vx{v}} (dashed black) repeated from panel (a). Panel (c) shows the 3D half-mass radius; the dashed line marks the half-mass radius at which the circular velocity of the enclosed stellar mass, \ifm{\vx{c,star}(\Rx{1/2})=\sqrt{G\Mx{\star}/(2\Rx{1/2})}}, equals \ifm{\vx{SN}}. Panel (d) shows \ifm{v/\sigma} for the star-forming gas (blue; the ratio of the quantities in panel (b)), for stars younger than \SI{10}{Myr} (green), and for all stars (orange); the horizontal dotted line marks \ifm{v/\sigma=1}. Panel (e) shows the outflow rates across \ifm{\Rx{1/2}}, \ifm{\Rx{gal}\equiv0.15\,\Rx{v}}, and \ifm{\Rx{v}}, in progressively darker shades of red, and the inflow rate across \ifm{\Rx{gal}} (teal), each divided by the SFR within \ifm{\Rx{gal}} averaged over the past \SI{50}{Myr}, i.e. the mass-loading factor with respect to the galaxy-wide SFR, together with the central star-formation fraction, \ifm{\mathrm{SFR}(<\Rx{1/2})/\mathrm{SFR}(<\Rx{gal})}, with both SFRs averaged over the past \SI{5}{Myr} (dashed orange). The annotated arrow marks the trend that outflows are gradually confined to smaller radii as stellar mass increases, where the rate across \ifm{\Rx{v}} remains below that across \ifm{\Rx{gal}}, which falls below that across \ifm{\Rx{1/2}} at \ifm{\Mx{\star}\simeq\SI{5e8}{\Msun}}. In all panels, curves show the median over the sample and shaded regions the 16th--84th percentile ranges; the shading terminates at \ifm{\Mx{\star}\gtrsim\SI{1.5e10}{\Msun}}, beyond which \run{MAGE01} is the only galaxy in the sample. The three vertical dashed lines in all panels denote the transition masses: (1) \ifm{\Mx{\star}\simeq\SI{1e8}{\Msun}}, where \ifm{\vx{v}>\vx{SN}}; (2) \ifm{\Mx{\star}\simeq\SI{5e8}{\Msun}}, where \ifm{\vx{c,tot}(\Rx{1/2})>\vx{SN}} and \ifm{\vx{c,star}(\Rx{1/2})>\vx{c,dm}(\Rx{1/2})}; and (3) \ifm{\Mx{\star}\simeq\SI{5e9}{\Msun}}, where \ifm{\vx{c,tot}(\Rx{1/2})>\vx{v}} and \ifm{\vx{rot,SFgas}} overtakes \ifm{\sigma_{\rm SFgas}}. These kinematic thresholds coincide with distinct structural transitions: the first marks the onset of steady half-mass radius growth; the second aligns with the mass where the outflow rate across \ifm{\Rx{1/2}} exceeds that across \ifm{\Rx{gal}} and size growth decelerates; and the third triggers the rapid emergence of rotational support in both the star-forming gas and young stars.
    }
    \label{fig:diskformation_centralsf}
\end{figure*}

\section{Disk Formation in Massive Galaxies}
\label{sec:diskformation}

We now connect the kinematics of the star-forming gas, the newly formed stars, and the full stellar population to size evolution, using the rotation-to-dispersion ratio \ifm{v/\sigma} defined in \cref{subsec:kinematics}. In the diagnostics below, we measure the inflow rate across \ifm{\Rx{gal}}, the outflow rates across \ifm{\Rx{1/2}}, \ifm{\Rx{gal}}, and \ifm{\Rx{v}}, and the SFRs within \ifm{\Rx{1/2}} and \ifm{\Rx{gal}}.

\Cref{fig:v2sigma_centrals} presents the time evolution of \ifm{v/\sigma} for the star-forming gas, for stars younger than \SI{10}{Myr}, and for the full stellar population, divided into the three final-mass groups defined in \cref{sec:in-situsf}. The low-mass group fluctuates strongly without a sustained increase. Apart from transiently high values at \ifm{t\lesssim\SI{330}{Myr}} (\ifm{z\gtrsim13}), the median \ifm{v/\sigma} of its young stars varies between \ifm{\simeq0.6} and \ifm{1.8}, with a temporary peak near \ifm{t\simeq\SI{500}{Myr}} that has fallen back to \ifm{\simeq1.1} by \ifm{z=9}, while the star-forming gas and the full stellar population remain at \ifm{v/\sigma\lesssim1}. High rotational support for young stars can therefore occur transiently in low-mass systems and does not by itself establish a lasting disk. The intermediate-mass group evolves more gradually: the median \ifm{v/\sigma} of its young stars remains near \ifm{\simeq0.8\text{--}1.4} for most of the interval and reaches \ifm{\simeq1.5} by \ifm{z=9}, the star-forming gas stays at \ifm{\simeq0.5\text{--}0.9} apart from isolated excursions and reaches \ifm{\simeq1\text{--}1.2} only after \ifm{t\simeq\SI{520}{Myr}}, and the full stellar population remains at \ifm{\simeq0.5\text{--}0.8}. Only the high-mass group exhibits a sharp, sustained kinematic transition near \ifm{t\simeq\SI{470}{Myr}} (\ifm{z\simeq10}). Before the transition, all three tracers remain below unity, mostly at \ifm{v/\sigma\simeq0.2\text{--}0.8}, for \ifm{\simeq\SI{150}{Myr}}. Within a few tens of Myr thereafter, the median \ifm{v/\sigma} of the young stars rises from \ifm{\lesssim1} to \ifm{\gtrsim2} and that of the star-forming gas from \ifm{\simeq1} to \ifm{\simeq1.6}. By \ifm{z=9}, the young stars reach \ifm{v/\sigma\simeq3} and the star-forming gas \ifm{\simeq1.6}, whereas the full stellar population responds more slowly, rising from \ifm{\simeq0.4} at the transition to \ifm{\simeq1} at \ifm{z=9}. Rotational support is thus established first in the gas and the newly formed stars and spreads only gradually to the stellar body as a whole. The transition occurs approximately \SI{30}{Myr} after the formation epoch implied by the median central stellar age, \ifm{t\simeq\SI{440}{Myr}} (\cref{fig:Insitu}, right panel), consistent with central stellar buildup preceding rapid rotational support. 

Panel (d) of \cref{fig:diskformation_centralsf} shows the mass dependence of \ifm{v/\sigma} for the star-forming gas, for stars younger than \SI{10}{Myr}, and for the full stellar population. The star-forming gas and the young stars track each other below the transition, with median \ifm{v/\sigma\simeq0.5\text{--}1}, and both rise above unity near \ifm{\Mx{\star}\simeq\SI{5e9}{\Msun}}. The young stars reach \ifm{v/\sigma\simeq2\text{--}3}, above the \ifm{\simeq1.5} of the gas. Part of this difference reflects the thermal and turbulent support included in \ifm{\sigma_{\rm SFgas}}, which the stars do not inherit. The median full-population ratio remains approximately at or below unity, with \ifm{v/\sigma\lesssim1} even at the highest masses, so the rotational support is confined to the newly formed component inside a stellar body that stays dispersion-dominated. Above \ifm{\Mx{\star}\simeq\SI{1.5e10}{\Msun}}, only \run{MAGE01} remains in the sample. It experiences a sharp decline in rotational support due to mergers, followed by a recovery, eventually reaching its peak \ifm{(v/\sigma)_{\star,10}\gtrsim3} at \ifm{\Mx{\star}>\SI{3e10}{\Msun}}. The rapid onset of rotational support near \ifm{\Mx{\star}\simeq\SI{5e9}{\Msun}} is therefore common to the gas and the young stars.

The stellar-mass threshold for disk formation coincides with the turnover in the half-mass radius (panel (c)). Although this panel isolates the 11 central galaxies without binning by redshift, the weak redshift dependence permits comparison with the broader morphological trends shown in the left panel of \cref{fig:sizemass}. Panel (b) shows the SFR-weighted rotation speed and velocity dispersion of the star-forming gas, together with the circular velocity at the half-mass radius and the halo virial velocity. The effective dispersion tracks the halo velocity scale over the full mass range, \ifm{\sigma_{\rm SFgas}\simeq0.6\,\vx{v}}, with no systematic change across the transition. The rotation speed behaves differently. Below the transition, \ifm{\vx{rot,SFgas}\simeq0.3\text{--}0.5\,\vx{c,tot}(\Rx{1/2})} and lies below \ifm{\sigma_{\rm SFgas}}. Near \ifm{\Mx{\star}\simeq\SI{5e9}{\Msun}} where \ifm{\vx{c,tot}} exceeds \ifm{\vx{v}}, it rises steeply to \ifm{\simeq0.6\text{--}0.7\,\vx{c,tot}(\Rx{1/2})} and overtakes \ifm{\sigma_{\rm SFgas}}. The rise in \ifm{v/\sigma} at the transition is therefore driven by the emergence of coherent rotation at a dispersion that continues to follow \ifm{\vx{v}}, rather than by a drop in dispersion. Even after the transition the gas does not rotate at the circular speed at \ifm{\Rx{1/2}}. \ifm{\vx{rot,SFgas}} is an SFR-weighted mean over the gas within \ifm{\Rx{gal}}, which includes non-circular and misaligned motions and gas at radii where the circular speed differs from its value at \ifm{\Rx{1/2}}, so the ratio should be read as a measure of coherent rotation rather than of the departure from circular orbits at a fixed radius.

To understand why \ifm{\vx{c,tot}} eventually exceeds \ifm{\vx{v}}, we examine the stellar, dark matter, and gas contributions in panel (a). This steep rise in \ifm{\vx{c,tot}} is driven by a concurrent surge in the stellar component, \ifm{\vx{c,star}} (orange line), which exceeds \ifm{\vx{c,dm}} from \ifm{\Mx{\star}\simeq\SI{5e8}{\Msun}} onward, signalling a phase of rapid central mass buildup prior to disk formation. We adopt a characteristic velocity \ifm{\vx{SN}}, the critical virial velocity of \citet{Dekel.Silk.86} shown by the horizontal dotted line. It is defined by equating the energy deposited in the interstellar gas by SNe during the first burst of star formation, after accounting for radiative losses from the remnants, with the gravitational binding energy of the gas in the halo. Haloes with \ifm{\vx{v}<\vx{SN}} can lose a substantial fraction of their gas to SN-driven winds, whereas deeper potential wells retain it. Because the deposited SN energy scales with the stellar mass formed, nearly independently of the gas density, the resulting threshold depends only weakly on the collapse epoch. We use \ifm{\vx{SN}} as a characteristic potential-depth scale associated with effective SN feedback, not as a measured wind launch speed or an exact escape criterion.

We find that the virial velocity \ifm{\vx{v}} and the central circular velocity \ifm{\vx{c,tot}} surpass \ifm{\vx{SN}} at \ifm{\Mx{\star}\simeq\SI{1e8}{\Msun}} and \ifm{\Mx{\star}\simeq\SI{5e8}{\Msun}}, respectively. Near \ifm{\Mx{\star}\simeq\SI{1e8}{\Msun}}, the half-mass radius begins to grow more rapidly as the halo virial speed exceeds \ifm{\vx{SN}\simeq\SI{100}{km/s}}. This is consistent with a scenario in which gas is ejected to the outer circumgalactic medium (CGM), mixes with high-angular-momentum gas, and falls back onto the outskirts of the galaxy. In this case, the angular momentum of the CGM gas may help the formation radius track halo growth, so the star-forming radius scales with the halo radius (\cref{sec:in-situsf}). Near \ifm{\Mx{\star}\simeq\SI{5e8}{\Msun}}, the total circular speed at \ifm{\Rx{1/2}} exceeds \ifm{\vx{SN}}, consistent with stronger retention of SN-ejected gas by the central potential. In this interpretation, faster replenishment sustains central star formation and cumulative stellar buildup, while the gas fraction declines and the stellar contribution overtakes that of dark matter within \ifm{\Rx{1/2}}. Size growth slows as the star-forming radius converges on the half-mass radius (\cref{fig:sizebirth}), until the final size turnover at \ifm{\Mx{\star}\simeq\SI{5e9}{\Msun}}. 

Panel (e) supports this picture through the outflow rates measured at \ifm{\Rx{1/2}}, \ifm{\Rx{gal}}, and \ifm{\Rx{v}}, together with the inflow rate at \ifm{\Rx{gal}}. We normalize these rates by the SFR within \ifm{\Rx{gal}}, averaged over the past \SI{50}{Myr}. The normalization expresses each rate relative to the galaxy-wide SFR and removes the common increase in the rates as the galaxy grows. The outflow rates include all gas moving outward through each shell (\cref{subsec:halofinder}), without tracing its origin or applying a velocity cut, so they include bound turbulent and orbital motions and, at the outer shells, gas associated with satellites and tidal debris that never belonged to the central galaxy. They are therefore crude measures of the competition between feedback and gravity, and we rely on their ordering and trends rather than on their absolute values. \zyn{The overall trend, highlighted by the arrow in panel (e), is towards less large-scale outflow at higher mass. Above \ifm{\Mx{\star}\simeq\SI{5e8}{\Msun}}, the rate across \ifm{\Rx{v}} drops faster than that across \ifm{\Rx{gal}}, whereas the rate across \ifm{\Rx{1/2}} rises rapidly and exceeds both. This ordering is consistent with outward-moving gas being increasingly confined to smaller radii as the potential deepens. The mass-loading factor across \ifm{\Rx{v}} declines from \ifm{\simeq1.5\text{--}2} at \ifm{\Mx{\star}\lesssim\SI{5e8}{\Msun}} to \ifm{\simeq0.4} at \ifm{\Mx{\star}\gtrsim\SI{5e9}{\Msun}}, as expected if a deepening potential retains a growing share of the ejected gas, whereas the mass-loading factor across \ifm{\Rx{1/2}} is \ifm{\simeq1} at \ifm{\Mx{\star}\gtrsim\SI{5e9}{\Msun}}. We take the difference \ifm{\dotM{out}(\Rx{1/2})-\dotM{out}(\Rx{gal})} as a proxy for the gas that leaves the centre but does not escape the galaxy, and is therefore available for recycling within it. Normalized by the SFR, this proxy grows with stellar mass and becomes comparable to the inflow mass-loading factor across \ifm{\Rx{gal}} at \ifm{\Mx{\star}\gtrsim\SI{5e9}{\Msun}}. This suggests that, at these masses, gas recycled within the galaxy supplies the centre at a rate comparable to that of gas entering the galaxy from outside.}

The central star-formation fraction, defined as the SFR within \ifm{\Rx{1/2}} divided by the SFR within \ifm{\Rx{gal}}, both averaged over the past \SI{5}{Myr} (dashed orange curve in panel (e)), stays at \ifm{\simeq0.4}, within \ifm{0.3\text{--}0.55}, up to and across the three transitions, and subsequently declines in the final bins at \ifm{\Mx{\star}\gtrsim\SI{2.5e10}{\Msun}}, where \ifm{\vx{c,gas}} shows a similar sharp decline. The decline of the central star-formation fraction, the low central gas fraction, and the younger outskirts in the age profiles of \cref{fig:Insitu} together suggest the onset of inside-out quenching, although this decline is traced only by \run{MAGE01}. We assemble these trends into a sequence of regimes in \cref{subsec:sequence} (\cref{fig:cartoon}).

\section{Discussion}
\label{sec:discuss}

\subsection{Connection to wet compaction events}
\label{subsec:compaction}

The central mass buildup, size turnover, disk formation, and inside-out quenching identified in \cref{sec:in-situsf,sec:diskformation} share several features with wet compaction at cosmic noon. In that picture, dissipative gas inflow builds a dense, star-forming core, often termed a ``blue nugget'', and changes both the central mass composition and the galaxy kinematics \citep{Dekel.Burkert.14,Zolotov.etal.15,Tacchella.etal.16,Lapiner.etal.23}. The end of compaction and the onset of disk formation occur at a characteristic stellar mass of \ifm{\sim\SI{1e10}{\Msun}} (\citealt{Lapiner.etal.23}\zyn{; blue} dashed line in the bottom panel of \cref{fig:host_gal_dists}), only a factor of two above the characteristic stellar mass at which we find rapid disk formation.

In these studies, compaction is typically triggered by wet mergers, counter-rotating streams, or violent disk instability, though \citet{Lapiner.etal.23} also identified gas recycling as a viable driver. Among the four massive galaxies in our sample that undergo compaction, \run{MAGE04} is likely driven by wet mergers (\cref{fig:MockImage_MAGE04}). In contrast, the remaining three systems (\run{MAGE01}--\run{MAGE03}) exhibit no evidence of prior major mergers, counter-rotating streams, or violent disk instability preceding disk formation. The compaction-like behavior in these \software{MAGE} galaxies is therefore likely driven by rapid internal gas recycling, aligning with the broader theoretical framework of compaction. Our interpretation emphasizes the internal conditions that confine outflows, permit rapid recycling, and allow central gas accumulation to persist long enough to become structurally significant, without attributing the transition to any particular inflow trigger. Stronger inflows induced by wet mergers, for instance, would accelerate central gas accumulation. The proposed sequence begins when \ifm{\vx{c,tot}(\Rx{1/2})>\vx{SN}}, marking a shift from halo-scale to galaxy-scale recycling, and is followed by rapid growth of the central potential. Cosmological accretion continues to supply gas, and interactions may shape individual bursts, but the proposed regulator of the characteristic transition is the evolving balance between central gravity and feedback.

In more detail, once \ifm{\vx{c,tot}(\Rx{1/2})} exceeds the characteristic feedback scale \ifm{\vx{SN}} adopted in \cref{sec:diskformation}, at \ifm{\Mx{\star}\simeq\SI{5e8}{\Msun}}, feedback becomes less effective at removing gas from the galaxy: the outflow rate across \ifm{\Rx{gal}} becomes lower than the rate across \ifm{\Rx{1/2}} (\cref{fig:diskformation_centralsf}). This ordering is consistent with outflows failing to escape the galaxy and recycling there, rather than reaching the CGM before returning to the outskirts, although shell fluxes alone do not establish those trajectories. \zyn{The proxy for gas recycled within the galaxy, \ifm{\dotM{out}(\Rx{1/2})-\dotM{out}(\Rx{gal})}, grows with stellar mass and becomes comparable to the inflow rate across \ifm{\Rx{gal}} near \ifm{\SI{5e9}{\Msun}} (\cref{sec:diskformation}), suggesting that recycling within the galaxy may supply the centre at a rate comparable to that of gas entering the galaxy from outside.} Note that bursty star formation in individual galaxies is not confined to this regime. Galaxies between \ifm{\SI{1e8}{\Msun}} and \ifm{\SI{5e8}{\Msun}} also undergo bursts, but the proposed scenario predicts that gas expelled from these galaxies returns on the longer halo-scale timescale, leading to longer intervals before the gas supply is restored and therefore a lower burst duty cycle. Recycling within the galaxy shortens the replenishment timescale, so the central gas supply is restored more quickly after each star-forming episode. The expected signature is a shorter burst cycle with a higher duty cycle: star formation recurs more frequently, and a gas supply maintained between bursts reduces the contrast between burst peaks and the inter-burst level without eliminating the central enhancements. A follow-up analysis of star-formation variability using \software{MAGE} indicates reduced burstiness above \ifm{\Mx{\star}\simeq\SI{5e8}{\Msun}} (Yao et al., in preparation), consistent with more effective gas confinement and shorter replenishment timescales. The accumulating stellar mass deepens the potential and further promotes gas retention, potentially sustaining efficient central star formation until star formation consumes the gas and/or redistribution reduces the central gas supply. A related consequence of this central buildup may be the suppression of global ``breathing modes'', in which gas coherently expands and contracts across the galaxy. A more centrally concentrated mass profile causes radial oscillations at different radii to lose coherence, promoting gas mixing and facilitating disk formation \citep{Hopkins.etal.23}.

Although individual galaxies undergo bursts in both regimes, the rapid stellar buildup near \ifm{\Mx{\star}\simeq\SI{5e9}{\Msun}} is the phase in our massive systems that coincides with both the size turnover and the sharp increase in rotational support. By this stage, the accumulated central mass has raised \ifm{\vx{c,tot}(\Rx{1/2})} above \ifm{\vx{v}}. The condition for retaining gas within the galaxy is therefore met before the condition associated with rapid disk formation: the central stellar buildup between \ifm{\SI{5e8}{\Msun}} and \ifm{\SI{5e9}{\Msun}} assembles the core, and the rapid buildup near \ifm{\SI{5e9}{\Msun}} marks the point at which the accumulated mass changes both the size evolution and the kinematics. In this picture, the compaction-like phase in \software{MAGE} may begin near \ifm{\Mx{\star}\simeq\SI{5e8}{\Msun}}, where the growth of \ifm{\Rx{1/2}} slows from rapid expansion. The size turnover near \ifm{\SI{5e9}{\Msun}} may therefore reflect the continued decrease in the size--mass slope during central buildup, rather than a separate event. Note that the characteristic mass of \ifm{\sim\SI{1e10}{\Msun}} in \citet{Lapiner.etal.23} marks the end of compaction, after which the galaxy's half-mass radius begins to increase again. Our simulations have not yet reached this phase of renewed size growth (\cref{subsubsec:gnz11}). The size turnover and the onset of rotational support in \software{MAGE} may thus be two consequences of the same central stellar buildup, although they need not occur at exactly the same stellar mass.

The role of gas recycling in this buildup remains an interpretation: our shell-flux measurements do not distinguish fresh accretion from returning gas or track its angular-momentum loss. Establishing the relative contributions of these processes requires a dedicated analysis of gas trajectories. The connection between central mass concentration and rotational support can nevertheless be examined directly, as we do next in \cref{subsec:diskcondition}.

\begin{figure*}
    \centering
    \includegraphics[width=\textwidth]{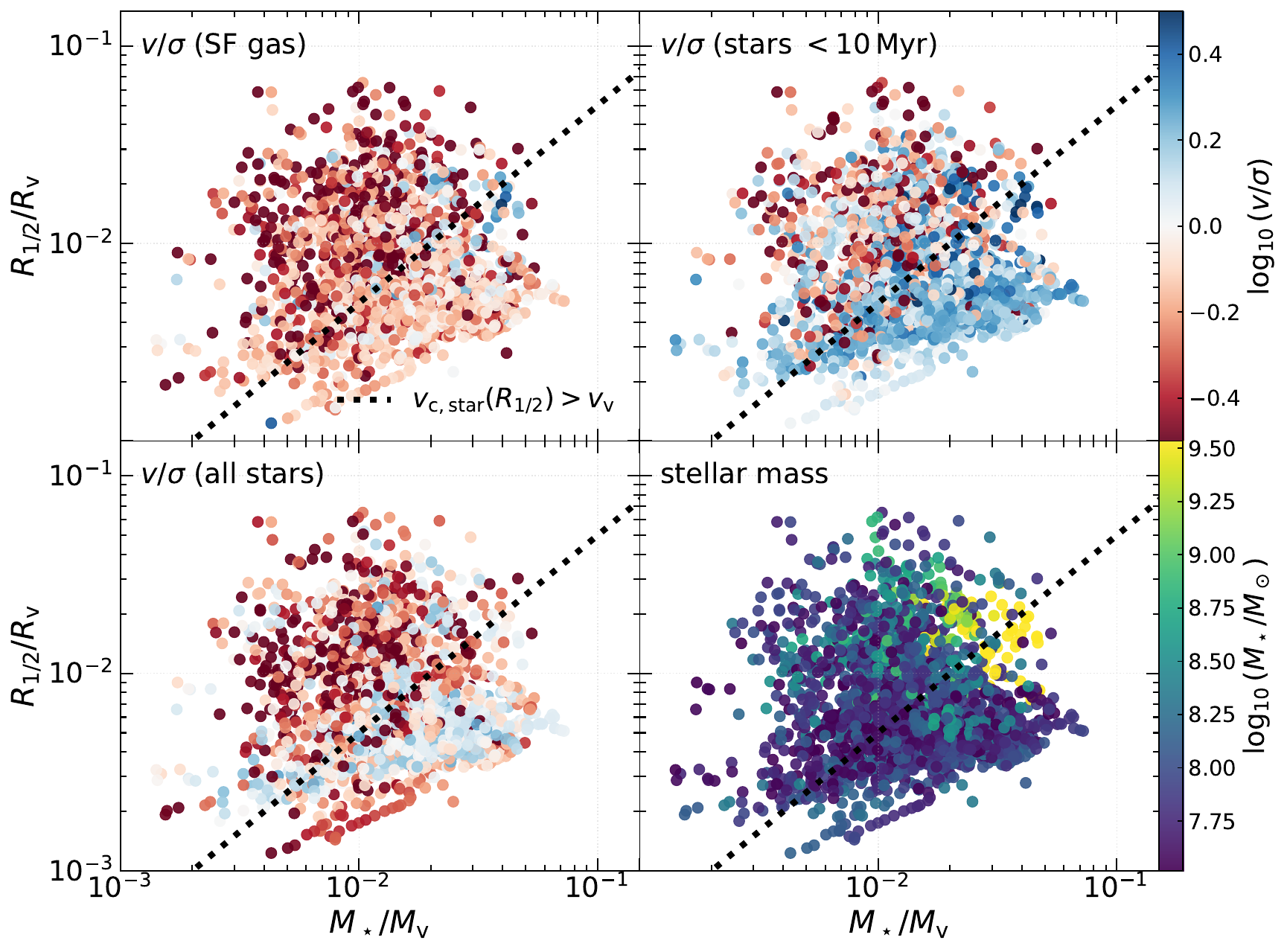}
   \caption{Galaxy size, expressed as the half-mass radius normalized by the virial radius (\ifm{\Rx{1/2}/\Rx{v}}), as a function of the stellar-to-halo mass ratio (\ifm{\Mx{\star}/\Mx{v}}). Each point represents a central galaxy from a single snapshot of the full sample across \ifm{9\lesssim z\lesssim15}. The points are colour-coded by \ifm{\log_{10}(v/\sigma)} of the star-forming gas (\zyn{\textit{top left}}), of stars younger than \SI{10}{Myr} (\zyn{\textit{top right}}), and of the full population of stars (\zyn{\textit{bottom left}}); blue indicates rotationally supported systems and red dispersion-dominated systems. \zyn{In the \textit{bottom right} panel the same points are colour-coded by stellar mass, \ifm{\log_{10}(\Mx{\star}/\si{\Msun})}, with the colour scale saturating at \ifm{\Mx{\star}=10^{9.5}\,\si{\Msun}}.} The black dotted line marks the reference boundary \ifm{\vx{c,star}(\Rx{1/2})=\vx{v}} of \cref{eq:diskformation}. The division into dispersion-dominated systems in the upper left and rotationally supported systems in the lower right is sharpest for the young stars. Rotational support in the star-forming gas occurs in only a small minority of snapshots, preferentially clustered at high \ifm{\Mx{\star}/\Mx{v}} and high \ifm{\Rx{1/2}/\Rx{v}}. \zyn{These snapshots belong to the most massive galaxies in the sample, with \ifm{\Mx{\star}\gtrsim10^{9.5}\,\si{\Msun}}, whereas the compact systems at low \ifm{\Rx{1/2}/\Rx{v}}, where young stars are rotationally supported but the star-forming gas is not, mostly have \ifm{\Mx{\star}\lesssim10^{8.5}\,\si{\Msun}}}.}
   \label{fig:diskformation_vc2vv}
\end{figure*}

\subsection{The conditions for disk formation}
\label{subsec:diskcondition}

In \cref{sec:diskformation}, the rapid emergence of rotational support in our massive galaxies coincides with \ifm{\vx{c,tot}(\Rx{1/2})>\vx{v}} at a stellar mass of approximately \ifm{\SI{5e9}{\Msun}}. This comparison of the central circular speed with the halo velocity scale provides an indirect diagnostic of disk formation. The connection depends on the actual rotation speed and velocity dispersion of the gas and stars. Writing \ifm{\vx{c}\equiv\vx{c,tot}(\Rx{1/2})}, the rotation-to-dispersion ratio can be written as the product
\begin{equation}
\frac{v}{\sigma}
=\frac{\vx{rot}}{\vx{c}}
\frac{\vx{c}}{\vx{v}}
\left(\frac{\sigma}{\vx{v}}\right)^{-1},
\label{eq:rotation_support_factors}
\end{equation}
where \ifm{\vx{rot}} denotes the mean rotation speed of the population whose dispersion is \ifm{\sigma}. For the star-forming gas, the relevant quantities are shown in panel (b) of \cref{fig:diskformation_centralsf}. The velocity dispersion remains at \ifm{\sigma_{\rm SFgas}\simeq0.6\,\vx{v}} across the mass range, while \ifm{\vx{rot,SFgas}/\vx{c}} rises from \ifm{\simeq0.4} below the transition to \ifm{\simeq0.7} above it. Once the disk has formed, \ifm{v/\sigma} is therefore approximately \ifm{\vx{c}/\vx{v}}, to within \ifm{\sim20} per cent. The substantially steeper rise in \ifm{\vx{rot,SFgas}} relative to \ifm{\vx{c,tot}} warrants further investigation. One plausible mechanism is the suppression of the breathing mode, which facilitates gas mixing so that \ifm{\vx{rot,SFgas}} approaches \ifm{\vx{c,tot}} \citep{Hopkins.etal.23}. A definitive numerical verification is deferred to future work. Under the approximate assumptions \ifm{\vx{rot}\sim\vx{c}} and \ifm{\sigma\sim\vx{v}}, the criterion for rotational support, \ifm{v/\sigma>1}, becomes \ifm{\vx{c}/\vx{v}>1}. This condition is met near the transition mass of \ifm{\Mx{\star}\simeq\SI{5e9}{\Msun}}. Since \ifm{\vx{c,tot}(\Rx{v})=\vx{v}}, this inequality implies a centrally peaked rotation curve that declines with radius. Because stars dominate the enclosed mass at these stellar masses, the more conservative diagnostic \ifm{\vx{c,star}(\Rx{1/2})>\vx{v}} can be used. Under spherical symmetry, this criterion becomes
\begin{equation}
\label{eq:diskformation}
\frac{\Mx{\star}}{\Mx{v}}>\frac{2\Rx{1/2}}{\Rx{v}}.
\end{equation}
This expression relates the stellar-to-halo mass ratio to the stellar-to-halo radius ratio, and identifies compact galaxies with a large stellar mass fraction as the systems most favourable to disk onset under these assumptions.

We evaluate this diagnostic against our statistical sample in \cref{fig:diskformation_vc2vv}, colouring each snapshot of a galaxy by the \ifm{v/\sigma} value for star-forming gas, stars younger than \SI{10}{Myr}, or the full stellar population\zyn{, and, in the bottom-right panel, by stellar mass}. The snapshots form a continuous distribution in the \ifm{\Rx{1/2}/\Rx{v}}--\ifm{\Mx{\star}/\Mx{v}} plane, with no clear structural bimodality. Young stars show a clear kinematic gradient across this plane: rotational support is more common in galaxies with a larger stellar mass fraction and a more centrally concentrated stellar component. The full stellar population shows a similar but weaker trend, consistent with the older stars retaining a broader range of orbital motions.

In contrast, the star-forming gas exhibits a distinct kinematic behavior. Numerous systems with rotation-supported young stars host dispersion-dominated star-forming gas—most notably at low stellar masses where the galaxy-to-halo size ratio, \ifm{\Rx{1/2}/\Rx{v}}, remains low. Conversely, rotation-supported gas is predominantly found in higher-mass systems that have developed a centrally concentrated stellar component. \zyn{The stellar-mass panel of \cref{fig:diskformation_vc2vv} makes this separation explicit: the snapshots with rotation-supported star-forming gas coincide with the most massive galaxies, \ifm{\Mx{\star}\gtrsim10^{9.5}\,\si{\Msun}}, consistent with the transition mass of \ifm{\simeq\SI{5e9}{\Msun}} identified in \cref{sec:diskformation}. The compact, low-\ifm{\Rx{1/2}/\Rx{v}} systems with rotation-supported young stars but dispersion-dominated gas instead mostly have \ifm{\Mx{\star}\lesssim10^{8.5}\,\si{\Msun}}, even though many of them lie well within the rotation-favoured side of the boundary of \cref{eq:diskformation}.} This difference arises mainly because young stars inherit the resolved velocities of their parent gas but do not retain its thermal or unresolved turbulent pressure support in our \ifm{v/\sigma} measurement. A coherent orbiting clump or interaction-triggered burst can therefore produce rotation-supported young stars even when the star-forming gas as a whole is not a settled disk. Young-star rotation alone may consequently overstate disk formation in low-mass galaxies. The full stellar population provides a complementary, longer-term measure, but its response is diluted by pre-existing stars and their accumulated orbital history.

This raises the separate question of long-term disk survival. \citet{Dekel.etal.20} argue that disks become long-lived only above \ifm{\Mx{v}\simeq\SI{2e11}{\Msun}}, where the merger timescale exceeds the orbital time (\ifm{\tx{mer}>\tx{orb}}). At our median stellar-to-halo mass ratio of \ifm{\Mx{\star}/\Mx{v}\simeq0.02}, this halo mass corresponds to \ifm{\Mx{\star}\simeq\SI{4e9}{\Msun}}, close to the transition mass identified empirically in \cref{sec:diskformation} \zyn{and in \ifm{(v/\sigma)_\mathrm{SFgas}} of \cref{fig:diskformation_vc2vv}. This coincidence suggests that the characteristic stellar mass for disk onset in our sample may also approximate the stellar-mass threshold for long-term disk survival, and that a centrally peaked rotation curve traced by young stars alone may overstate short-term disk formation in low-mass galaxies.}

The approximation \ifm{v/\sigma\sim\vx{c}/\vx{v}} fails when \ifm{\vx{rot}\ll\vx{c}} or \ifm{\sigma\ll\vx{v}} (\cref{eq:rotation_support_factors}). An observational example for the first case is found in the stellar dynamical models of CALIFA galaxies by \citet{Kalinova.etal.17}, where galaxies with sharply peaked circular-speed curves are slowly rotating and pressure-supported systems (their Fig.~9). A central peak in the inferred circular speed therefore need not correspond to strong stellar rotation. Gas can reach \ifm{\vx{rot}\simeq\vx{c}} more readily than a pre-existing collisionless stellar population because shocks and radiative cooling dissipate non-circular motions, provided that the gas has coherent angular momentum. However, collisionality alone does not ensure this outcome, since pressure support, continuing inflow, and feedback can prevent the gas from settling. A deep central potential thus facilitates rotation without guaranteeing it.

The second case, \ifm{\sigma\ll\vx{v}}, usually applies at later epochs, after a dynamically cold, thin disk has settled. \citet{Ubler.etal.19} measure a decline in the mean intrinsic ionized-gas dispersion from \ifm{\sim\SI{45}{km/s}} at \ifm{z\sim2.3} to \ifm{\sim\SI{30}{km/s}} at \ifm{z\sim0.9}, well below the virial velocity of the host haloes, and \citet{Tamburro.etal.09} find a characteristic \ion{H}{i} dispersion of \ifm{10\pm2\,\si{km/s}} at the optical radius of 11 nearby disk galaxies. These measurements trace different gas phases and should not be equated directly, but together they illustrate a dynamically colder regime in which \ifm{\sigma\ll\vx{v}} and rotational support can be high even without a centrally peaked rotation curve. At early epochs, before disk formation, gas velocity dispersion can instead be comparable to the virial velocity, as shown in simulations \citep{Goldner.etal.26}. In our sample \ifm{\sigma_{\rm SFgas}\simeq0.6\,\vx{v}} both before and after the transition (panel (b) of \cref{fig:diskformation_centralsf}).

Because disk formation criteria are sensitive to the chosen definition of \ifm{v/\sigma} and underlying theoretical assumptions, simulations diverge on the precise criteria. Specifically, young stellar tracers systematically exhibit stronger rotational support than the underlying star-forming gas, and standard kinematic approximations such as \ifm{\vx{rot}\simeq\vx{c}} and \ifm{\sigma\simeq\vx{v}} could break down. In the \software{ARTEMIS} zoom-in simulations of Milky Way-mass halos, \citet{Dillamore.etal.23} find that stellar disks generally form at a halo virial mass threshold of \ifm{\simeq(6\pm3)\times10^{11}\,\si{\Msun}}, while the circular velocity profile is still rising between one and three stellar half-mass radii (their Figs.~4 and 5). Similarly, using the \software{FIRE-2} suite, \citet{Myrtag.etal.26} show that their \run{Romeo} system develops a young stellar disk while the circular-velocity slope diagnostic remains positive. They attribute this disk formation to the cessation of baryonic mass sloshing relative to the center of mass within the host potential, noting that the velocity slope becomes negative only several gigayears after spin-up (their Fig.~20). In \software{MAGE}, we instead identify disk formation using star-forming gas weighted by SFR. Under this metric, reaching a sufficiently high stellar mass alongside an established, centrally concentrated stellar core provides a more robust criterion for disk formation. We note that the disks formed here are kinematically thick rather than thin Milky Way analogs, as we discuss in \cref{subsubsec:diskcomparison}, meaning that \ifm{\sigma} remains comparable to the halo virial velocity.

\begin{figure}
    \centering	\includegraphics[width=\columnwidth]{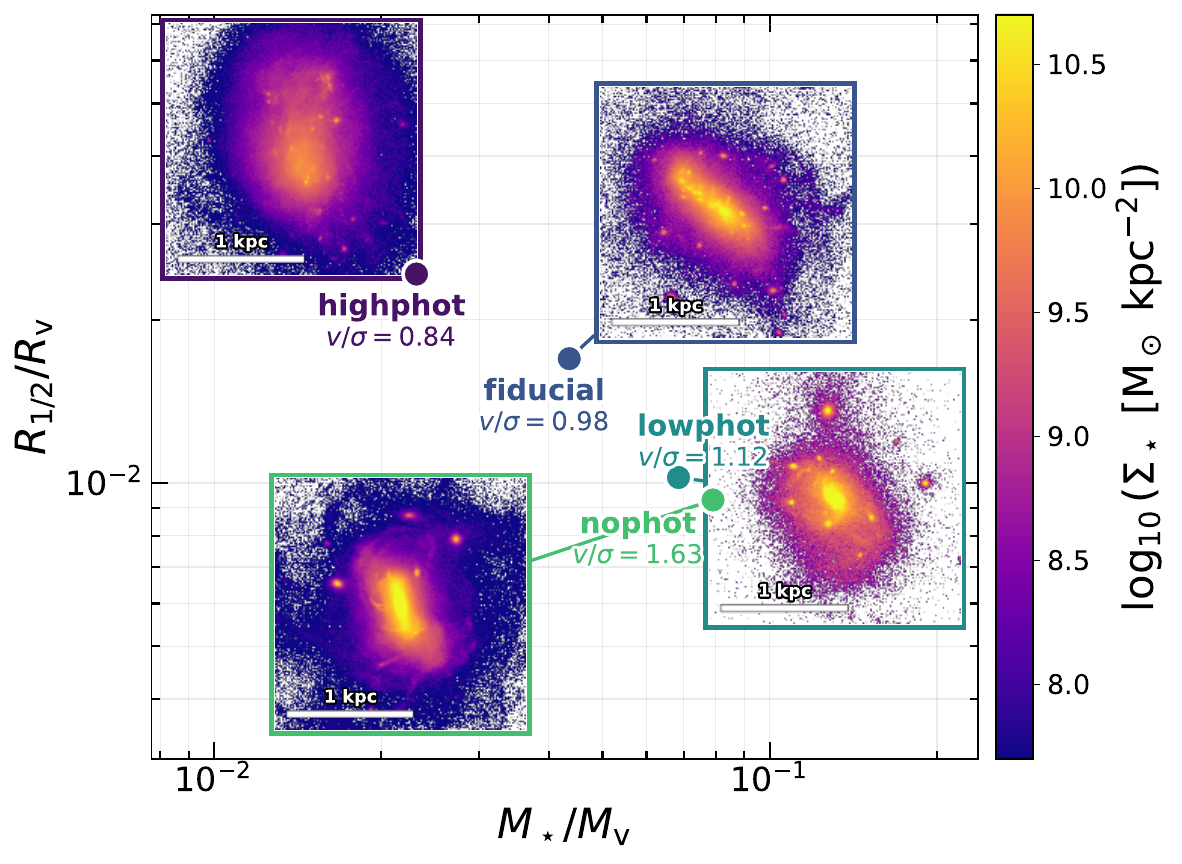}
   \caption{The impact of photoionization feedback strength on galaxy sizes and kinematics. The 3D stellar half-mass radius normalized by the virial radius, \ifm{\Rx{1/2}/\Rx{v}}, is plotted against the stellar-to-halo mass ratio, \ifm{\Mx{\star}/\Mx{v}}, at \ifm{z=9} for four simulations of the same halo from \citet{Andalman.etal.25} that differ only in the strength of photoionizing early stellar feedback: \run{highphot}, \run{fiducial}, \run{lowphot}, and \run{nophot}, in order of decreasing photoionization strength. \zyn{The \run{fiducial} run uses the same initial conditions and physical parameters as our \run{MAGE07} simulation (see \cref{subsec:phot}).} The axes match those of \cref{fig:diskformation_vc2vv}; because the four runs share the same halo, the normalized quantities are proportional to \ifm{\Mx{\star}} and \ifm{\Rx{1/2}}. \zyn{The SFR-weighted rotation-to-dispersion ratios of the star-forming gas, \ifm{(v/\sigma)_{\rm SFgas}} (\cref{eq:kinematic_definition}), are annotated next to the points: \ifm{v/\sigma=0.84}, 0.98, 1.12, and 1.63 in order of decreasing photoionization strength.} Each point is accompanied by a stellar surface density map of the corresponding galaxy, projected along the \ifm{z}-axis of the simulation box, \SI{2}{kpc} wide and framed in the same colour, with a common logarithmic colour scale for \ifm{\Sigma_\star} (colour bar). White scale bars mark \SI{1}{kpc}. As the photoionization feedback weakens, the galaxies become more massive, with \ifm{\Mx{\star}/\Mx{v}} rising from \ifm{\simeq0.02} to \ifm{\simeq0.08} (\ifm{\Mx{\star}\simeq\SI{2e9}{\Msun}} to \ifm{\simeq\SI{6e9}{\Msun}}), more centrally concentrated, with \ifm{\Rx{1/2}/\Rx{v}} shrinking from \ifm{\simeq0.024} to \ifm{\simeq0.009} (\ifm{\Rx{1/2}\simeq\SI{350}{pc}} to \ifm{\simeq\SI{130}{pc}}), and more rotationally supported, with \zyn{\ifm{(v/\sigma)_{\rm SFgas}} rising monotonically from \ifm{\simeq0.8} to \ifm{\simeq1.6}}. In these axes the boundary of \cref{eq:diskformation}, \ifm{\Mx{\star}/\Mx{v}=2\Rx{1/2}/\Rx{v}} (the dotted line in \cref{fig:diskformation_vc2vv}), separates \run{highphot}, which lies on the dispersion-dominated side, from the three weaker-feedback runs.}
   \label{fig:phot}
\end{figure}

\subsection{Impact of feedback strength on galaxy sizes and kinematics}
\label{subsec:phot}
We now assess the sensitivity of our results to the adopted feedback model, with particular emphasis on photoionization heating. \cref{fig:phot} compares four simulations of the \run{MAGE07} halo at \ifm{z=9} that differ only in the strength of early stellar feedback from photoionization: \run{highphot}, \run{fiducial} (\zyn{same set-up as \run{MAGE07}})\footnote{\zyn{The \run{fiducial} run uses the same initial conditions and physical parameters as \run{MAGE07}, but was performed with an earlier version of the code, so the two are not identical realizations. At \ifm{z=9} their host haloes agree closely (\ifm{\Mx{v}\simeq}\zyn{\ifm{\SI{0.95e11}{\Msun}}}), but the galaxies are caught in different phases of their bursty star formation, and the star-forming gas has \ifm{(v/\sigma)_{\rm SFgas}=0.98} in the \run{fiducial} run versus 1.26 in \run{MAGE07} (\cref{tab:sims}). This difference is within the short-timescale variability of \run{MAGE07}, whose \ifm{(v/\sigma)_{\rm SFgas}} spans \ifm{\simeq0.9\text{--}1.3} (16th--84th percentiles) over the surrounding \ifm{\pm\SI{30}{Myr}}. We therefore compare the photoionization runs only with one another.}}, \run{lowphot}, and \run{nophot}, in order of decreasing photoionization strength. The model parameters are described in \citet{Andalman.etal.25}. For each run, we compare the galaxy-to-halo size ratio (\ifm{\Rx{1/2}/\Rx{v}}), the stellar-to-halo mass ratio (\ifm{\Mx{\star}/\Mx{v}}), and \zyn{the rotational support of the star-forming gas, \ifm{(v/\sigma)_{\rm SFgas}}.}

The figure shows that varying photoionization strength produces systematic shifts in the normalized size--mass plane. From \run{highphot} to \run{nophot}, the stellar-to-halo mass ratio, \ifm{\Mx{\star}/\Mx{v}}, increases from \ifm{\simeq0.02} to \ifm{\simeq0.08}, while the galaxy-to-halo size ratio, \ifm{\Rx{1/2}/\Rx{v}}, decreases from \ifm{\simeq0.024} to \ifm{\simeq0.009}. The two models with the weakest photoionization, \run{lowphot} and \run{nophot}, have similar stellar-to-halo mass ratios and small normalized sizes, with size ratios of approximately \ifm{0.010} and \ifm{0.009}, respectively. \zyn{The rotational support of the star-forming gas also increases as photoionization weakens: \ifm{(v/\sigma)_{\rm SFgas}} rises monotonically from \ifm{\simeq0.8} in \run{highphot} to \ifm{\simeq1.0} in \run{fiducial}, \ifm{\simeq1.1} in \run{lowphot}, and \ifm{\simeq1.6} in \run{nophot}. The largest increase occurs from \run{lowphot} to \run{nophot}, although their mass and size ratios change comparatively little.} The stellar surface-density maps illustrate the corresponding structural differences. With strong photoionization, the stellar body is extended and diffuse, spanning most of the \SI{2}{kpc} field, whereas with weak or no photoionization the stellar distribution is concentrated in a compact, high-surface-density core surrounded by a low-density envelope and satellite clumps. The compact cores in the models with weak or no photoionization also appear elongated in projection, consistent with their higher \ifm{v/\sigma}.

The joint trend in size, mass, and kinematics is consistent with the evolutionary sequence identified in \cref{sec:diskformation}. Weaker photoionization allows gas to cool and collapse more efficiently in the centre, which enhances central star formation, deepens the central potential, and raises \ifm{\vx{c,tot}(\Rx{1/2})} relative to \ifm{\vx{v}}. The weaker-photoionization runs therefore lie further along the compaction sequence of \cref{fig:diskformation_centralsf}, towards the compact, more rotationally supported regime. Stronger photoionization, acting together with supernova feedback, may favour more spatially extended star formation and larger stellar half-mass radii. While \run{highphot} remains extended and \zyn{dispersion-dominated, with \ifm{(v/\sigma)_{\rm SFgas}<1}, both \run{lowphot} and \run{nophot} host compact galaxies with rotation-dominated star-forming gas, reaching \ifm{(v/\sigma)_{\rm SFgas}\simeq1.6} in \run{nophot}, comparable to the \ifm{\simeq1.5} of the \software{MAGE} galaxies above the transition (\cref{fig:diskformation_centralsf}), and} \ifm{\Mx{\star}\simeq\SI{6e9}{\Msun}}, above the disk-formation mass of \ifm{\simeq\SI{5e9}{\Msun}}. According to the criterion in \cref{eq:diskformation}, both ratios change in ways that favour disk formation as photoionization weakens: at fixed halo mass, the stellar-to-halo mass ratio increases while \ifm{\Rx{1/2}/\Rx{v}} decreases.

These experiments vary photoionization while keeping the SN prescription fixed, so they do not establish the relative importance of photoionization and SN feedback. They demonstrate that physical galaxy sizes, stellar masses, and \zyn{the rotational support of star-forming gas depend} on the feedback prescription. This sensitivity is not specific to cosmic dawn. Zoom-in simulations of Milky Way-mass haloes at lower redshift have shown that the strength and implementation of stellar feedback change galaxy sizes at fixed halo mass by factors of a few and determine whether an extended disk forms at all \citep{Sales.etal.10,Ubler.etal.14,Agertz.Kravtsov.16}. Galaxy sizes have accordingly served as a calibration target for subgrid feedback in large-volume simulations \citep{Crain.etal.15}.

\subsection{Comparison to previous theoretical work}

\subsubsection{Compact sizes at the low-mass end}
\label{subsubsec:lowmasscompact}
\zyn{At \ifm{\Mx{\star}\simeq\SI{1e8}{\Msun}}, MAGE galaxies have stellar half-mass radii of approximately \ifm{20\text{--}30\,\mathrm{pc}}, over an order of magnitude smaller than those of the extended galaxies in \software{THESAN-ZOOM}, \software{FirstLight}, and \software{Lumina} (\cref{fig:sizemass,fig:sizeevolution3d}). The \software{FIRE-2} galaxies are similarly extended: at \ifm{z=10}, their projected half-mass radii are \ifm{\simeq0.3\text{--}0.5\,\si{kpc}} at \ifm{\Mx{\star}\simeq10^{7}\text{--}10^{8}\,\si{\Msun}} \citep{Ma.etal.18}, approximately half those in \software{THESAN-ZOOM} but an order of magnitude larger than ours\footnote{We do not show them in \cref{fig:sizemass} because only six of their \ifm{z=10} galaxies lie above \ifm{\Mx{\star}=10^{7.5}\,\si{\Msun}}.}. This represents a substantial difference in stellar concentration. The compactness in \software{MAGE} suggests an analogy with compact stellar systems, although our approximately \ifm{\SI{10}{pc}} spatial resolution does not resolve individual star clusters or establish their internal structure. There are other suites, however, that do form compact systems.} In \software{SERRA} \citep{Pallottini.etal.22}, central galaxies at \ifm{z=7.7} with \ifm{\Mx{\star}\lesssim\SI{1e9}{\Msun}} also have small reported sizes, with dust-attenuated UV half-light radii \ifm{\lesssim\SI{100}{pc}}; these cannot be compared directly with our intrinsic sizes. \zyn{This compactness extends into cosmic dawn: the \software{SERRA} galaxy Amaryllis has a dust-attenuated UV half-light radius of \ifm{\simeq\SI{63}{pc}} at \ifm{z\simeq11.5} \citep{Kohandel.etal.25}, although this is not directly comparable to our intrinsic sizes.} The recent \software{VESTRAL} simulations also produce compact galaxies at high redshift (Nelson et al., in preparation; Ravishankar et al., in preparation), and future comparisons will allow us to assess similarities and differences between their results and ours.

Such differences are commonly attributed to the sub-grid star formation and feedback models. These models certainly affect galaxy sizes: varying effective photoionization feedback changes \ifm{\Rx{1/2}/\Rx{v}} of \run{MAGE07} by a factor of \ifm{\simeq3} (\cref{subsec:phot}). It is nevertheless difficult to attribute an offset of more than an order of magnitude to this effect alone. \software{SERRA} provides a direct test because it also uses \software{RAMSES}, but adopts an otherwise independent set of sub-grid models, including a Schmidt--Kennicutt law applied to molecular hydrogen with a fixed efficiency of 10 per cent per free-fall time; SN, stellar-wind, and radiation-pressure feedback with blast-wave energy dissipation; non-equilibrium chemistry with \software{KROME}; and on-the-fly radiative transfer with \software{RAMSES-RT}. Despite these differences, its central galaxies at \ifm{z=7.7} have sizes within a factor of a few of ours.

Numerical factors beyond the sub-grid models may therefore matter. The \software{AGORA} code comparison at \ifm{z\geq10} \citep{Kim.etal.26} finds that \software{RAMSES} produces noticeably higher stellar masses and star-formation rates at \ifm{z\gtrsim13\text{--}15} than both the particle-based codes and \software{ENZO}, regardless of halo mass, before converging with the other codes to within \ifm{\sim1\,\si{dex}} by \ifm{z\sim10}, once the haloes collapse and the star-forming gas becomes dense in all codes. They attribute the early boost to the refinement strategy of \software{RAMSES}, which extends the maximally refined region around dense cells to avoid abrupt resolution jumps between dense and neighbouring lower-density regions. The resulting expansion of the highly refined star-forming region allows \software{RAMSES} \zyn{to reach \ifm{\nx{H}\sim3\times10^3\,{\rm cm^{-3}}}, the densities at which local star-cluster formation becomes efficient \citep{Grudic.etal.18,Menon.etal.24}, in every halo}, whereas the other codes reach such densities only in the most massive systems. This suggests that numerical factors, in addition to the sub-grid prescriptions, may contribute to higher star-forming gas densities. Such effects could also occur in \software{MAGE} and \software{SERRA}, depending on their refinement settings. Denser star-forming gas favours more concentrated star formation and a deeper central potential, promoting the formation of star clusters. However, whether the refinement strategy also affects stellar sizes remains to be tested, as \citet{Kim.etal.26} do not report these measurements.

Sample selection effects may also play a role. Because both \software{MAGE} and \software{SERRA} target massive galaxies at cosmic dawn via cosmological zoom-in simulations, our galaxy size measurements are likely influenced by the overdense environments sampled by these zoom-in simulations, particularly at the low-mass end. According to the FFB scenario \citep{Dekel.etal.23}, galaxies exceeding the threshold denoted by the red line in \cref{fig:host_gal_dists} are expected to be more compact and host bound star clusters. At \ifm{z>12}, this threshold corresponds to a halo mass of \ifm{\Mx{v}\simeq\SI{1e10}{\Msun}} and a stellar mass of a few times \ifm{\SI{1e8}{\Msun}}, which could naturally account for the observed compactness of low-mass systems at these early epochs. However, our low-mass galaxies at \ifm{z=9\text{--}11} remain comparably compact, and the three target galaxies lying below the FFB threshold at \ifm{z=9} show no qualitative structural differences from those above it, suggesting that the transition between the FFB and standard regimes is gradual rather than sharp. In addition, the large-size tail in \software{SERRA} is driven predominantly by non-central galaxies undergoing mergers or having recently merged. Because our sample is restricted to central galaxies and explicitly excludes systems with multiple dense stellar cores, obvious major mergers are filtered out (\cref{subsec:halofinder}). We note, however, that including merger systems primarily inflates the size scatter without substantially altering the median relation.

\subsubsection{Turnover of the size--mass relation}
\label{subsubsec:turnovercomparison}

The turnover of the intrinsic stellar size--mass relation occurs at different masses across simulation suites. In \software{THESAN-1}, \citet{Shen.etal.24} find a maximum in the stellar half-mass radius near \ifm{\Mx{\star}\sim\SI{1e8}{\Msun}} at \ifm{z\gtrsim6}. In \software{FirstLight}, \citet{Cataldi.etal.25} identify a negative slope during a compaction phase spanning \ifm{\Mx{\star}\sim10^{8.5}\text{--}10^{9.5}\,\si{\Msun}} over \ifm{5\lesssim z\lesssim9}. Negative intrinsic slopes have also been reported at lower redshifts for \ifm{\Mx{\star}\gtrsim\SI{1e9}{\Msun}} in \software{BLUETIDES} \citep{Marshall.etal.22}, \software{IllustrisTNG} \citep{Genel.etal.18}, and \software{FLARES} \citep{Roper.etal.22}. Our turnover near \ifm{\Mx{\star}\simeq\SI{5e9}{\Msun}} therefore occurs at a higher mass, with an additional difference in size normalization (\cref{fig:sizemass}). We note that these comparisons concern intrinsic stellar half-mass radii. A turnover in the intrinsic half-mass size--mass relation need not persist in the relation measured using dust-attenuated UV half-light radii (\cref{subsubsec:observedsize}).

The proposed mechanisms all link central gas inflow to stellar mass buildup. \citet{Cataldi.etal.25} attribute the negative slope to compaction triggered by inflows that strengthen as the galactic centre becomes baryon-dominated, although they do not identify the physical mechanism that triggers these inflows. \citet{Shen.etal.24} instead associate rapid compaction in massive galaxies with internal disk instability. In \software{THESAN-ZOOM}, \citet{McClymont.etal.25} find repeated central starbursts followed by inside-out quenching and renewed size growth, while the population retains a positive intrinsic size--mass relation. Thus, central bursts and individual contraction episodes do not necessarily produce a turnover in the population relation. Their cumulative effect on the central mass distribution, and the subsequent spatial distribution of star formation, determine whether size growth resumes or remains suppressed.

Our feedback experiments demonstrate that changes in the size normalization cannot be interpreted independently of changes in stellar mass. Increasing photoionization feedback in the same halo produces a less massive, more extended stellar body at \ifm{z=9} (\cref{subsec:phot}), moving the galaxy towards the larger sizes found in the comparison samples. However, this also reduces the stellar contribution to the central potential, since \ifm{\vx{c,star}^{2}(\Rx{1/2})=G\Mx{\star}/(2\Rx{1/2})}, which can delay the development of a central potential deep enough to overcome the effects of feedback. A shift towards lower stellar mass and larger radius induced by stronger feedback therefore does not imply that the turnover itself moves to lower mass. The four feedback variants establish the sensitivity of one halo's structure to photoionization, but do not measure how the population's transition mass changes with feedback strength.

A further consideration is that \software{MAGE} does not model black-hole growth or AGN feedback. At lower redshift, zoom-in simulations that switch AGN feedback on and off in the same haloes find that it enlarges massive galaxies and removes the flattening or turnover of the size--mass relation at high mass. In \software{RAMSES} zoom-ins of massive haloes evolved to \ifm{z=0}, AGN jets turn overly massive, compact, and rotation-dominated central galaxies into extended, dispersion-dominated systems that follow the observed size--mass relation \citep{Dubois.etal.13}. \citet{Choi.etal.18} find that, without AGN feedback, the 30 massive galaxies have present-day sizes approximately one-fifth of those observed at fixed stellar mass and follow a shallow growth track in the size--mass plane, whereas mechanical and radiative AGN feedback quenches in-situ star formation, so that subsequent gas-poor mergers and the expulsion of central gas puff up the stellar body along the observed relation. In \software{FIRE}, \citet{Cochrane.etal.23} re-simulate a \ifm{\Mx{\star}\sim\SI{1e11}{\Msun}} galaxy at \ifm{z\simeq2.3} and show that, without AGN winds, a rapid central starburst shrinks the half-mass radius, whereas mechanical winds lower the central gas surface density, suppress the compact starburst, and hold the half-mass radius approximately constant. In \software{FIRE-3} runs evolved to \ifm{z\simeq0}, the most massive galaxies are an order of magnitude too compact without AGN feedback, and the stronger feedback model recovers the observed size--mass relation \citep{Byrne.etal.24}. Even at \ifm{z\simeq6}, bursty AGN feedback in simulated quasar hosts expands compact bulges by an order of magnitude by \ifm{z\simeq3} through fluctuations of the gravitational potential \citep{vanderVlugt.Costa.19}. The processes identified in these studies affect the physical ingredients underlying our turnover: efficient central star formation, the retention of gas in a deepening central potential, and the resulting rapid central stellar buildup.

At cosmic dawn, however, a direct test in one of our haloes finds that AGN feedback has a comparatively weak effect. \citet{Sunseri.etal.26} introduced a central black hole into the \run{MAGE07} halo, systematically varying the seed mass (\ifm{10^{4}\text{--}10^{5}\,\si{\Msun}}), the accretion limit (one to three times the Eddington rate, as well as unlimited accretion), the thermal AGN feedback coupling efficiency, and the strength of stellar feedback. Because stellar feedback maintains a turbulent, multiphase nuclear ISM, the central black hole remains starved for roughly half the time after the onset of star formation. Even when AGN feedback becomes self-regulating, the star formation history, the final stellar mass at \ifm{z\simeq9}, and the thick, partly rotationally supported disk remain essentially unchanged relative to the black-hole-free baseline. Furthermore, while \software{THESAN-1} includes AGN feedback and still exhibits a negative slope, \software{THESAN-ZOOM} lacks AGN feedback yet displays a positive relation. AGN feedback alone therefore does not guarantee a positive slope. If early black holes grow more rapidly than assumed in these numerical experiments and cosmological simulations, as suggested by the proposed AGN interpretation of GN-z11 \citep{Maiolino.etal.24}, stronger feedback could potentially flatten or erase the negative high-mass slope. Consequently, the turnover mass reported here should be interpreted strictly within a stellar-feedback-only framework, although the parameter explorations of \citet{Sunseri.etal.26} and relevant simulations suggest that the impact of omitting AGN feedback may be modest for haloes assembling as rapidly as ours at \ifm{z\gtrsim9}.

\subsubsection{Disk formation}
\label{subsubsec:diskcomparison}

Comparing disk formation across simulations requires distinguishing the emergence of coherent rotation from subsequent settling into a dynamically cold, thin disk. In the \software{FIRE-2} simulations, \citet{Yu.etal.23} and \citet{McCluskey.etal.24} identify successive phases in which stars form first in a dispersion-dominated protogalaxy, then in a turbulent rotating disk, and finally in a colder disk. The transition to thin-disk formation is associated with steadier star formation and virialization of the inner CGM \citep{Stern.etal.21,Hafen.etal.22,Gurvich.etal.23}. The inner CGM virializes when the cooling time of shocked gas at \ifm{\sim0.1\,\Rx{v}} exceeds the local free-fall time, so that cool, supersonic inflows and outflows with large pressure fluctuations give way to a hot, subsonic, quasi-static phase near the virial temperature. Accreting gas can then align its angular momentum with that of the galaxy disk before joining the galaxy. The gas velocity dispersion drops sharply and can fall well below the virial velocity.

The connection between central stellar mass buildup and rotational support in \software{MAGE} is qualitatively consistent with the controlled experiments of \citet{Hopkins.etal.23}. They show that a centrally concentrated mass profile promotes disk formation by providing a well-defined dynamical centre and suppressing large-scale motions that disrupt coherent rotation. In \software{TNG50}, \citet{Semenov.etal.24} find that disk spin-up accompanies bulge growth, the inflow of prealigned gas, outflow recycling, and hot-halo formation. Their analysis does not establish which process causes the transition. Here, ``bulge'' refers to a central mass concentration and does not necessarily imply dispersion support: \citet{Luo.etal.26} find that, at \ifm{z=3}, the central structures formed through compaction in massive \software{TNG50} galaxies are mostly rotation-supported nuclear disks and bars, which become dispersion dominated only at \ifm{z\lesssim1}. The analytical model of \citet{Semenov.26} links disk formation to the steepening of the potential as baryons accumulate and predicts a corresponding change in size regulation. However, as discussed in \cref{subsec:diskcondition}, the \software{ARTEMIS} and \software{FIRE-2} results show that a centrally peaked rotation curve need not precede the formation of a stellar disk \citep{Dillamore.etal.23,Myrtag.etal.26}.

Simulations resolving the cold interstellar medium also demonstrate that disks can form well before the disk-formation epochs reported for Milky Way progenitors in many earlier studies. \citet{Semenov.etal.25b} find that a disk forms at \ifm{z\sim6\text{--}7} in an early-forming Milky Way analogue. Disk formation is aided by rapid central mass growth and reduced gas pressure support. Companion experiments by the same authors show that a turbulence-dependent star-formation efficiency can allow both bursty star formation at early times and disk survival \citep{Semenov.etal.25a}. At \ifm{z>7}, \citet{Mayer.etal.25} find compact, self-gravitating gas disks that fragment into massive star-forming clumps in an overdense environment. \zyn{\citet{Kohandel.etal.25} find that the \software{SERRA} galaxy ``Amaryllis'' develops a rotationally supported gas disk with \ifm{v/\sigma \simeq 4\text{--}6} by \ifm{z \simeq 11}, while its stellar component remains dispersion-dominated.} These studies provide precedents for early rotating systems.

An alternative route to the early formation of massive disks invokes cosmology rather than internal galaxy physics. \citet{Shen.etal.26b} present cosmological hydrodynamic simulations with Early Dark Energy, in which the modified pre-recombination expansion history accelerates early structure formation and raises the number density of stellar and gaseous disks by \ifm{\simeq0.5} dex at \ifm{z\simeq6\text{--}7}. This enhancement primarily reflects the higher abundance of massive galaxies. At \ifm{z\lesssim3}, the model predictions approach those of \ifm{\Lambda}CDM.

Other simulations using different subgrid models for star formation and stellar feedback also suggest an association between compaction and disk formation. In \software{THESAN-1}, \citet{Shen.etal.24} find that more massive, compact systems have a larger fraction of their gas mass in the disk component, consistent with our results. By contrast, in \software{THESAN-ZOOM}, \citet{McClymont.etal.25} report a positive intrinsic size--mass relation, while \citet{Wang.etal.25} find that galaxies in the same suite have not yet developed well-defined disks. Thus, even if our size turnover and disk formation are not linked by a direct causal relationship, the broader association between compactness and disk formation appears in several simulation frameworks.

\subsection{Observational implications}
\label{subsec:observations}

The size evolution and the connection between central stellar mass buildup and disk formation in \software{MAGE} can be tested through the joint structural, star-formation, and kinematic properties of high-redshift galaxy observations.

\subsubsection{Size--mass and size--magnitude relations}
\label{subsubsec:observedsize}

M24 find a positive UV size--mass relation, \ifm{\Rx{e}\propto\Mx{\star}^{0.19\pm0.03}}, and relatively weak size evolution, \ifm{\Rx{e}\propto(1+z)^{-0.4\pm0.2}}, across \ifm{5<z<14}. Their sample also contains a compact subpopulation (\ifm{\simeq13} per cent of their galaxies) with intense star formation, which is qualitatively consistent with our positive low-mass branch and weak evolution over \ifm{9\lesssim z\lesssim15} (\cref{fig:sizemass,fig:sizeevolution3d}). The COSMOS-Web sample of \citet{Yang.etal.25}, which extends to lower redshift, instead yields stronger evolution of optical sizes over \ifm{2<z<10}, with \ifm{\Rx{e}\propto(1+z)^{-1.21\pm0.05}} at a reference mass of \ifm{\SI{5e10}{\Msun}}. The size--mass slopes are poorly constrained above \ifm{z\sim8}. They also report very compact blue sources that remain unresolved at \ifm{z>5}, even after excluding X-ray AGN and little red dots. At later epochs, \citet{ChenLongyue.etal.26} find that the rest-frame optical size--mass relation of star-forming galaxies at \ifm{0.5<z<6} flattens above a nearly redshift-independent pivot mass of \ifm{\Mx{\star}\sim\SI{1e10}{\Msun}}, within a factor of two of our turnover mass of \ifm{\simeq\SI{5e9}{\Msun}}. They interpret this as evidence that galaxy size growth tracks halo growth below the pivot, whereas central mass buildup decouples the size growth of more massive galaxies from halo growth, in line with the saturation of stellar birth radii and the decline of \ifm{\Rx{1/2}/\Rx{v}} in \software{MAGE} (\cref{fig:sizebirth,fig:sizemass,subsec:compaction}).

In the size--magnitude relation shown in \cref{fig:sizemag}, all eight \ion{C}{iv}-strong sources are labelled, and all but GHZ4 have half-light radii \ifm{\lesssim100\text{--}150\,\si{pc}}, with most sources remaining unresolved \citep{Roberts-Borsani.etal.26}. \zyn{At the faint end they coincide with our simulated galaxies, which populate the compact tail of the observed size distribution (\cref{subsec:size-magnitude})}, and because most of their reported sizes are upper limits, their intrinsic sizes may be smaller still. Their strong rest-frame UV lines, including \ion{C}{iv} and \ion{N}{iv}] in GHZ2 and MoM-z14 \citep{Castellano.etal.24,Naidu.etal.25}, are indicative of hard ionizing radiation fields from intense, compact star formation and/or active galactic nucleus (AGN) activity \citep{Harikane.etal.25,Roberts-Borsani.etal.26}. The evidence for a stellar origin of these lines is discussed in \cref{subsubsec:gnz11}. The \ion{C}{iv}-weak sources, in contrast, range from similarly compact, unresolved systems to extended galaxies with \ifm{\Rx{e}\simeq\SI{1}{kpc}}. Compactness is therefore not always accompanied by strong lines. U37126 is a clear example: it is similarly compact (\ifm{\Rx{e}\simeq\SI{61}{pc}}; see footnote~\ref{fn:u37126}) and extremely blue (\ifm{\beta\simeq-2.9}) but shows no detectable nebular emission. \citet{Marques-Chaves.etal.26} attribute this absence to a near-unity Lyman-continuum escape fraction after feedback cleared gas from the stellar core of U37126. Interestingly, GN-z11, the brightest source in the sample, is classified as \ion{C}{iv}-weak in the prism data, but its spectrum shows strong \ion{N}{iv}], \ion{N}{iii}], and \ion{C}{iii}] emission and a possible P-Cygni \ion{C}{iv} feature \citep{Bunker.etal.23}. Its extreme compactness and strong emission lines may reflect an intense, weakly obscured central starburst \citep{ChenZuyi.etal.26,Nakane.Ouchi.26} and/or a luminous unresolved AGN that enhances the central UV surface brightness \citep{Maiolino.etal.24,Scholtz.etal.24}.

Although our simulation results match the observed compact systems, we do not reproduce the more extended population indicated by the \zyn{grey} bands in \cref{fig:sizemass,fig:sizemag}. Dust attenuation represents one of the primary systematic uncertainties in comparing intrinsic UV sizes with observed size relations. Including dust attenuation in \software{THESAN-1} \citep{Shen.etal.24} and \software{FirstLight} \citep{Ceverino.etal.26} changes their intrinsically negative size--mass slopes to flat or positive slopes, bringing them into closer agreement with observations. This behavior arises because intrinsically compact, massive galaxies frequently host concentrated central dust columns. Preferential attenuation of the luminous core consequently expands the emergent UV half-light radius. By contrast, \software{THESAN-ZOOM} shows a negligible effect of dust on UV half-light radii, a result mirrored by the \software{Astrid} simulation for rest-frame optical sizes at lower redshifts \citep[\ifm{z=3\text{--}6};][]{LaChance.etal.25}. The extent to which dust alters UV sizes therefore depends on the spatial distribution of dust and its survival under stellar feedback. Rapid dust destruction in star-forming environments minimizes attenuation \citep{Garaldi.etal.26}.

Observationally, the compact systems with strong emission lines identified by \citet{Roberts-Borsani.etal.26} have sizes overlapping those in our simulations and blue UV continuum slopes, \ifm{\beta<-2.5}. These blue slopes suggest little UV attenuation, potentially because of radiation-pressure-driven dust clearing or spatial segregation of dust from star-forming complexes \citep{Ferrara.etal.23,Ziparo.etal.23}. In contrast, typical systems with weak emission lines exhibit redder slopes between \ifm{\beta\simeq-2.5} and \ifm{-1.5}, leaving room for moderate size expansion. For \software{MAGE}, preliminary radiative transfer calculations suggest that dust attenuation can increase the effective UV half-light radius by a factor of a few in our massive galaxies (Nakazato et al., in preparation). In the size--magnitude plane, central obscuration increases the half-light radius while simultaneously diminishing the total UV luminosity. Given that \citet{Kim.etal.26} report significant UV attenuation even at a dust-to-metal ratio of \ifm{0.01}, dust shifts galaxies along both axes of \cref{fig:sizemag}. However, the mean UV continuum slope of the \ifm{z>9} population becomes bluer with increasing redshift, changing from \ifm{\langle\beta\rangle\simeq-2.2} at \ifm{z\simeq9.5} to \ifm{\langle\beta\rangle\simeq-2.6} at \ifm{z\simeq11.5}, approaching the theoretical limit for dust-free stellar populations \citep{Topping.etal.24,Cullen.etal.24,Morales.etal.24,Austin.etal.25}. In this interpretation, dust-induced size inflation may be more important for a dust-rich minority at the lower-redshift end of our sample, whereas observed UV half-light radii at \ifm{z\gtrsim11} should be close to their intrinsic values. Future work will determine the net shift in the \software{MAGE} size--magnitude relation and test whether such attenuation can be reconciled with the blue UV slopes at \ifm{z\gtrsim11}.

Moreover, at low stellar masses, where the discrepancy between our results and observations is more severe, the effect of dust is limited. Another factor is bias arising from limited observational resolution. The NIRCam F200W PSF has a FWHM of \ifm{\simeq0.066\arcsec}, corresponding to \ifm{\simeq250\text{--}300\,\si{pc}} at \ifm{z=9\text{--}12}, whereas the UV half-light radii of our galaxies with \ifm{\Mx{\star}\lesssim10^{8.5}\,\si{\Msun}} are \ifm{\Rx{e}\simeq10\text{--}40\,\si{pc}}, an order of magnitude smaller. Profile fitting with a PSF model can recover sizes below the FWHM, but sizes far below it are poorly constrained (\citealp{Ono.etal.23}; M24), so such galaxies enter observed size--mass and size--magnitude relations with size upper limits or measurements scattered upwards towards the resolution floor. The compact subpopulation of M24 is defined partly by this limit, as it includes gravitationally lensed sources with \ifm{\Rx{e}} below half the PSF FWHM. Detecting additional compact systems at the low-mass end could steepen the inferred size--mass relation and bring it into closer agreement with our results.

Although higher-resolution observations may yield smaller median sizes, our simulations produce no galaxies with intrinsic UV half-light radii approaching \ifm{\simeq\SI{1}{kpc}}, even among the most massive systems. On the observational side, a fraction of seemingly extended sources may be close merging pairs that cannot be cleanly deblended in projection. On the simulation side, our sample selection preferentially targets overdense regions as mentioned in \cref{subsubsec:lowmasscompact}. Because our low-mass systems are close neighbors of massive zoom-in halos rather than typical field galaxies, they assemble earlier and reside in halos that are more concentrated at fixed mass than those probed by blank-field surveys. Consequently, our suite underrepresents the broader field population, which could artificially compress the predicted scatter in galaxy size. A rigorous quantitative comparison will require forward-modeled mock observations with matched prescriptions for dust attenuation, PSF convolution, surface brightness noise, source selection, and profile fitting, which we defer to future work.

\subsubsection{In-situ star formation and merger-driven growth}
\label{subsubsec:insitucomparison}

\zyn{JWST observations reveal evidence for mergers in early galaxies, including the clumpy, bursty systems studied by \citet{Harikane.etal.25} and the resolved major-merger systems Himiko and CR7 studied by \citet{Kiyota.etal.25}. Statistical studies nevertheless infer different contributions from mergers to early stellar mass growth. For a mass-selected JADES sample at \ifm{z\simeq3\text{--}9}, \citet{Puskas.etal.25} estimate that stellar accretion through major mergers contributes approximately 3--13 per cent of the total growth, with in-situ star formation dominating. Using close pairs at \ifm{4.5<z<11.5}, \citet{Duan.etal.25} infer a larger contribution: major-merger stellar accretion amounts to \ifm{71\pm25} per cent of the growth through star formation. Both analyses infer accretion statistically from pair counts and merger time-scales, rather than directly identifying the birthplaces of observed stars. \citet{Puskas.etal.25} note that their merger rates agree with those of \citet{Duan.etal.25}, and suggest that different estimates of the specific SFR help explain the discrepant growth fractions.}

\zyn{Clump demographics provide a complementary constraint on the processes driving this growth. \citet{Puskas.etal.26} find that the clumpy fraction increases with stellar mass: major mergers could account for many low-mass clumpy galaxies at high redshift, whereas the higher fractions in more massive galaxies require an additional in-situ clump-formation channel. \citet{delaVega.etal.26} similarly favour disk fragmentation in intermediate-to-high-mass galaxies at later epochs and merger-induced gas compression at higher redshift.}

\zyn{Simulations suggest that the availability of a sufficiently massive, self-gravitating disk helps determine which channel operates. \citet{Mandelker.etal.17} find that gas-rich disks form giant clumps through gravitational instability, while \citet{Nakazato.etal.24} find predominantly merger-induced clumps in \software{FirstLight} at high redshift and argue that fragmentation requires a massive disk with a large disk mass fraction. Predominantly local stellar assembly is also found in \software{THESAN-ZOOM}, where \citet{Shen.etal.26a} report high in-situ stellar mass fractions, although their classification uses the halo virial radius rather than our galaxy aperture.}

\zyn{\software{MAGE} provides a possible connection between disk assembly and the origin of clumpy structure. Rotational support in the star-forming gas rises rapidly near \ifm{\Mx{\star}\simeq\SI{5e9}{\Msun}}, and the more massive systems exhibit richer clumpy structure, predominantly within the in-situ component (\cref{fig:proj_insitu}). Together, these findings are consistent with fragmentation becoming important once a massive disk has assembled, as in \citet{Mandelker.etal.17}. The high central in-situ fractions---approximately \ifm{0.8\text{--}1} in our four high-mass systems (\cref{fig:Insitu})---and the close correspondence between the total and in-situ half-mass radii (\cref{fig:sizegrowth}) further support predominantly local stellar growth. Interactions can nevertheless trigger this growth: \run{MAGE04} retains a high central in-situ fraction despite recent major mergers, illustrating that merger-triggered star formation need not entail predominantly ex-situ stellar assembly.}

\subsubsection{Early disk formation}
\label{subsubsec:observeddisks}

Observational evidence for early disks spans structural and kinematic diagnostics. \citet{Ono.etal.25} find that galaxies at \ifm{z\sim10\text{--}16} have a log-normal size distribution, stellar-to-halo size ratios of \ifm{\Rx{e}/\Rx{v}\simeq0.015}, a mean S\'{e}rsic index close to unity, and a nearly uniform axis-ratio distribution, which they interpret as consistent with disk formation. At \ifm{4<z<7.7}, \citet{Mirzoyan.etal.26} use a denoising neural-network morphological classifier to classify \ifm{\simeq34} per cent of 100 JWST/NIRCam galaxies as disk-like across a range of stellar masses. At \ifm{z=7.43}, \citet{Baker.etal.25} identify a compact stellar core with a half-light radius of \ifm{\sim\SI{80}{pc}} surrounded by a star-forming component with a half-light radius of \ifm{\sim\SI{400}{pc}}. The specific SFR increases with radius, providing evidence for inside-out growth around an already dense centre. These observations support the possibility that compact central buildup and subsequent disk growth can occur early, as in our simulations.

Structural similarities alone do not establish the degree of rotational support. Gas kinematics reveal a range of dynamical states at these epochs. \citet{Parlanti.etal.23} find that ALMA [\ion{C}{ii}] and [\ion{O}{iii}] observations of 22 star-forming galaxies at \ifm{z>5} are consistent with turbulent rotating disks, although the available resolution does not uniquely distinguish disks from closely separated mergers. Using JWST/NIRSpec spectroscopy, \citet{deGraaff.etal.24} infer both rotation- and dispersion-dominated configurations among six low-mass galaxies at \ifm{5.5<z<7.4}. A broader population study by \citet{Danhaive.etal.25} finds that \ifm{34\pm5} per cent of \Halpha-selected galaxies with \ifm{10^9<\Mx{\star}/\si{\Msun}<10^{10}} have \ifm{v/\sigma_0>1} in both the \ifm{z\sim4.5} and \ifm{z\sim5.5} bins, where \ifm{\sigma_0} is the intrinsic ionized-gas velocity dispersion. In the Twister-z5 galaxy at \ifm{z=5.4}, \citet{Nelson.etal.24} infer rapid ionized-gas rotation, with \ifm{v=305\pm70\,\si{km/s}} and \ifm{v/\sigma\sim2\text{--}3}. The increase in \Halpha\ equivalent width with radius is consistent with inside-out growth. Together, these observations show that ordered rotation can coexist with substantial gas turbulence.

\zyn{Dynamically cold gas disks are also observed, primarily through [\ion{C}{ii}] observations.} \citet{Rizzo.etal.21} infer \ifm{v/\sigma\simeq7\text{--}15} from ALMA [\ion{C}{ii}] kinematics of strongly lensed dusty star-forming galaxies at \ifm{z\sim4\text{--}5}. \citet{Lelli.etal.26} find regular rotation in all 16 massive star-forming galaxies (\ifm{\Mx{\star}\simeq10^{10.5}\text{--}10^{12}\,\si{\Msun}}) of the TRICEPS sample at \ifm{z=4\text{--}5}, observed in [\ion{C}{ii}] at \ifm{\simeq\SI{0.8}{kpc}} resolution. Regular rotation is present even though half of the galaxies belong to pairs, suggesting that rotational support was widespread among massive systems by that epoch. Mock [\ion{C}{ii}] observations of massive \software{TNG50} galaxies at the same redshifts yield \ifm{v/\sigma\simeq2\text{--}3} in most cases \citep{HeYi.etal.26}, below the observed values, so reproducing such cold disks remains a challenge for current cosmological simulations. At still earlier epochs, \citet{Rowland.etal.24} establish the presence of a dynamically cold disk in REBELS-25 at \ifm{z=7.31} using ALMA [\ion{C}{ii}] observations, while \citet{Rowland.etal.26} find that [\ion{C}{ii}] and [\ion{O}{iii}] \SI{88}{\micro\metre} trace a common rotating structure in REBELS-25, despite the galaxy's clumpy rest-frame UV morphology. Similarly, the strongly lensed Cosmic Grapes galaxy at \ifm{z=6.072} contains at least 15 star-forming clumps embedded in a rotating gas disk \citep{Fujimoto.etal.25}. Such systems show that clumpy emission can coexist with ordered rotation and does not by itself imply a merger. \zyn{We note that our dynamically hotter intrinsic kinematics for the star-forming gas and young stars do not necessarily imply tension with the colder [\ion{C}{ii}] disks reported in observations. Indeed, using synthetic observations of \software{SERRA} galaxies, \citet{Kohandel.etal.24} demonstrated that \Halpha\ velocity dispersions can exceed those inferred from [\ion{C}{ii}] by more than a factor of two. Establishing a robust, quantitative mapping between these tracers will ultimately require matched forward modeling.}

Tentative kinematic evidence extends into the redshift range probed by \software{MAGE}. For GN-z11 at \ifm{z=10.6}, \citet{Xu.etal.24} model the \ion{C}{iii}] velocity gradient as disk rotation and infer \ifm{v/\sigma=2.83^{+1.82}_{-1.41}}, while noting that an outflow can also explain the observations. At \ifm{z=14.18}, \citet{Scholtz.etal.25} report a tentative detection of a velocity gradient at \ifm{3\sigma} significance in [\ion{O}{iii}] \SI{88}{\micro\metre} emission from JADES-GS-z14-0. Their disk model gives \ifm{v/\sigma>2.5}. These candidates suggest that appreciable rotational support may emerge very early, but higher-resolution observations are needed to test whether the observed velocity gradients arise from disk rotation.

\subsection{A proposed evolutionary sequence and its observational implications}
\label{subsec:sequence}

\Cref{fig:cartoon} combines the trends of \cref{sec:in-situsf,sec:diskformation} and the interpretation developed above into a proposed evolutionary sequence, ordered by stellar mass, together with its possible observational counterparts.

\subsubsection{Physical sequence}

The sequence is organized by the relative magnitudes of three velocity scales: the halo virial speed \ifm{\vx{v}}, the total circular speed at the half-mass radius \ifm{\vx{c,tot}(\Rx{1/2})}, and the characteristic SN-feedback scale \ifm{\vx{SN}\simeq\SI{100}{km/s}} (\cref{sec:diskformation}). We interpret the
trends as four approximate regimes, \zyn{corresponding} to the shallow-potential, halo-recycling, galactic-recycling, and disk-formation regimes in \cref{fig:cartoon}\zyn{, followed by a fifth phase of later disk growth that lies beyond our simulated range}. Panel labels below refer to \cref{fig:diskformation_centralsf}.
\begin{enumerate}[label=(\roman*),itemindent=1em,align=left]
 \item \textit{\zyn{Shallow potential, gas escape, and compact stellar bodies}}, \ifm{\Mx{\star}\lesssim\SI{1e8}{\Msun}}: \ifm{\vx{v}<\vx{SN}}, so SN-driven winds can remove a substantial fraction of the gas from the halo itself \citep{Dekel.Silk.86}, and dark matter dominates the enclosed mass at \ifm{\Rx{1/2}} (panel (a)). The star-forming gas and the young stars remain dispersion-dominated, with median \ifm{v/\sigma\simeq0.5\text{--}1} (panel (d)) and only transient excursions to higher values (\cref{fig:v2sigma_centrals}). Stars form predominantly outside the half-mass radius, at \ifm{\Rx{birth}\simeq2\,\Rx{1/2}}, with birth radii varying by more than an order of magnitude between galaxies, consistent with the chaotic assembly of the low-mass progenitors (\cref{fig:sizebirth}). The galaxies remain compact, with half-mass radii only a few times the spatial resolution of \ifm{\simeq\SI{10}{pc}}.
 \item \textit{Halo-scale retention and outward star formation}, \ifm{\SI{1e8}{\Msun}\lesssim\Mx{\star}\lesssim\SI{5e8}{\Msun}}: \ifm{\vx{v}>\vx{SN}} while \ifm{\vx{c,tot}(\Rx{1/2})<\vx{SN}}, and dark matter still dominates the enclosed mass at \ifm{\Rx{1/2}} (panel (a)). SN-driven outflows are expected to be retained by the halo but not by the galaxy. In this interpretation, ejected gas may mix with high-angular-momentum CGM gas before returning to the galactic outskirts. The measured median birth radius is \ifm{\Rx{birth}\simeq2\,\Rx{1/2}\simeq0.02\,\Rx{v}} (\cref{fig:sizebirth}), and \ifm{\Rx{1/2}} grows steeply with stellar mass (panel (c)). \zyn{Gas entering the galaxy from outside appears to dominate the central gas supply: the outflow rate across \ifm{\Rx{gal}} still exceeds that across \ifm{\Rx{1/2}}, so the proxy for gas recycled within the galaxy, \ifm{\dotM{out}(\Rx{1/2})-\dotM{out}(\Rx{gal})}, is small or negative (panel (e)).}
 \item \textit{Galaxy-scale retention and central baryonic buildup}, \ifm{\SI{5e8}{\Msun}\lesssim\Mx{\star}\lesssim\SI{5e9}{\Msun}}: \ifm{\vx{c,tot}(\Rx{1/2})>\vx{SN}}, and stars dominate the enclosed mass at \ifm{\Rx{1/2}}. We interpret this transition as a shift from halo-scale to galactic-scale recycling. The outflow rate across \ifm{\Rx{gal}} falls below that across \ifm{\Rx{1/2}} (panel (e)). This ordering is consistent with failed outflows and recycling within the galaxy that could sustain the buildup of central stellar mass. Size growth slows, and the central gas fraction declines as the stellar contribution grows.
 \item \textit{\zyn{Size turnover, disk formation, and inside-out quenching}}, \ifm{\Mx{\star}\gtrsim\SI{5e9}{\Msun}}: \ifm{\vx{c,tot}(\Rx{1/2})>\vx{v}}, driven by the stellar component, while \ifm{\vx{rot,SFgas}/\vx{c,tot}(\Rx{1/2})} rises from \ifm{\simeq0.4} to \ifm{\simeq0.7} and \ifm{\sigma_{\rm SFgas}} continues to track \ifm{\vx{v}} (panel (b)). The rapid central stellar buildup at \ifm{\Mx{\star}\simeq\SI{5e9}{\Msun}} coincides with the turnover of \ifm{\Rx{1/2}} (panel (c)) and a rise in rotational support: both the star-forming gas and stars younger than \SI{10}{Myr} reach \ifm{v/\sigma\gtrsim1.5}, while the full stellar population remains dispersion-dominated (panel (d)). \zyn{The proxy for gas recycled within the galaxy becomes comparable to the inflow rate across \ifm{\Rx{gal}}, suggesting that recycled gas may supply the centre at a rate comparable to that of gas entering the galaxy from outside (panel (e)).} The central star-formation fraction remains at \ifm{\simeq0.4} up to the turnover. At the highest masses sampled, \ifm{\Mx{\star}\gtrsim\SI{2.5e10}{\Msun}}, \run{MAGE01} shows a decline in this fraction, while gas contributes only a few per cent of the central enclosed mass (panels (a) and (e)). This decline, the low central gas fraction, and the younger outskirts (\cref{fig:Insitu}) suggest the onset of inside-out quenching.
 \item \zyn{\textit{Later disk growth (expected)}, beyond the sampled range: our runs end at \ifm{z\simeq9}, shortly after the turnover, so this phase is not simulated here. By analogy with the post-compaction evolution in \software{FirstLight} \citep{Cataldi.etal.25} and \software{VELA} \citep{Lapiner.etal.23}, continued accretion onto the outskirts of a quenching core is expected to build an extended star-forming disk and resume the growth of \ifm{\Rx{1/2}}. This phase corresponds to the dashed regime outline in \cref{fig:cartoon} and is an extrapolation rather than a result of our simulations (\cref{subsubsec:gnz11}).}
\end{enumerate}

\begin{figure*}
    \centering
    \includegraphics[width=\textwidth]{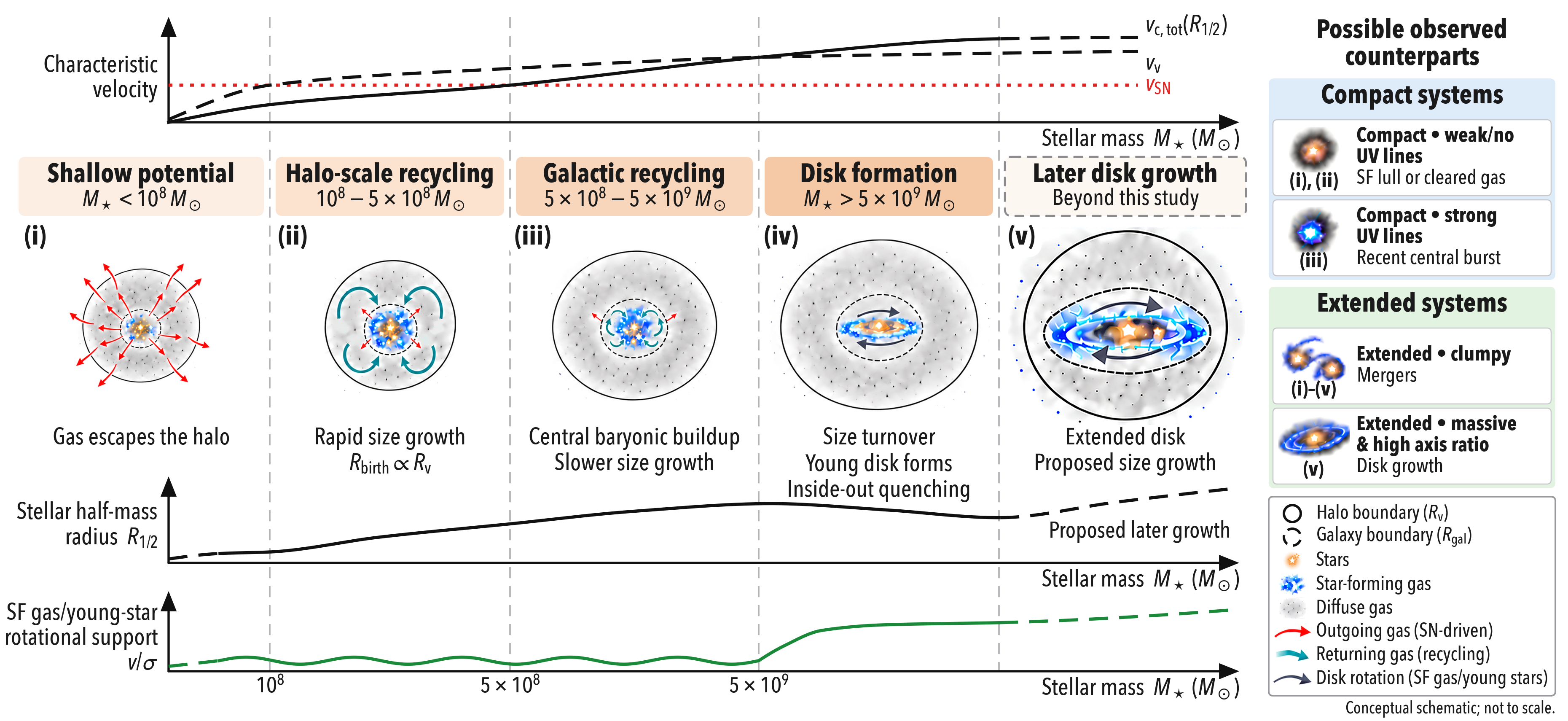}
    \caption{Schematic of the proposed evolutionary sequence for massive galaxies at cosmic dawn. The upper curves compare the total circular speed at the stellar half-mass radius, \ifm{\vx{c,tot}(\Rx{1/2})} (solid black), the halo virial speed, \ifm{\vx{v}} (dashed black), and the characteristic SN-feedback scale, \ifm{\vx{SN}\simeq\SI{100}{km/s}} (dotted red). The vertical guides mark their crossings near \ifm{\Mx{\star}\simeq\SI{1e8}{\Msun}}, \ifm{\SI{5e8}{\Msun}}, and \ifm{\SI{5e9}{\Msun}}, separating the proposed shallow-potential, halo-recycling, galactic-recycling, and disk-formation regimes. Solid and dashed circles denote \ifm{\Rx{v}} and \ifm{\Rx{gal}}; the legend at the bottom right identifies the stellar and gas components, outflows, returning gas, and rotation. The lower curves illustrate the accompanying size turnover and increase in the rotational support of the star-forming gas and young stars. The fourth, unlabelled guide separates the sampled sequence from later disk growth; dashed continuations and the dashed regime outline indicate expectations beyond this study. Roman numerals (i)--(v) label the five phases. The right-hand column gives possible observational counterparts; the numerals on each card mark the phases in which that type of system is preferentially, but not exclusively, found: compact systems with weak or strong UV lines (lulls or bursts), extended clumpy systems (mergers), and extended massive systems with high axis ratios (disk growth). These associations describe favoured phases rather than measured mass classes or exclusive assignments (\cref{subsubsec:gnz11}). All curves are illustrative, the transition masses characterize our simulated sample, and the drawings are not to scale.}
    \label{fig:cartoon}
\end{figure*}

\subsubsection{Observational counterparts}
\label{subsubsec:gnz11}

The sequence offers a possible connection between the compact and extended populations introduced in \cref{sec:intro}. Compact galaxies with strong high-ionization lines may be observed near central burst peaks, preferentially during the galactic-recycling phase, when the duty cycle is high and youngest stars dominate the UV light from a region of a few tens of parsecs \citep{Harikane.etal.25,Roberts-Borsani.etal.26}. For the \ifm{z>10} sample of \citet{Roberts-Borsani.etal.26}, individual and composite spectra support bursty star formation as the primary explanation for the spectral diversity. The strong \ion{C}{iv} emitters have the bluest UV slopes (\ifm{\beta<-2.5}) and the highest inferred star-formation surface densities, consistent with young, concentrated bursts.

GN-z11 at \ifm{z=10.6} provides a plausible example of this phase and may be at the transition to disk formation. Its UV light is dominated by an unresolved component embedded in more extended emission, with a combined half-light radius of \ifm{\Rx{e}=64\pm20\,\si{pc}} \citep{Tacchella.etal.23}. Stellar-wind features indicate a recent burst with an age of \ifm{\lesssim\SI{3}{Myr}} \citep{ChenZuyi.etal.26}. \zyn{Its Ly\ifm{\alpha} emission, detected well before the end of reionization, implies that GN-z11 has already carved out a local ionized bubble \citep{Bunker.etal.23}, consistent with a recent episode of intense star formation.} Stellar models better reproduce its UV absorption and emission diagnostics than the AGN models considered \citep{Nakane.Ouchi.26}, and narrow rest-frame optical lines support vigorous star formation \citep{Alvarez-Marquez.etal.25}. An accreting black hole has nevertheless been proposed based on the reported [\ion{Ne}{iv}] and \ion{C}{ii*} detections, high gas densities, and a fast ionized outflow \citep{Maiolino.etal.24,Scholtz.etal.24}. The tentative rotation inferred by \citet{Xu.etal.24} suggests that central star formation may coexist with rotational support. Gas recycling may contribute to the metal enrichment invoked in models of its chemical abundances \citep{Citro.etal.26}. Its estimated stellar mass of order \ifm{\SI{1e9}{\Msun}} \citep{Tacchella.etal.23} falls naturally within this galactic recycling regime, reinforcing our proposed framework.

Compact galaxies with weak or undetected lines may be the same systems caught between bursts, when SN feedback has cleared the gas and nebular lines fade within \ifm{\sim\SI{10}{Myr}} while the UV continuum remains concentrated in the stellar core. These lulls should be longer below \ifm{\SI{5e8}{\Msun}}, where expelled gas returns through the halo, than above it, where galactic recycling restores the supply more quickly (\cref{subsec:compaction}). Weak lines can also occur soon after a burst if its gas has already been removed: U37126 has a very young, extremely blue stellar population and a half-light radius of \ifm{\simeq\SI{61}{pc}} but shows no detectable nebular lines. This absence is attributed to a near-unity Lyman-continuum escape fraction after gas clearing \citep{Marques-Chaves.etal.26}. If longer replenishment times produce a lower burst duty cycle, lower-mass galaxies may more frequently show weak or undetected nebular lines. Quantifying this prediction requires detailed emission-line modelling and the observational selection function.

More extended galaxies may correspond to later disk growth around a less active core, after inside-out quenching suppresses central star formation (\cref{subsec:compaction,subsubsec:observedsize}). Their light profiles are consistent with this possibility: typical star-forming galaxies at \ifm{z\simeq1\text{--}6} have median S\'{e}rsic indices of \ifm{n\simeq1\text{--}1.5} \citep{Shibuya.etal.15}, while the luminous galaxies of \citet{Ono.etal.25} at \ifm{z\sim10\text{--}16} have \ifm{n=1.3\pm0.6} and typical sizes of \ifm{100\text{--}300\,\si{pc}}. The size and axis-ratio distributions reported by Ono et al. provide additional morphological evidence for disks (\cref{subsubsec:observeddisks}). \zyn{At lower redshift, \citet{Xiao.etal.25} report Zh\'{u}l\'{o}ng (Torch Dragon), a massive spiral galaxy at a photometric redshift of \ifm{z_{\rm phot}\simeq5.2}, with \ifm{\Mx{\star}\simeq\SI{1e11}{\Msun}} and a quiescent core surrounded by an extended star-forming stellar disk. This structure is consistent with the proposed progression from central stellar buildup to inside-out quenching and subsequent disk growth. Systems such as Zh\'{u}l\'{o}ng could therefore be descendants of the massive young disks in MAGE, provided that continued accretion sustains disk growth at later times.} However, this mapping is an extrapolation: our runs end at \ifm{z\simeq9}, shortly after the turnover, and the simulated half-mass radii do not re-grow within the sampled range. Renewed size growth, seen after compaction in \software{FirstLight} \citep{Cataldi.etal.25} and \software{VELA} \citep{Lapiner.etal.23}, is therefore shown in \cref{fig:cartoon} as an expectation rather than a result of our simulations.

Mergers offer another route to extended morphologies. In \software{FOREVER22}, the stellar component becomes temporarily larger during a major merger and shrinks after coalescence and angular-momentum loss \citep{Yajima.etal.22}. \citet{Ono.etal.25} likewise invoke mergers or tidal interactions for some of their largest galaxies. Our main-branch tracks show transient enlargement during the major merger of \run{MAGE04} (\cref{fig:MockImage_MAGE04}), but the statistical sample excludes snapshots with multiple spatially separated stellar components (\cref{subsec:halofinder}). Such interacting systems therefore have no counterpart in our size--mass relation.

These associations apply most directly when star formation powers the compact emission. The observed spectral diagnostics do not uniquely establish the source of nebular ionization: for GHZ2 at \ifm{z=12.34}, UV line ratios are consistent with both AGN photoionization and low-metallicity star formation, although the limit on [\ion{Ne}{iv}]/\ion{N}{iv}] favours stellar photoionization in some model grids \citep{Castellano.etal.24}. More generally, UV line ratios from metal-poor star formation at high ionization parameter overlap with those of AGN, and narrow lines do not exclude an obscured nucleus. Because our simulations omit black-hole growth (\cref{subsec:caveats}), an unresolved AGN contribution must be separated before comparing observed sizes with our intrinsic stellar UV sizes. Such a contribution can enhance the central surface brightness and reduce the measured half-light radius (\cref{subsec:size-magnitude}).

The proposed counterparts suggest distinct observational tests. Later disks should preferentially be massive systems with ordered rotation and older central populations. Compact weak-line galaxies should retain young or intermediate-age populations in gas-cleared cores, while interacting systems should show multiple components or tidal features. Resolved stellar ages, emission-line maps, and kinematics spanning the compact and extended populations would test whether they include different phases of the proposed sequence. In GN-z11, such measurements could distinguish a brief central burst from sustained stellar buildup. These associations are tentative and need not identify a unique evolutionary phase. The limitations of the sample and physical model are discussed in \cref{subsec:caveats}.

\subsection{Caveats}
\label{subsec:caveats}

\textit{Sample and selection.} The main limitations are the small number of independent haloes and the selection of massive haloes in overdense regions. The low-mass galaxies in these zoom regions inhabit the overdense environments around the massive target haloes and do not constitute a representative field sample. This environmental difference may contribute to their compactness relative to observed low-mass galaxies. To measure the structure of individual stellar systems, we impose a cut on the outer stellar-mass fraction to exclude active merger phases, during which sizes are temporarily inflated. Consequently, this selection introduces a bias against systems with multiple spatially separated stellar components. The high-mass turnover is driven primarily by four systems, and only \run{MAGE01} contributes at the largest stellar masses. These systems cross the transition mass only shortly before \ifm{z=9}, so the phase after disk onset is sampled briefly and the half-mass radii do not re-grow within the simulated range. The identification of extended observed systems with later disk growth in \cref{subsec:sequence} is therefore an extrapolation. Repeated snapshots are correlated, and different progenitor tracks can share an environment or a descendant, so regression errors that treat snapshots as independent may underestimate the full sampling uncertainty, particularly for the high-mass slope. A larger sample is needed to determine the intrinsic scatter and any redshift dependence of the characteristic stellar mass inferred from the combined structural and kinematic trends.

\textit{Feedback model.} Absolute sizes and stellar masses depend on the adopted feedback model. Four feedback variants of one halo demonstrate sensitivity to photoionization but do not establish how population slopes or transition masses depend on feedback strength. The simulations also omit black-hole growth and AGN feedback, as well as radiation pressure and stellar winds. These processes could modify the central stellar buildup and gas retention identified here, particularly in the most massive systems. The omission of AGN feedback bears most directly on the high-mass turnover: at lower redshift, simulations run with and without AGN feedback show that it can suppress compact central starbursts, enlarge massive galaxies, and remove the flattening or turnover of the size--mass relation at high mass \citep{Dubois.etal.13,Choi.etal.18,vanderVlugt.Costa.19,Cochrane.etal.23,Byrne.etal.24}. The negative high-mass slope and the turnover mass should therefore be regarded as predictions of a model with stellar feedback alone (\cref{subsubsec:turnovercomparison}).

\textit{Radiation pressure.} The neglect of radiation pressure deserves particular emphasis. Our early feedback model is limited to photoionization heating (\cref{sec:methods}), yet analytic and radiation-hydrodynamic studies of massive star-cluster formation find that radiation pressure, rather than the thermal pressure of photoionized gas, becomes the dominant pre-SN feedback channel at the high gas surface densities and cluster masses characteristic of starburst environments \citep{Krumholz.Matzner.09,Murray.etal.10,Fall.etal.10,Kim.etal.18,Krumholz.etal.19}. Radiation-hydrodynamic simulations of cluster-forming clouds under the dense (\ifm{\Sigma\gtrsim10^{3}\,\si{\Msun}\,\mathrm{pc^{-2}}}), metal-poor conditions expected at cosmic dawn find that direct radiation pressure on dust, rather than reprocessed infrared radiation, limits the star-formation efficiency and launches outflows from such clouds \citep{Menon.etal.22,Menon.etal.24,Menon.etal.25}, with Lyman-\ifm{\alpha} radiation pressure potentially contributing where dust is scarce \citep{Ferrara.etal.25,Menon.Smith.26}. The omission of radiation pressure may therefore cause \software{MAGE} to underestimate the strength of early stellar feedback in the dense, compact star-forming regions that host the rapid central star formation identified here. By analogy with the photoionization variants of \cref{subsec:phot}, stronger early feedback could lower stellar masses and enlarge radii at fixed halo mass. However, the same cloud-scale studies find that the star-formation efficiency remains high at low dust-to-gas ratios. The net effect on the transition mass identified here is therefore uncertain.

\textit{Numerical resolution.} The smallest stellar radii approach the numerical resolution, making the sizes of the compact low-mass galaxies of \cref{subsubsec:lowmasscompact} potentially sensitive to resolution. Resolution tests are needed to assess the sensitivity of the main results.

\textit{Correspondence to observed quantities.} The quantities presented here are intrinsic properties of the simulated galaxies and are not directly equivalent to observational measurements. The stellar half-mass radii are not directly observable, and the UV sizes we present here omit dust attenuation and instrumental effects such as PSF convolution and noise. Likewise, neither the star-forming gas disks nor the stellar disks in \cref{sec:diskformation} correspond directly to observationally inferred early disks. Such observed systems are typically traced by cool or ionized gas via emission lines such as [\ion{C}{ii}], [\ion{O}{iii}], \Halpha, and \ion{C}{iii}]. Measurements of these systems depend on aperture choices, beam smearing, and the adopted velocity-dispersion definitions. Consequently, our gas or stellar \ifm{v/\sigma} values should not be directly equated with observationally inferred gas \ifm{v/\sigma}. A more direct comparison with observations therefore requires additional post-processing: radiative-transfer calculations for the dust-attenuated UV light, emission-line modelling of the gas kinematics, and mock observations with matched apertures, PSF, source selection, and fitting procedures.

\textit{Gas flows and stellar redistribution.} Shell fluxes do not distinguish returning gas from fresh accretion and do not account for the motion of the aperture boundary. They therefore support a recycling interpretation without measuring individual recycling times or the complete mass balance of a moving region. The birth-radius statistics identify where stars form but do not quantify how subsequent orbital redistribution changes galaxy sizes. Tracking gas trajectories would constrain recycling, while following fixed stellar cohorts would quantify stellar redistribution.

\section{Summary}
\label{sec:summary}

We have investigated galaxy size evolution and disk formation over \ifm{9\lesssim z\lesssim15} using \software{MAGE}, a new suite of 11 cosmological zoom-in simulations performed with \software{RAMSES} at \ifm{\sim\SI{10}{pc}} spatial resolution using the star-formation and feedback model of \citet{Andalman.etal.25}. The target galaxies span \ifm{\Mx{\star}\simeq2\times10^{8}\text{--}3\times10^{10}\,\si{\Msun}} at \ifm{z=9}. Our main conclusions are as follows.

\begin{enumerate}[label=(\roman*),itemindent=1em,align=left]
\item Galaxy sizes increase steeply with stellar mass at low masses and evolve weakly with redshift at fixed mass. Near \ifm{\Mx{\star}\simeq\SI{5e9}{\Msun}}, the stellar half-mass size--mass relation turns over, while the UV half-light size--mass relation becomes approximately flat (\cref{fig:sizemass}). The ratio \ifm{\Rx{1/2}/\Rx{v}} reaches \ifm{\simeq0.02} near the transition and subsequently declines. Our intrinsic UV sizes overlap the compact subpopulation of M24 and the compact end of the observed size--magnitude distribution (\cref{fig:sizemass,fig:sizemag}), but do not reproduce the more extended observed galaxies.

\item In-situ stars dominate the central stellar mass, and the half-mass radius of the in-situ component closely tracks that of the full stellar population. At \ifm{z=9}, the in-situ fractions within \ifm{\Rx{1/2}} are \ifm{\gtrsim0.7} in all but \run{MAGE08}, which experienced a major merger within the preceding \ifm{\SI{50}{Myr}} (\cref{fig:proj_insitu,fig:Insitu}). Characteristic stellar birth radii remain near \ifm{0.02\,\Rx{v}} below \ifm{\Mx{\star}\simeq\SI{5e9}{\Msun}} and saturate around \SI{400}{pc} above this mass, suggesting a transition from size growth that tracks halo growth to stronger internal regulation (\cref{fig:sizegrowth,fig:sizebirth}).

\item Comparing the characteristic SN-feedback scale \ifm{\vx{SN}\simeq\SI{100}{km/s}} \citep{Dekel.Silk.86} with the halo and central velocity scales identifies two characteristic masses below the size turnover: the virial velocity exceeds \ifm{\vx{SN}} near \ifm{\Mx{\star}\simeq\SI{1e8}{\Msun}}, and the total circular speed at \ifm{\Rx{1/2}} exceeds it near \ifm{\Mx{\star}\simeq\SI{5e8}{\Msun}}. The outflow rates measured across \ifm{\Rx{1/2}}, \ifm{\Rx{gal}}, and \ifm{\Rx{v}} show a radial ordering consistent with increasing confinement of outward-moving gas as the potential deepens. The SFR within \ifm{\Rx{1/2}} remains at \ifm{\simeq0.3\text{--}0.55} of the galaxy-wide SFR until \ifm{\Mx{\star}\gtrsim\SI{2.5e10}{\Msun}}, where this fraction declines in \run{MAGE01}, accompanied by a decline in \ifm{\vx{c,gas}}. Together with the younger outskirts, these changes suggest the onset of inside-out quenching (\cref{fig:diskformation_centralsf,fig:Insitu}).

\item The rapid central stellar buildup drives \ifm{\vx{c,tot}(\Rx{1/2})} above \ifm{\vx{v}} near \ifm{\Mx{\star}\simeq\SI{5e9}{\Msun}} and coincides with the size turnover. Rotational support rises rapidly in the star-forming gas and young stars, with both the gas and stars younger than \SI{10}{Myr} reaching \ifm{v/\sigma\simeq1.5\text{--}3} at \ifm{z\simeq9} (\cref{fig:v2sigma_centrals,fig:diskformation_centralsf}). For the star-forming gas, the rise in \ifm{v/\sigma} is driven primarily by increasing rotation speed: \ifm{\vx{rot,SFgas}/\vx{c,tot}(\Rx{1/2})} increases from \ifm{\simeq0.4} to \ifm{\simeq0.7}, while its effective dispersion stays near \ifm{0.6\,\vx{v}}. The full stellar population remains dispersion-dominated. Together with the preceding central buildup, this joint structural and kinematic transition shares key features with wet compaction events at cosmic noon.

\item Varying photoionization feedback in the same halo changes both size and stellar mass by factors of a few. Weaker photoionization feedback produces a more massive, compact stellar body and increases \zyn{the rotational support of the star-forming gas from \ifm{(v/\sigma)_{\rm SFgas}\simeq0.8} to \ifm{\simeq1.6}}, indicating a correlation between central concentration and rotation in these experiments (\cref{fig:phot}).
\end{enumerate}

\Cref{fig:cartoon} and \cref{subsec:sequence} summarize the proposed evolutionary sequence, ordered by stellar mass, in which the deepening central potential increasingly confines SN-driven outflows. At cosmic dawn, GN-z11 combines evidence for compact central star formation \citep{ChenZuyi.etal.26,Nakane.Ouchi.26} with tentative evidence for rotational support from its \ion{C}{iii}] velocity gradient \citep{Xu.etal.24}. This makes it a promising candidate for exploring the connection between central stellar buildup and early disk formation. The inferred transition masses depend on the adopted feedback model and limited sample, while direct observational tests require matched modelling of dust attenuation and kinematic tracers, which we leave to future work. The Nancy Grace Roman Space Telescope's tiered High-Latitude Wide-Area Survey could identify tens to thousands of galaxies at \ifm{z\sim13} \citep{Yung.etal.23}, depending on the bright end of the UV luminosity function \citep{Harikane.etal.22}. Such galaxies could provide counterparts to our massive systems, and the brightest could be targets for JWST and ALMA follow-up of their sizes, star-formation profiles, and kinematics.

\section*{Acknowledgements}
We dedicate this paper to the memory of Avishai Dekel, our co-author, mentor and friend. May his memory be a blessing. We thank Fangzhou Jiang, Hyeonyong Kim, Sharon Lapiner, Jinning Liang, Rongrong Liu, Yurina Nakazato, Dylan Nelson, Anirudh Ravishankar, Xuejian Shen, and Jonathan Stern for helpful discussions. We thank the \software{VESTRAL} and \software{Lumina} teams for sharing results ahead of publication, and the \software{FirstLight} team for providing dust-free UV half-light radii. This material is based upon work supported by the National Science Foundation (NSF) and the
U.S.-Israel Binational Science Foundation (BSF) under Award Number 2406558 and Award Title ``The Origin of the Excess of Bright Galaxies at Cosmic Dawn''. ZY and NM acknowledge support from BSF grant 2022736 (from the NSF-BSF program) and from BSF grant 2022281. The authors gratefully acknowledge the Scientific Computing Core at the Flatiron Institute for providing the high-performance computing resources and storage that enabled the research results reported within this paper. The Flatiron Institute is supported by the Simons Foundation. Artificial intelligence tools assisted with code development, language editing, literature verification, and the rendering of the artwork in the conceptual schematic of \cref{fig:cartoon}, whose content, layout, and annotations were specified and verified by the authors. The authors retain responsibility for the scientific analysis, interpretation, conclusions, and accuracy of the manuscript.

\section*{Data Availability}

The simulation data underlying this article will be shared on reasonable request to the corresponding author.



\bibliographystyle{mnras}
\bibliography{SizeEvolution} 




\appendix

\bsp	
\label{lastpage}
\end{document}
